\documentclass[PhD,oneside,english]{sapthesis}

\title{The hard-sphere model of strongly interacting  fermion systems}
\author{Angela Mecca}
\IDnumber{696270}
\course[Physics]{Fisica}
\courseorganizer{Scuola di Dottorato Vito Volterra}
\cycle{XXVIII}
\submitdate{December 2015}
\AcademicYear{December 2015}
\copyyear{2015}
\advisor{Prof. Omar Benhar}
\coadvisor{Dr. Alessandro Lovato}
\authoremail{lelamecca@gmail.com}
\examdate{22 January 2016}
\examiner{Prof. Luciano Maria Barone}
\examiner{Prof. Cinzia Da Vi\`{a}}
\examiner{Prof. Teresa Rodrigo}

\usepackage[pagebackref]{hyperref}
\hypersetup{colorlinks=true,
	linkcolor=blue,
	anchorcolor=blue,
	citecolor=blue,
	urlcolor=blue,
	pdftitle={PhD Thesis},
	pdfauthor={Angela Mecca}
	backref=true,
	}
\usepackage{verbatim}    

\usepackage[english]{babel} 
\usepackage[fixlanguage]{babelbib}
\selectbiblanguage{english}

\usepackage{cite}
\usepackage{feynmf}
\usepackage{mathrsfs}
\usepackage{amssymb}
\usepackage{pstricks}
\usepackage{color}
\usepackage{enumitem}
\usepackage{subfig}
\usepackage{fnbreak}

\DeclareMathAlphabet{\mathpzc}{OT1}{pzc}{m}{it}

\def\lsim{\buildrel < \over {_{\sim}}}

\newcommand{\myc}{  k_F a }
\newcommand{\mycs}{ (k_F a_0) }
\newcommand{\mycp}{ (k_F a_1) }

\newcommand{\ketcbf}[1]{ \left |  \left . {#1} \right. \right ) }
\newcommand{\bracbf}[1]{ \left (  \left . {#1} \right . \right | }
\newcommand{\braketcbf}[2]{ \left (  {#1}  |  {#2} \right) }

\newcommand{\veff}{v_{\rm eff}}
\newcommand{\RE}{\operatorname{Re}}
\newcommand{\IM}{\operatorname{Im}}

\begin{document}

\frontmatter

\maketitle
\abstract

The formalism based  on Correlated Basis Functions (CBF) and the cluster-expansion technique 
has been recently  employed to derive an effective interaction  from a realistic nuclear Hamiltonian.
One of the main objectives of the work described in this  Thesis is establishing the accuracy of  
this novel approach\textemdash that allows to  combine the flexibility of perturbation theory in the basis of eigenstates 
 of the noninteracting system with a realistic description of short-range correlations in coordinate space\textemdash 
 by focusing on the  hard-sphere fermion  system.

 The properties of the hard-sphere fluid\textemdash
which has long been recognized as a valuable model for investigating concepts and 
approximations employed in the study of strongly correlated systems, whose structure and dynamics 
are largely driven by the presence of a short-ranged and strongly repulsive interaction\textemdash
have been extensively  analyzed 
 within perturbative approaches yielding exact results in the low-density limit. 
To gauge the reliability of CBF effective interaction scheme, we have performed a systematic comparison between the results of   its application  to the Fermi hard-sphere systems and the predictions obtained
 form low-density expansions, as well as form other many body techniques.


As a first application of the formalism, the quasiparticle properties of hard spheres 
of degeneracy four have been  determined from the  two-point Green's function. 
The calculation has been performed carrying out a  perturbative expansion
of the self-energy, up to the second order in the CBF effective interaction.
The main results of this study are the  momentum distributions, the quasiparticle spectra and their description in terms of effective mass. 
The analysis of these properties shows that the effective interaction approach is quite accurate, 
thus suggesting that it may be employed to achieve a consistent description of the structure and the dynamics of 
nuclear matter in the density region relevant to astrophysical applications.
 
The investigation of the hard-sphere fermion fluid has been extended  to study the shear viscosity and thermal 
conductivity coefficients of the system with degeneracy two, that can be regarded as a model of pure 
neutron matter. The resulting transport coefficients, evaluated taking into account perturbative contributions up to  second order in the CBF effective interaction,  show a strong sensitivity to the quasiparticle effective mass, 
reflecting  the effect of second order contributions to the self-energy which are not taken into account in nuclear matter studies available in the literature. The  difference between first and second order results   is likely to play an important role in astrophysical applications and needs to be carefully investigated extending our analysis to nuclear matter.

\tableofcontents

\include{intro}

\mainmatter

\chapter[Theory of strongly interacting fermion systems]{\huge Theory of strongly interacting fermion systems }
\label{chap1}

The theoretical treatment of the many-body problem, which amounts to describing a quantum mechanical system of $N$ interacting non relativistic particles,
involves daunting challenges.  The Schr\"odinger equation associated with the Hamiltonian
\begin{equation}
H = \sum_{i } ^{} t (i) + \sum_{j > i } ^ {} v (ij) \ , 
\end{equation}
where $t(i) = -\nabla_i^2/2m$, $m$ being the particle mass,  is  the kinetic energy operator,  and  $v(ij)$ denotes the interaction potential,
can be solved exactly, using deterministic methods, only for $N<4$ and selected interactions.

In addition to the computational issues associated with the treatment of a large number of particles, in strongly correlated systems---such as liquid helium and
nuclear matter---one has to confront the difficulties arising from the strongly repulsive nature of the interaction, which makes the use of 
standard approximation methods, based on perturbation theory, highly problematic. In order to make perturbative calculations feasible, one has 
to either replace the bare potential with a well behaved interaction, such as the scattering matrix, or replace the basis of eigenfunctions of  the 
non interacting system with a basis of states suitably modified to take into account the effects of the repulsive core of the potential. In this chapter, we provide a brief 
outline of both approaches.

Section \ref{sec:lowdensity} will be devoted to the discussion of the perturbative methods, leading to low-density expansions which have been widely 
employed to study the properties of  the fermion hard-spheres  system. The alternative approach based on formalism of correlated basis functions,  
and the cluster expansion technique needed to compute matrix elements involving correlated states, will be  discussed in Sections \ref{sec:CBF}  and \ref{sec:cluster}, 
respectively.
 
 
\section{Low-density expansions}
\label{sec:lowdensity}

A low-density, a Fermi gas consists of  point-like spin one-half particles interacting via a  strongly repulsive pair potential, which can be chosen  to be the infinite hard-core potential defined in the Introduction.
The restriction to a purely repulsive interaction prevents the possible formation of Cooper pairs, leading to the appearance of a superconducting or superfluid phase.
The hard-sphere model provides a reasonable approximation of systems such as nuclear matter and liquid $^3$He over a broad range of density, and has been analysed by 
several authors using different methods.  
 
A seminal study was carried out in the 1950s by Huang, Lee and Yang within the framework of Fermi's pseudopotential method, which amounts to solving a Schr\"odinger equation for the wave function, in which the interaction is replaced  by suitable boundary conditions \cite{HuangYang1957,LeeYang1957}.  These authors obtained an expression for the ground state energy in  terms of the low-energy parameters, \emph{e.g.} scattering length and effective range. 

In his pioneering work of Ref. \cite{Galitskii1958}, Galitskii applied the methods of quantum field theory to determine the energy spectrum of the system, \emph{i.e.} the energy and lifetime of quasiparticle states, described by the complex poles of the two-point Green's function. 
The approach of Ref.\cite{Galitskii1958} is based on an expansion in powers of the dimensionless parameter $c = k_F a$, where $k_F$ is the Fermi momentum---trivially related 
to the particle density $\rho$ through $k_F=(6 \pi^2 \rho/\nu)^{1/3}$, $\nu$ being the degeneracy of the momentum eigenstates---and  $a$ is the hard-core radius\footnote{In the original formulation,  the dimensionless parameter was defined as the product between $k_F$ and the  real part of the scattering amplitude at small 
momentum, $f_0$, which for the hard-sphere systems is given by the hard-sphere radius $a$.}.
The resulting ground state energy and the spectrum of quasiparticles of degeneracy $\nu = 2$ carrying momenta close to the Fermi momentum,  computed including terms of order up to $c^2$, 
are given by
\begin{align}
E_0  = \frac{3}{5}    \frac{k_F^2}{2m} \left[  1 + \frac{10}{ 9 \pi}c + \frac{4}{21 \pi^2} \left( 11 - 2 \ln 2 \right) c^2  \right] \ ,
\end{align}
and 
\begin{equation}
\label{eq:speGal}
\frac{e(k)}{k_F^2} =  \frac{1}{2} x^2 + \frac{2}{3 	\pi} c + \frac{2}{15 \pi^2 } c^2  (11 - 2 \ln 2) - \frac{8}{15 \pi^2} c^2 (7 \ln 2 - 1) (x-1)  \ ,
\end{equation}
where $x = k / k_F$. The  knowledge of the single particle spectrum  allows one to obtain the effective mass, defined as
\begin{align}
\label{def:mstar}
m^\star(k) =  \left( \frac{1}{k} \frac{ d e}{d k} \right)^{-1} \ .
\end{align}
At second order in $c$ one finds
\begin{align}
\label{mstar_gal}
\frac{m^\star(k_F)}{m} = 1 + \frac{ 8 }{ 15 \pi^2 } c^2 ( 7 \ln 2 - 1 )  \ .
\end{align}
 Similar expressions  have been derived  for all quasiparticle properties, ranging from the lifetime to the chemical potential and the Green's function renormalisation constant 
 $Z$, to be identified with the discontinuity  of the momentum distribution at the Fermi surface \cite{migdal1957}. These results agree with  those obtained by  Abrikosov and Khalatnikov in Ref. \cite{AbriKala1958}.

The same procedure has been applied   in Ref. \cite{czyz,belyakov}  to  obtain the expansion of the momentum distribution, describing the occupation probability of single particle states of momentum $k$, for  a broad range of  momenta. The analytic results of Ref.\cite{belyakov}, including terms of order up to $c^2$,   have   been carefully   investigated by Mahaux and Sartor in 1980 \cite{mahaux, mahaux_erratum}. The explicit expression reported in  Ref.~\cite{mahaux, mahaux_erratum} can be found in Appendix \ref{nkApp}.

In the 1970s, Bishop carried out a systematic analysis of the existing results using different computational schemes \cite{Bishop1973}.  
He analysed  two-body scattering \emph{in vacuum}, as well as scattering in the presence of  a filled Fermi sea using both Goldstone time ordered diagrams 
(Goldstone method) and Feynman diagrams (Green's function method). The expression of the ground state energy was obtained including the first four terms 
of the expansion in powers of the  dimensionless parameter $c$.
 It turned out that the first three terms can be completely derived in terms of the low-energy scattering parameters \cite{Galitskii1958,AbriKala1958,DeDominicisMartin,EfimovAmusia}, while the effect of including three-particle collisions is the appearance of a term  logarithmically dependent on  $c$ \cite{Baker1965, Efimov1965,Efimov1966,AmusiaEfimov1968}.
 
 The analysis of  two-body scattering in free space demonstrates that for  highly repulsive interactions the perturbation series  in powers of the potential  requires a large number of terms to  be at best  asymptotically  convergent,  as shown by the author of Ref. \cite{Baker}. 
 However, in the presence of a singular hard-core potential this procedure is not viable.
In analogy with scattering in free space, where the bare interaction is replaced by the $t$-matrix describing the entire Born series of  multiple scattering processes,  
 the formalism for the calculation of the ground-state energy requires the  rearrangement of the perturbation series in terms of suitable new operators. 
 These are the $K$- and $T$-matrix,  representing   the sum of \emph{ladder} diagrams associated with scattering in the  Fermi sea,  obtained using  time ordered (Goldstone)   or standard (Feynman) perturbation theory,  respectively.

\begin{figure}[htbp]
\begin{center}
\includegraphics[width=0.8\linewidth]{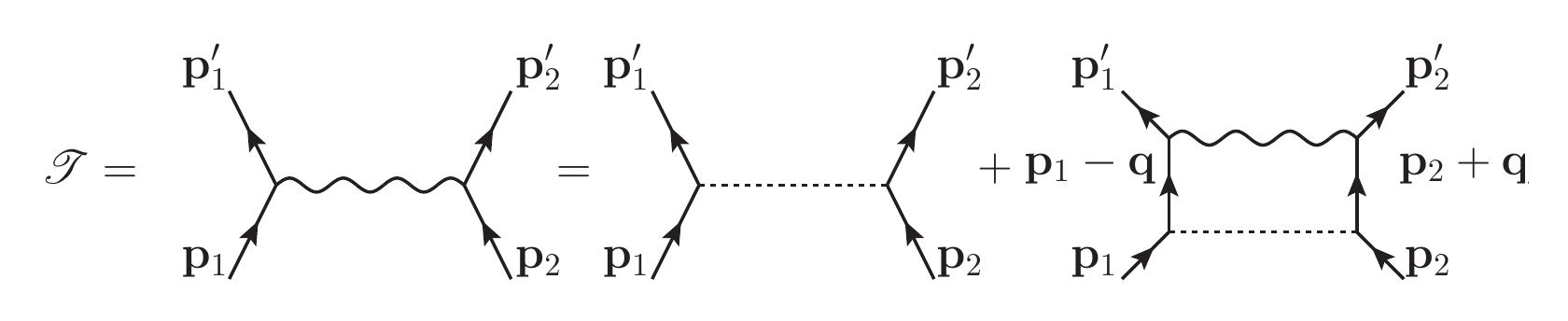} 		
\caption[Diagrammatic  representation of the $\mathscr{T}$ matrix]{Diagrammatic representation of the generic $\mathscr{T}$ matrix, sum of  \emph{ladder} diagrams in free space or in the Fermi sea.}
\label{fig}
\end{center}
\end{figure}
	
The diagrammatic representation of the generic matrix $\mathscr{T}$ (hereafter, $\mathscr{T}$ denotes $t$, $K$, or $T$) is illustrated in  Fig.  \ref{fig}.
 The only difference between the three cases arises from  the interpretation of  the internal lines, representing the particle propagators, which determines the  explicit form of the integral equation defining the three matrices. 
 
 In free space the propagators are written as
 \begin{subequations}
 \begin{equation}
\mathcal{G}^{0 }  _{ free } (p) = \left( E -  {\bf{p} }^2 - i \eta \right)^{-1} \ , 
 \end{equation}
 with $\eta=0^+$. If the  presence of the filled Fermi sea is taken into account,  the Goldstone  propagator, describing particles outside the Fermi sea, reads
\begin{equation}
\mathcal{G}_{  G }^{0} ( p ) =  \theta(p - k_F )\left( E -  {\bf{p} }^2 - i \eta \right)^{-1} \ , 
\end{equation} 
while  the internal lines in  Feynman  diagrams, describing both particles outside the Fermi sea and holes inside the Fermi sea, correspond to
\begin{equation}
\mathcal{G}_{ F } ^{0} (p) =  \left[ E -  {\bf{p} }^2 + i \eta \  {\rm sgn} (p-k_F) \right]^{-1} \ .
\end{equation} 
\end{subequations}

Summing up  \emph{ladder} diagrams in free space is equivalent to solving the  Lippman-Schwinger equation.  
The   same procedure with time-ordered  and Feynman diagrams leads to the Bethe-Goldstone and Bethe-Salpeter equations, respectively. These equations have both  exact solutions, well behaved even in the case of singular interactions, provided the potential is replaced by the $t$-matrix.

The relevant $\mathscr{T}$  matrices are the key elements of the diagrammatic perturbative approach for the calculation of the ground-state energy. To derive the expansion of the energy  to  order $c^3$, diagrams containing up to three $\mathscr{T}$ interactions must be taken into account. In such diagrams no more than two particles are interacting at a given time and only two body scattering parameters are involved in the final expression, while in diagrams with more than four $\mathscr{T}$ interactions intermediate states with three or more than three particles are allowed.

 Some of the diagrams containing more than three $\mathscr{T}$ interactions involve divergent integrations.  Since the singularities  come
from high momenta  ($k \gg k_F$) and the hole lines  are restricted to momenta less than $k_F$, 
 the leading  divergence will show up  in the
 diagrams containing the maximum number of particle lines.
It turns out that diagrams containing four $\mathscr{T}$ interactions and minimal number of hole lines are logarithmically  dependent on the density of the system, or equivalently on the parameter $c  = \myc $.

The final result for the  first four terms of the expansion  of the ground-state energy in the presence of a  generic repulsive potential  can be expressed in terms of 
 the two body scattering parameters $a_0$ and   $a_1$, describing the $S$-  and $P$-wave scattering lengths,  and the $S$-wave effective range $r_0$ \cite{collisionbook}
  \begin{align}
 \frac{E}{N} =  &  \frac{k_F^2}{2 m}
  \left \{   \frac{3}{5} + \left( \nu -1 \right )  
 \left[  \frac{2}{3 	\pi } \mycs + \frac{4}{35 \pi^2}  \left ( 11 - 2 \log 2 \right ) \mycs^2   
  \right . \right.  \nonumber \\
 \left.    \left . \right. \right . &  \left.    \left .  
 +  \frac{1}{10 \pi } \left(k_F r_0 \right) \mycs^2 +   \right (  0.076 + 0.057 \left ( \nu - 3 \right) \left ) \mycs^3  + \frac{1}{5 \pi} \left ( \nu + 1 \right ) \mycp ^3     \right]  
 \right.    \nonumber \\
&  \left.  \left .   
+ \left (\nu -1 \right ) \left (\nu - 2  \right ) \frac{16}{27 \pi^3} \left (4 \pi - 3 \sqrt{3}  \right) \mycs^4 \log \mycs  + \ldots \ .
\right . \right \} \ \ , 
 \end{align}
For the hard-sphere case, where the scattering lengths and the effective range  are related to the hard-core radius $a$ through the relations $a_0 = a_1 = a$ and  $r_0 = 2a/3$, the final expressions for the energy per particle, in terms of the dimensionless parameter $c  \equiv     \left ( k_F a \right) $,  read

\begin{subequations}
\begin{alignat}{5}
\label{eq:E0nu2}
 \nu = & 2 \ \ :  \ \ \   &&        \frac{E}{N}  =  \frac{k_F^2}{2 m} \left [  \frac{3}{5}    +   \frac{2}{3 \pi } c  \right.  &&    \left.+       \frac{4}{35 \pi ^ 2 }       \right.  &&    \left.    \left ( 11 - 2 \log 2  \right ) c ^ 2+ 0.230 c^3       \right.  &&    \left. + O (c^4)\right ]  \ \ , 
\nonumber \\
\\
\label{eq:E0nu4}
 \nu = & 4 \ \ :  \ \  \ &&      \frac{E}{N} =  \frac{k_F^2}{2 m} \left [  \frac{3}{5}   +    ^{} \frac{2}{  \pi } c^{}   \right.  &&    \left. +       \frac{12}{35 \pi ^ 2 }        \right.    &&  \left.       \left ( 11 - 2 \log 2  \right ) c ^ 2   +  0.780 c^3      \right.  \nonumber \\ 
   \left.      \right.  &  && && &&  \left.+   \frac{32}{9 \pi^3} \left  (4 	\pi - 3 \sqrt{3}  \right)  c^4 \log  c  \right.  &&    \left.+  O (c^4)   \right ] 
\ \ . 
 \nonumber \\
 \end{alignat}
\end {subequations}

In the above expansions,  the linear term describes the effects of forward scattering, the quadratic term takes into account Pauli's exclusion principle and the higher-order terms arise  from the occurrence of processes involving al least three particles. 
Note that the logarithmic term vanishes identically for $\nu < 3$. This term, in fact, arises from three body correlations at high momentum, or small relative distance.  
Because for a system of degeneracy $\nu = 2$
at least two of the three particles must be identical,  three body correlations at small relative distances can only appear at higher degeneracy,  $\nu > 2$.

As for the effective mass, the generalization  of Eq. \eqref{mstar_gal} for a system of degeneracy $\nu$ can be written as
\begin{align}
\label{pert_mstar}
\frac{m^\star(k_F)}{m} = 1 + \frac{ 8(\nu - 1) }{ 15 \pi^2 }  c^2 ( 7 \ln 2 - 1 ) \ . 
\end{align}

\section{Correlated Basis Functions formalism}
\label{sec:CBF}

Conceptually, low-density expansions are perturbative solutions of the many-body problem. 
A well established alternative scheme, originally proposed by Jastrow in 1955 \cite{jastrow},  is based on a variational treatment in which the trial 
ground-state wave function is built using two-body {\em correlation functions}. 

The idea underlying Jastrow's approach is that, in the presence of a potential exhibiting a strongly repulsive core of radius $r_c$, 
the ground state wave function, $\Psi_0$, must be such that,  for any $i,j \in \{ 1, \ldots, N \}$
\begin{align}
\label{jastrow_0}
r_{ij} = |{\bf r}_i - {\bf r}_j| < r_c  \Longrightarrow | \Psi_0({\bf r}_1,\ldots,{\bf r}_N)|^2 \approx 0   \ ,
\end{align}
implying that the probability of finding any two particles within a distance $r_c$ of one another is negligibly small. The above condition can be 
easily fulfilled writing the trial wave function in the form
\begin{align}
\label{jastrow_1}
\Psi_0({\bf r}_1,\ldots,{\bf r}_N) \propto \prod_{j>i=1}^N f(r_{ij}) \ ,
\end{align}
where the two-particle correlation function is defined in such a way that $f(r) \approx 0$ at $r < r_c$.

In principle, the expectation value of the Hamiltonian in the state described by the {\em correlated} wave function  provides an upper bound to the ground state energy 
of the system. However, the corresponding $3N$-dimensional integration is not factorisable into integrals involving the coordinates of only one-particle. Therefore, its calculation involves severe difficulties and, in general, requires the use of approximations that may spoil the upper bound property of the result.

In spite of  the above difficulty, the variational approach and the formalism based on correlated wave functions have reached a remarkable degree of accuracy, and 
have been widely and successfully employed to study the properties of a variety of  interacting many-body systems, from liquid helium  to neutron star matter.

In the following sections we will briefly discuss the elements of the variational treatment and its generalisation, based on an extension of the Jastrow
{\em ansatz} allowing one to build  a complete set of correlated states. The main tenet underlying this approach, the validity of which needs to 
be thoroughly investigated, is that the correlation structure of the ground and excited states be the same.  

\subsection{Variational method}

 A variety of many body systems, e.g. liquid helium and atomic nuclei, 
are characterised by strong correlations between their constituents,  that cannot be taken into account within the mean field (MF) approximation 
at the basis of the independent particle model.  For example, there is ample experimental evidence 
that, as pointed out  by the authors of Ref. \cite{correlations},  nucleon-nucleon correlations lead to a sizeable depletion of the occupation probability of the shell model orbitals. 

The correlated ground-state wave function is defined through the transformation (see, e.g., Ref.~\cite{Feenberg})
\begin{equation}
\label{CBF:GS}
 \ketcbf{ 0}     \equiv     \frac{ F | \Phi_0 ] }{ [ \Phi_0 | F^{\dagger} F | \Phi_0 ] ^ {1/2} } \ ,
\end{equation}
where   the {\em model function} $ | \Phi_0 ] $ describes the system ground-state in the absence of correlations, 
which are taken into account by the operator $F$.
In translationally invariant fermion systems $|\Phi_0 ]$  is a Slater determinant of single particle states comprising a plane wave and Pauli spinors 
describing the spin and  isospin degrees of freedom. All energy levels corresponding to $|{\bf k}| < k_F$, belonging to the Fermi sea,  are occupied with unit probability.

The accuracy of the variational estimate provided by the expectation value 
\begin{equation}
\label{eq:E0var}
E^{v}_0 = \bracbf{0}  H \ketcbf{0}  \ , 
\end{equation}
depends on the choice of the correlation operator determining the form of the trial wave function. Its main role is producing an excluded region in configuration  
space,  in which the particles penetrate with small probability because of the strong repulsive core of the interaction potential.
 
The operator $F$ is usually defined as the  product of pair correlation operators, $F_{ij}$, according to
\begin{equation}
F= \mathcal{S} \prod_{j>i} F_{ij}\, .
\end{equation}
The structure of $F_{ij}$ must reflect the properties of the potential. Hence, for spin-isospin dependent interactions $[F_{ij},F_{ik}] \neq 0$, and 
the right hand side of the above equation needs to be properly symmetrized  through the action of the operator $\mathcal{S}$.
In the case of spherically symmetric and spin-isospin-independent interactions, on the other hand, the two-particle correlation function depends on the interparticle distance only,  and Eq.\eqref{jastrow_1} is recovered.
The simple radial correlation function is usually referred as  (Bijil-Dingle)-Jastrow correlated wave function, or DBJ \emph{ansatz}  \cite{jastrow,Bijl,Dingle,  drell_huang}.
Its shape  is often determined by functional minimization of the expectation value  of the Hamiltonian in the correlated ground state.

As mentioned above,  however, the calculation of $E^{v}_0$ 
involves serious difficulties.  For a system of $N$ particles, the right side of  Eq.\eqref{eq:E0var} includes $3N$-dimensional 
integrations, whose evaluation requires a computational effort that rapidly increases with $N$. A viable option to overcome this problem is the use of the
Variational Monte Carlo method (VMC) \cite{VMCHe4,VMCfermion}, a stochastic technique allowing---at least in principle---to perform exact  calculations of the 
Hamiltonian expectation value. However, in practice VMC is affected by the intrinsic uncertainty arising from the fact that the system is modelled as a collection of a 
 finite number of particles enclosed in a box of finite size.
 
Very large uniform systems are often treated within a formalism derived from the approach originally proposed by Jastrow \cite{jastrow}, which
essentially amounts to  expanding the Hamiltonian expectation value in powers of the particle density. The terms of the resulting series can be conveniently 
represented by diagrams, that can be classified according to their topological structure and  summed up to all orders solving a system of
integral equations dubbed Fermi Hyper-Netted Chain (FHNC) equations. While not being exact---because it does not included all 
topological classes---the FHNC summation scheme has been shown  to provide accurate results for many strongly-interacting fermion systems  \cite{fhnc1,fhnc2}.

\subsection{Correlated Basis Functions (CBF)}

The Correlated Basis Functions (CBF) approach \cite{fabrocinifantoniCBF,clarkwesthausCBF} extends the basic idea underlying the variational 
method, and uses the correlation operator $F$ determined by the variational calculation of $E^{v}_0$ to generate not only the correlated ground state, but a complete set of  
basis functions spanning the $N$-particle  Hilbert space. The transformation
\begin{equation}
\label{eq:correlstat}
\ketcbf{ n}  \equiv     \frac{ F | \Phi_n ] }{ [ \Phi_n | F^{\dagger} F | \Phi_n ] ^ {1/2} }   \ .
\end{equation}
establishes a direct correspondence  between   the  true  excited states of the system $\ketcbf{n} $ and the uncorrelated states  $ | \Phi_n ]  $ 
 constructed by moving $n$ particles from states belonging to the Fermi sea to excited states, corresponding to momenta larger than $k_F$. 
These states are referred to as $n$-particle-$n$-hole ($n$p-$n$h) states. 
\enlargethispage{\baselineskip}
By using the second quantization formalism, creation and annihilation operators of correlated states can be defined 
and the full Fock space of correlated states can be constructed, with $n$p-$n$h correlated states  mapped onto the $n$p-$n$h
uncorrelated ones\footnote{
Following Ref. \cite{krotscheck}, creation and annihilation operators of correlated states $\alpha^{\dagger}_k  ,  \alpha^{}_k$   are defined by their action  on the basis states 
$ | m )$ through the following relations
\begin{align}
\label{eq:creation}
\alpha ^{\dagger}_k   | m ) =  \frac { F_{N+1} a^{\dagger}_k  | \Phi_m ]  }   {  [ \Phi_m  | a^{}_k  F^{\dagger}_{N+1} F^{}_{N+1} a^{\dagger} _ k | \Phi_m ] ^{1/2}  }   \ \ \ , \ \ \ 
\alpha ^{}_k  | m )                =  \frac { F_{N-1} a^{}_k  | \Phi_m ]  }   
{  [ \Phi_m  | a^{\dagger}_k  F^{\dagger}_{N-1} F^{}_{N-1} a^{} _ k | \Phi_m ]^{1/2}  } \ .
\end{align}
They obey the same anticommutation rules as the uncorrelated ones $a^{}_k ,  a^{\dagger}_k$, but are not Hermitian conjugates of one another, as the  states in Eqs. \eqref{eq:creation}  
involve a $(N+1)$ and a $(N-1)$ correlation operator, respectively. As a consequence the number operator is  not hermitian and its eigenstates corresponding to  different eigenvalues,  \emph{i.e.}  the Fock states,  are not mutually orthogonal. }.

The diagonal matrix elements of the Hamiltonian between correlated states
\begin{equation}
\label{eq:env}
H_{nn} \equiv  \bracbf{  n}   H  \ketcbf{n}   \equiv E _ n ^ v,
\end{equation}
are referred to  as variational energies, $E^v_n$, although only the ground state  energy has been  variationally estimated.  They all  are  of order $N$, while the excitations energies $E_{n}^ {v}  - E_0^ v $ are of order $1$.

\subsection{CBF perturbation theory}

The variational energies of Eq.~\eqref{eq:env},  the off-diagonal elements of Hamiltonian  
\begin{align}
H_{nm} & \equiv    \bracbf {n}  H  \ketcbf{ m }   \ ,
\end{align} 
and the off-diagonal elements   of the unit operator, or metric matrix
\begin{align}
\label{eq:metric}
N_{nm} &\equiv    \braketcbf{n}{m}    	\ ,
\end{align}
are the building blocks of  CBF perturbation  theory. 
This approach is based on the observation that, if the correlation function is determined in such a way that correlated states
have large overlaps with the eigenstates of the Hamiltonian,   $H$ is nearly diagonal in the correlated basis. 
As a consequence, the off-diagonal matrix elements can play the role of small parameters of a perturbative expansion. 

Following the scheme of standard perturbation theory,  the Hamiltonian is decomposed into  the sum of unperturbed  and interaction contributions,
defined in terms of their matrix elements according to  
\begin{equation}
H = H_0 + H_I  \ , 
\end{equation}
with
\begin{align}
\label{eq:H0nm}
{H}_{0, nm}   =   \bracbf {n}  H_0  \ketcbf{ m }   \equiv  \bracbf {n}  H   \ketcbf{ n }   \delta_{nm}     =   E^{v}_n \   \delta_{nm}    \ ,
\end{align}
and 
\begin{equation}
\label{eq:HInm}
{H} _ {I, nm} =  \bracbf{n} H_I \ketcbf{ m } =  \left ( 1 - 	\delta _{nm } \right )   H_{nm} \ .
\end{equation}

If the correlated states are close to the true eigenstates of the Hamiltonian, the quantities 
\begin{equation}
\label{eq:hnm}
W_{mn} (n)  =  \bracbf{m}   H - E_n \ketcbf{n} = H _{mn } - E_n N_{mn} \ ,
\end{equation}
can be treated as a perturbation. The resulting expansion  can be employed to calculate  corrections to the variational estimate of the energies, 
which in this context can be seen as zeroth-order  approximations.

The energy shift between CBF and variational energies can be written in the form \cite{fabrocinifantoniCBF}
\begin{align}
\label{eq:deltaen}
\Delta E _ n = &  \  E ^{} _ {n} - E^{v}_{ n}    \nonumber \\ 
 = & \sum _ {p \neq n }  \frac{ W_{np} (n) W _{pn} (n) } { E_n - E^v_p }  + \sum_{p \neq q \neq n }  \frac{ W_{np} (n) W _{pq} (n)  W _{qn} (n) } {  ( E^{} _n - E^v_p)(E^{}_n - E^v_q) } + \ldots \  . 
\end{align}

Substitution of
\begin{align}
W_{mn} (n)  =&  \  H _{mn } - E^v_n N_{mn}   - \left (  E^{}_{n } - E^{v}_n  \right ) N_{mn}  \equiv  W^v_{mn} (n) - \Delta E ^v_n N_{mn}  \ ,  
\end{align}
and 
\begin{align}
\frac{1}{E_n - E ^v_ p} = &\    \frac{1}{E^v_n - E ^v_ p + \Delta E^v_n}  = \frac{1}{E^v_n - E ^v_ p} \sum _{i}   \left( -   \frac{    \Delta E _n} {E^v_n - E ^v_ p} \right)^i , 
\end{align}
in Eq.~\eqref{eq:deltaen} allows one to identify  perturbative corrections order by order. 
Note that the appearance of an additional dependence on $E_n$ in  the perturbative series, besides the one arising from the energy denominators, is a peculiar
feature of perturbation theory in a non-orthogonal basis. 

CBF perturbation theory has been extensively applied to  nuclear matter, to obtain  
second order corrections to the ground state  energy \cite{fantonifrimanpandha_eCBF, fabrocinifantoni_eCBF},  the real and imaginary part of the energy dependent optical potential \cite{fantonifrimanpandha_eCBF, fabrocinifantoni_eCBF},  the momentum distribution \cite{momentumCBF} and the two-point 
Green's function \cite{spectral,green}.  A major difficulty associated with these calculations  is the presence of spurious terms, arising from the non orthogonality
of the basis states. A fully consistent---although quite demanding from the computational point of view---procedure to generate a basis of orthogonalised correlated states has been developed in  Ref. \cite {fantonipandhaOCB}.

\section{Cluster expansion formalism}
\label{sec:cluster}

The use of correlated states and the resulting CBF formalism entails the issue of the numerical calculation of matrix elements of many body operators
in the correlated basis. The problem of evaluating  multidimensional integrals  for a large number of particles has been effectively dealt with in the study of the classical imperfect gas, in which the partition function is developed in cluster integrals, each one defined on a subsystem of increasing number of particles. If the density is not too high and the correlation range is short enough, an accurate evaluation of, e.g.,  the energy of the system can be obtained in terms of few clusters.
Having this picture in mind, in this section we outline the main elements of the  cluster expansion formalism and discuss its application to the evaluation of the energy. 

Both the diagonal and off-diagonal elements of the Hamiltonian and metric matrices are built form  matrix elements of the operators $F^{\dagger} H F ^{}$ and  $F^{\dagger}  F ^{}$ between uncorrelated Fermi gas states.
We assume that the correlation operator $F$ is symmetric in its argument and translationally invariant. In addition, because of the short-range nature of the interaction,  it is required to exhibit the  cluster decomposition property, implying that 
if any subset, say $i_1 \ldots i_m$,  of the particles is moved away from the rest,
$F$ decomposes into a product of two factors according to 
\begin{equation}
F ( 1 	\ldots N ) = F_m (i_1 \ldots  i_m) F_{N-m} ( i_{m+1} \ldots i_N)  \ . 
\end{equation}

If, for the sake of simplicity, we limit ourselves to the case of Jastrow correlations, the operators 
$F^{\dagger} H F ^{}$ and  $F^{\dagger}  F ^{}$  can expanded in terms of the quantity $h(r_{ij})
$\footnote{In nuclear matter, where  the correlation operators $F_{ij}$ and $F_{ik}$ do not commute,  one needs to carefully take into account the ordering of correlations.} 
\begin{equation}
\label{eq:hcorrel}
h(r_{ij} ) = f^2(r_{ij}) - 1  \ ,
\end{equation}
 and the one-body density matrix
 \begin{equation}
 \label{eq:onebodydens}
 \rho (i, j) =   \sum_{n \in \{F\}} \phi^{\star}_{ n} (i) \phi^{}_{ n} (j)  \ .
\end{equation}
where $\phi^{}_{ n} $ are the single particle wave functions and the sum is restricted to the occupied states within the Fermi sea. 
The label $i$ refers to both space and spin coordinates. 

The cluster expansion formalism and its application for the calculation of the ground state energy will be discussed in the next section.
Note that 
we will use  {\em non normalized}  correlated states $ \ketcbf{n} = F | \Phi_n ]$.

\subsection{Ground-state energy}
\label{gse}

The main quantity needed to obtain the cluster expansion of the ground state energy is the generalized normalization integral,  
defined as
\begin{equation}
 I   \left ( \beta \right ) = \bracbf{0} \exp \left [ \beta \left(   H - T _ F  \right) \right]  \ketcbf{0} \ ,
\end{equation}
where $T_F$ is the ground state energy of the non interacting Fermi Gas. 
The desired energy expectation value may be recovered via the well-known  formula
\begin{equation}
\label{eq:E0}
 E_ 0  =  T_F +   \left . \frac{\partial }{ \partial \beta }   \log  I ( \beta)     \right | _ {\beta = 0} \ .
\end{equation} 
This first derivation---referred to as IY cluster expansion---was carried out  by Iwamoto and Yamada rearranging the terms of the expansion according to the powers  of 
the  smallness parameter $\eta $, defined as  \cite{YwaIama}
\begin{equation}
\eta = \frac{1}{N} \sum_{j>i}  [ ij | F_2^{\dagger}(12) F^{}_2(12) - 1 | i j ] _ a \ , 
\end{equation}
$F_2(ij)$ being the two-particle correlation operator.
Hereafter, the symbol   $| i ]$ will denote a  Fermi gas state, and $| i j ] _ a =( | i j ] - | ji ] )/\sqrt{2}$.  

We will focus on a slightly  different formulation, known as Factorized-Ywamoto-Yamada (FYI)\cite{clarkwesthausCBF}, where the $n$-th  cluster collects all contributions involving, in a {\em linked} manner, $n$ Fermi sea orbitals.

The starting point is the definition, for each $n$-particle subsystem,  of a set of $N ! / (N - n) ! n! $ {\em subnormalization integrals}  defined as
\begin{alignat}{3}
I_{i} \left ( \beta \right )   &  = [ i | \exp   \big  \{    \beta [ t (1) -     e^0_i ] \big  \} | i ]  =  1  \ , &&      \nonumber  \\
I_{ij} \left ( \beta \right )  &  =  [ i j |   F_2 ^{\dagger} (12) \exp  \big  \{     \beta [ t (1) + t(2) \ +  &&    v(12) -  e^0_i - e^0_j ]   \big \}   F_2^{}(12)  | i j ]_a       \ ,  \nonumber \\
I_{ijk} \left ( \beta \right )  & =   [ i j k |   F_3 ^{\dagger} (123) \exp    \big\{     \beta [ t (1) + t( 2 ) &&  +  \  t( 3)  +  v(12)+ v(23) +v(31)   \nonumber \\
& &&    -  e^0_i - e^0_j -e^0_k] \big \}   F_3^{}(123)  | i j k ]_a       \ ,  \nonumber \\
\vdots    \nonumber  \\ 
I_{i_1 i_2 i_3  \ldots  i_N} & =  I   \left ( \beta \right )  \ , 
\end{alignat}
where $e^0_i$ is the kinetic energy of  the Fermi gas state $| i ]  $,  $t(i)= -\nabla_i^2/2m$ is the kinetic energy operator and $v (ij)$ is the two-body potential.

The observation that in the absence of  interactions and correlations $I_{ij}$ reduces to  the product $I_{ij} = I _ i I _ j $   suggests
 to introduce the deviations from this simple expression within a multiplicative scheme, which amounts to writing  $I_{ij} = I_i I_j Y_{ij}$  \cite{ristig}. 
Following this procedure one obtains
\begin{align}
I_{i} \left ( \beta \right )     &=   Y _ i    \ ,     \nonumber  \\
I_{ij} \left ( \beta \right )    &=   Y _i  Y _ j Y _ {ij}     \ ,  \nonumber \\
I_{ijk} \left ( \beta \right )  &=    Y_i  Y _ j Y _ k  Y _ {ij}  Y _ {ik}  Y _ {jk}         \  \ldots      \ ,   \nonumber  \\
\vdots \\
I_{i_1 i_2 i_3  \ldots  i_N} & =  \prod_i   Y_ i \prod_{i<i}   Y_ {ij}  \prod_{i < j < k } Y _{ijk}  \   \ldots \   Y _ {1 \ldots N}\ . 
\end{align}
%
Within the factor-cluster decomposition,   $\log {I (\beta)}$ can be expressed  as a sum of logarithms,  each of them involving exactly $n$ orbitals 
in a connected manner, that can therefore be identified as  $n$-body cluster contributions. 

Collecting the above results, the expectation value of the Hamiltonian can be written as an expansion in the number  of  the correlated particles 
\begin{align}
\label{eq:expansionE}
E_0 =  T _ F +  \sum_{n = 2 } ^ {N } \left(  \Delta E  \right)_n  \ , 
\end{align} 
with
 \begin{align}
\label{eq:deltaEn}
    \left(  \Delta E  \right)_n = \sum_{i_1 <	\ldots < i_n }   \left.  \frac{\partial}{\partial \beta} \log Y _{i_1 \ldots i_n}     \right| _ {\beta = 0} \ .
\end{align}
Substituting the expression of the $Y$'s in terms on the subnormalization integrals 
\begin{equation}
Y _{i_1 \ldots i_n} = {I _{i_1 \ldots i_n} }   /    {  \prod_{i} I _ i \prod _{i<j} I _{ij} \ldots \prod{I_ {i_1 \ldots i_n}} } \ ,
\end{equation}
the $n$-body cluster contribution to the energy can be finally cast in the form
\begin{align}
\label{eq:deltaEnbis}
\left( \Delta E \right) _ n =  \sum_{i_1<\ldots <  i_n}  \left [    \frac{1}{I_{i_1 \ldots i_n} } \frac{\partial}{\partial \beta} I _{i_1 \ldots i_n}  - \sum_{k= 1}^{n}  \frac{1}{I_{i_1 \ldots i_k }  }    \frac{\partial}{\partial \beta} I _{i_1 \ldots i_k}       \right ] _{\beta = 0}\ .
\end{align}

We note that the terms accounting for the deviation of $   \left . I_{i_1 \ldots i_n} \right | _{\beta = 0}$ from unity are all $	\mathcal{O} \left( 1/ N  \right)$, 
or smaller, in the $N \to \infty$ limit.  
Substituting the expression for the $I$'s, expanding in powers of $( 1/N)$ and retaining the leading term, one obtains the linked cluster expansion for the energy in the thermodynamic limit.	

By way of  example, we report the  two-body cluster  contribution
\begin{align}
\label{eq:eff2b}
 \left( \Delta E \right) _ 2  =  \frac{1}{2} \sum_{ij}      [ ij |   w_{2} | ij ] _ a    \ ,  
 \end{align} 
 where
 \begin{align}
  w_2  =  \frac{1}{2}   F_2^{\dagger}(12)    \Big [ t(1) + t(2) , F_2^{}(12)  \Big]   &  +      \Big [F_2^{\dagger}(12)   ,  t(1) + t(2)    \Big ]   F_2^{}(12)   \\ 
  \nonumber 
  & +  F_2^{\dagger}(12) v(12) F_2(12)  \ .
  \end{align}

Each term of the expansion \eqref{eq:expansionE} can be represented by a diagram with $n$ vertices, representing the particles in the cluster, connected by lines corresponding the dynamical and statistical correlations. As pointed out above, the terms in the resulting diagrammatic expansion can be classified according to their topological structure, and selected classes of diagrams can be summed up to all orders solving  the system of FHNC integral equations \cite{fhnc1, fhnc2}.

 \chapter{The CBF effective interaction}
\label{chap2}

The approaches described  in Chapter \ref{chap1} can be merged. An effective interaction can in fact be derived within the formalism of 
correlated basis functions,  and employed to perform perturbative calculations in the basis of eigenstates of the non interacting system.
This procedure allows to properly take into account correlation effects in a simple manner, avoiding at the same time the problems 
arising from the use of non orthogonal perturbation theory.

In Section \ref{sec:defveff} we will discuss the derivation of the CBF effective interaction and the central assumptions involved in  this approach,
while  Section~\ref{sec:veff}  will be  focused on the explicit derivation of the effective interaction for the hard-sphere system.

\section{Definition of the CBF effective interaction}
\label{sec:defveff}
The formalism based on correlated basis functions and the cluster expansion technique, discussed in the previous chapter, has been recently employed to derive an effective interaction from a realistic nuclear  Hamiltonian\footnote{
Realistic nucleon-nucleon potentials provide an accurate description of the properties of the two-particle system, in both bound and scattering states.}.


As pointed out in Chapter \ref{chap1}, correlated states provide  a non-orthogonal basis, that can be orthogonalized using a complex procedure at the cost of 
introducing  a number of additional terms in the evaluation of matrix elements.  The  strategy underlying the effective interaction  approach, aimed at 
 simplifying the calculations and circumventing the  problem of non orthogonality corrections, amounts to  exploiting the correlated states to construct a 
 well-behaved  effective interaction, whose matrix elements  between states of the non interacting system can be used in perturbation theory.

The CBF effective interaction 
\begin{equation}
\label{eq:veff}
{\rm V _{eff }} = \sum_{j>i} v _{\rm eff} (r_{ij} ) \ , 
\end{equation}
is defined through the relation
\begin{align}
\label{def:veff}
\frac{1}{N} \langle H \rangle & \approx \frac{1}{N} {( 0 | H | 0 )} \equiv
\ T_{F} \ + \ \frac{1}{N}[ 0_{FG} | V_{{\rm eff}}| 0_{FG} ]  \ ,
\end{align}
where $N$ denotes the particle number, $\langle H \rangle$ is the {\em true} ground-state energy of the system the  
and $|0)$ is the correlated ground state defined in Section \ref{sec:CBF}. The first contribution 
to the right-hand side of Eq.\eqref{def:veff} is the expectation value of the kinetic energy in the non interacting ground state, which in 
translationally invariant systems reduces to a fully degenerate Fermi gas. In this case,  $T_{F} = 3 k_F^2/10m$.

The effective interaction, defined by Eq. \eqref{def:veff}, is designed to obtain the ground state 
expectation value of the  Hamiltonian at first order of perturbation theory in the Fermi gas basis. 
The procedure to construct $v_{\rm eff}$  is based on the tenets that:
\begin{itemize}

\item[{(i)}] an accurate estimate of $\langle H \rangle/N$ can be obtained using an advanced many-body technique, e.g.  the 
FHNC summation scheme;

\item[{(ii)}] the FHNC results, identified with $\langle H \rangle/N$, can be reproduced expanding $( 0 | H | 0 )$ 
at two-body cluster level. 

\end{itemize}

In the next section we will show that  under the above conditions Eq. \eqref{def:veff} is fulfilled by construction.

The definition of $v_{\rm eff}$, requiring the equivalence between matrix elements, implies that the effective interaction incorporates the 
effects of correlations. As a consequence, unlike the bare potential, $v_{\rm eff}$ is finite and well-behaved.  It has to be pointed out, 
however, that Eq. \eqref{def:veff} defines the CBF effective interaction {\em not} in operator form,  but in terms of 
its  expectation value in the FG ground state. 

The calculations discussed in this Thesis are largely
based on the assumption---that will be ultimately tested comparing our results to those obtained from alternative many-body approaches---that perturbative calculations
involving matrix elements of  between Fermi gas states provide accurate estimates of {\em all} properties of the Fermi hard-sphere system.





\section{CBF effective interaction for the fermion hard-sphere systems}
\label{sec:veff} 

In this section we describe the derivation of the effective interaction of the Fermi hard-sphere system obtained in Ref. \cite{mecca_etal_1}, 
which will then be applied to the calculation of a variety of properties, including the energy per particle, 
the self-energy, the effective mass, the momentum distribution and the transport coefficients.

\subsection{The CBF effective interaction  \emph{ansatz} }

Because the hard-core potential only depends on the magnitude of the distance between the interacting particles, 
an accurate description of correlation effects can be achieved using 
a simple Jastrow-type wave function, defined as in  Eq.~\eqref{CBF:GS} with
\begin{align}
F = \prod_{j>i=1}^N f(r_{ij}) \ , 
\label{eq:jastrow_central}
\end{align}
where $f(r)$ satisfies the boundary conditions
\begin{equation}
\label{eq:shape}
f(r_{ij} \leq a) = 0 \ \ \ , \ \ \ \lim_{r_{ij} \to \infty} f(r_{ij}) = 1 \ .
\end{equation}

The effective interaction is derived following the procedure originally proposed in Refs. \cite{shannon, BV}.
The expectation value of the  Hamiltonian in the correlated ground state appearing in left-hand side of Eq. \eqref{def:veff} is expanded keeping the two-body cluster contribution only, which amounts to setting 
 \begin{align}
\label{H:2body1}
\frac{1}{N} {( 0 | H | 0 )} =  T_F  + (\Delta E)_2  \ . 
\end{align}
Since the correlation operator is assumed to be hermitian,  the two-body operator appearing in   Eq. \eqref{eq:eff2b}  reduces to 
\begin{align}
 w_2 = &  \frac{1}{2}  \Big [   F_2(12)  ,   \left [ t(1) + t(2) , F_2(12)  \right]   \Big]  +    {F_2}^2(12) v(12) \ ,
\end{align}
implying ($F_2(12) = f(r)$ , $r = r_{12}$)
\begin{align}
\label{H:2body2}
(\Delta E)_2 =  \frac{\rho}{2} \int d^3 r \   \frac{1}{m}  \left  [   \left[ {\boldsymbol \nabla f(r)} \right]^2     + v(r) f^2(r) \right ] \left[ 1  -  \frac{1}{\nu} \ell^2(k_Fr) \right]  ,
\end{align}
where the Slater function $\ell(x)$---trivially related to the density matrix of the non interacting Fermi gas, defined according to 
Eq. \eqref{eq:onebodydens}---is given by
\begin{align}
\label{eq:slater}
\ell(x)~=~\frac{3}{x^3} \   \left( \sin x - x \cos x \right) \ .
\end{align}
The details of the calculations leading to Eq. \eqref{H:2body2} are given in Appendix \ref{app:2bc}. 

On the other hand, the expectation value of the effective potential $V_{\rm eff}$ in the FG ground-state,  the calculation of which is also discussed in  Appendix \ref{app:2bc}, can be written in the form
\begin{align}
\label{eq:2bcveff}
\frac{1}{N} [ 0_{FG} | &  V_{\rm eff}  | 0 _ {FG} ]    = \frac  {\rho}{2}  \int d^3r  \ v_{\rm eff}(r)
\left [  1 - \frac{1}{\nu} \ell^2(k_F r ) \right]  \ .
\end{align}

From Eqs. \eqref{H:2body1}, \eqref{H:2body2} and \eqref{eq:2bcveff}, it follows that, to the extent to which for 
a suitable choice of the correlation function
\begin{align}
\frac{1}{N}\langle  H \rangle = T_F  + (\Delta E)_2  \ , 
\end{align}
Eq.~\eqref{def:veff}, can be fulfilled with $v_{\rm eff}$ given by 
\begin{equation}
\label{veff:final1}
v_{\rm eff}(r) = \frac{1}{m} \left[ {\boldsymbol \nabla f(r)} \right]^2  +  v(r) f^2(r) \ .
\end{equation}
In the case of the hard-sphere system, because the correlation function is nonzero only in the region where $v(r) = 0 $,   the 
above equation  reduces to
\begin{equation}
\label{veff:final2}
v_{\rm eff}(r) = \frac{1}{m} \left[ {\boldsymbol \nabla f(r)} \right]^2  \ \  ,  \ \ \ r>a \ .
\end{equation}

We emphasise again that the procedure leading to Eq.\eqref{veff:final2} rests on the premise that the {\em true} ground state energy of the system 
can be accurately evaluated using some advanced many-body technique. 


In the following, $\langle H \rangle$ will be obtained using the variational FHNC scheme or Monte Carlo (MC) techniques.
Within the FHNC approach, based on the cluster expansion formalism, the calculation of $\langle H \rangle$ requires 
the solution of a set of integral equations taking into account the relevant cluster contributions to all orders. The MC approach, 
on the other hand, allows for a brute-force calculation of the Hamiltonian expectation value in the correlated ground state
(Variational Monte Carlo, or VMC) as well for the determination of the {\em true} ground-state energy (Diffusion Monte Carlo, 
or DMC). The main features of the VMC and DMC schemes will be outlined  in Chapter~\ref{chap4}.

The advantage of the effective interaction is that it can be used to obtain a variety 
of properties whose calculations within the FHNC approach involves severe difficulties.

\subsection{Determination of the correlation function}

Equation \eqref{veff:final2} clearly shows that the effective interaction of the hard-sphere system  is  completely determined by the correlation function $f(r)$. 

The shape of  $f(r)$ is obtained by functional minimization of the expectation value of the  Hamiltonian in the correlated ground state. Within the  two-body cluster approximation, this procedure yields an Euler\textendash Lagrange equation, to be solved with the boundary conditions dictated by the hard-core 
potential, as well as by the requirement that correlation effects vanish for large separation distances
\begin{align}
\label{eq:boundaryfa}
f(r \le a) = 0  \ \ \ , \ \ \ f( r \geq d ) = 1  \ .
\end{align}
The additional constraint
\begin{align}
\label{eq:boundaryfprime}
f^{\prime} (d)  = 0
\end{align}
that can be fulfilled introducing a Lagrange multiplier $\lambda$, 
enforces continuity of the derivative of the correlation function at $r = d$.
 
The details of the procedure of functional minimization can be found in Appendix \ref{app:euler}.
Here, we report the resulting Euler\textendash Lagrange equation
\begin{equation}
\label{ELeq}
g^{\prime \prime}(r) - g(r) \left[ \frac{\Phi^{\prime \prime}(r)}{\Phi(r)}  + m \lambda \right] = 0 \ , 
\end{equation}
where 
\begin{align}
\label{eq:grel}
g(r) = f(r) \Phi(r) \ , 
\end{align}
with
\begin{align}
\label{eq:phiel}
\Phi(r)   \equiv r \sqrt{1 - \frac{1}{\nu} \ell^2 ( k_F r) } \ . 
\end{align}
For any given values of density and correlation range, $d$, 
Eq.\eqref{ELeq} can be solved numerically to obtain the  correlation function $f(r)$,  with the lagrange multiplier $\lambda $ adjusted 
so that Eq.\eqref{eq:boundaryfprime} is satisfied. The interaction range $d $ is the only free parameter. It is  usually referred to as \emph{healing distance}, 
since at $r=d$ the two-particle wave-function ``heals'', smoothly reducing to the wave function describing non interacting particles.  

Within the variational approach, $d$ is determined minimising the FHNC ground state energy. This scheme has provided accurate upper bounds to the 
energies of a variety of interacting many-body systems, including liquid helium~\cite{FFM}, nuclear and neutron 
matter~\cite{LBFS} and the Fermi hard-sphere system~\cite{FFPR,AMBP}. 
Based on these results, we have have carried out FHNC  calculations of the expectation value $( 0 | H | 0)$ of the hard-sphere system, to 
be used for the determination of the corresponding CBF effective interaction from Eq.\eqref{def:veff}. 

We have considered a system of particles of mass $m = 1 \ {\rm fm}^{-1}$ and  degeneracy  $\nu=4$, and set the hard core radius to $a=1 \ {\rm fm}$.

\begin{figure}[htbp]
\begin{center}
 \includegraphics[width= 0.7\textwidth]{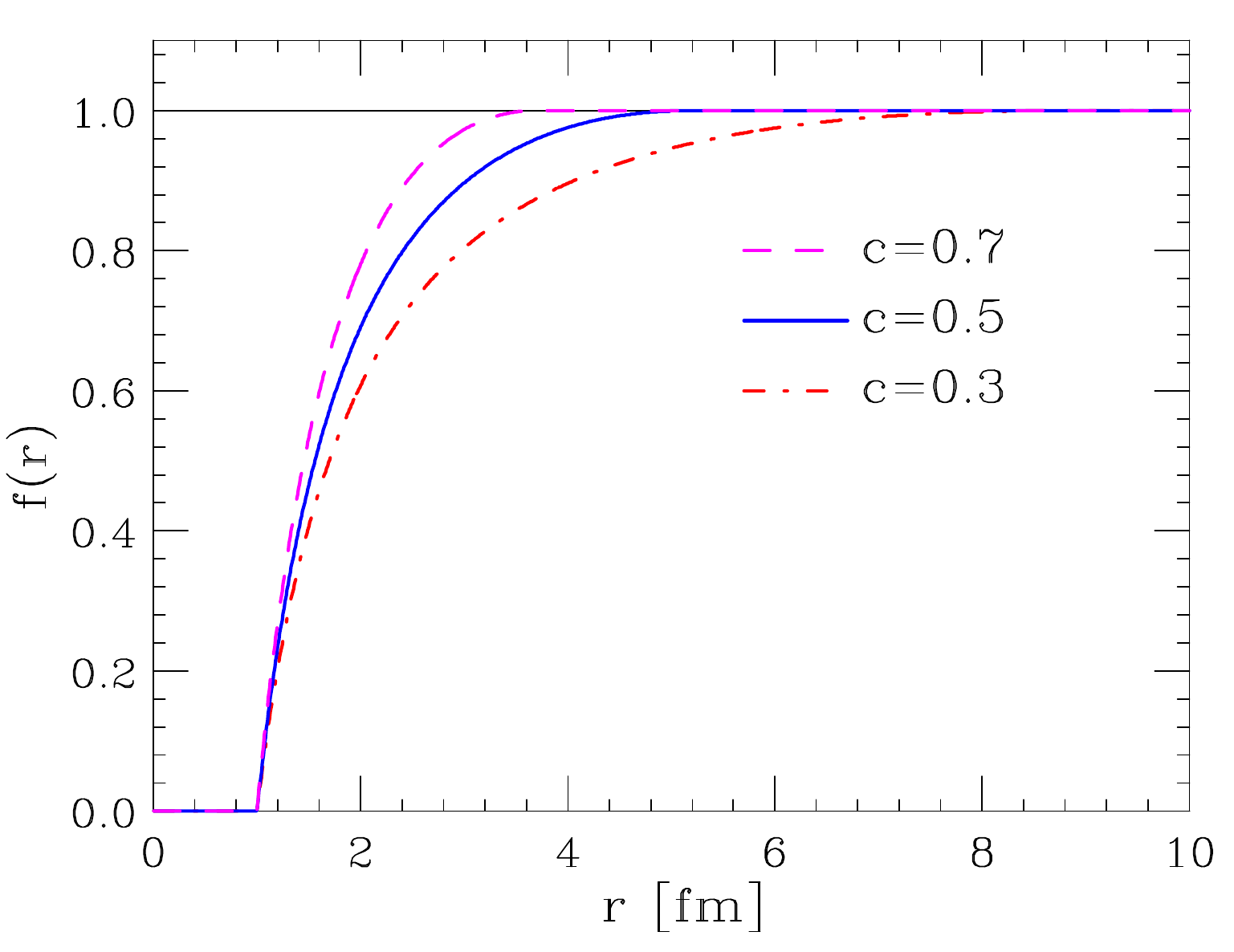}
  \caption[Correlation functions]{ Radial dependence of the correlation functions obtained from the solution of the Euler-Lagrange equation \eqref{ELeq}. 
The solid, dashed and dot-dash lines correspond to $c = k_F a =  $0.3, 0.5 and 0.7, respectively.}
\label{fig:fcel}
\end{center}
\end{figure}

Figure \ref{fig:fcel} shows the radial dependence of the correlation functions obtained from minimisation of the FHNC
ground state energy at different densities, corresponding 
to $c=k_F a=$0.3, 0.5 and 0.7, respectively. It clearly appears that, as it was to be expected, the correlation range $d$ is a decreasing function of
density.

As mentioned above, the ground state energy of the hard-sphere system has been evaluated solving the FHNC integral equations, in which the contribution of
the class of diagrams referred to as 	``elementary diagrams'' is neglected.   

Following Ref.~\cite{FFPR},  the ground state  energy  can be  conveniently written in terms of the dimensionless quantity $\zeta$, parametrizing the deviation from the corresponding FG result, defined through the equation
\begin{equation}
\label{def:z}
E_0 = \frac{3 k_F^2}{10m} \left( 1 + \zeta \right) \ .
\end{equation}

\begin{figure}[htbp]
\begin{center}
 \includegraphics[width= 0.7\textwidth]{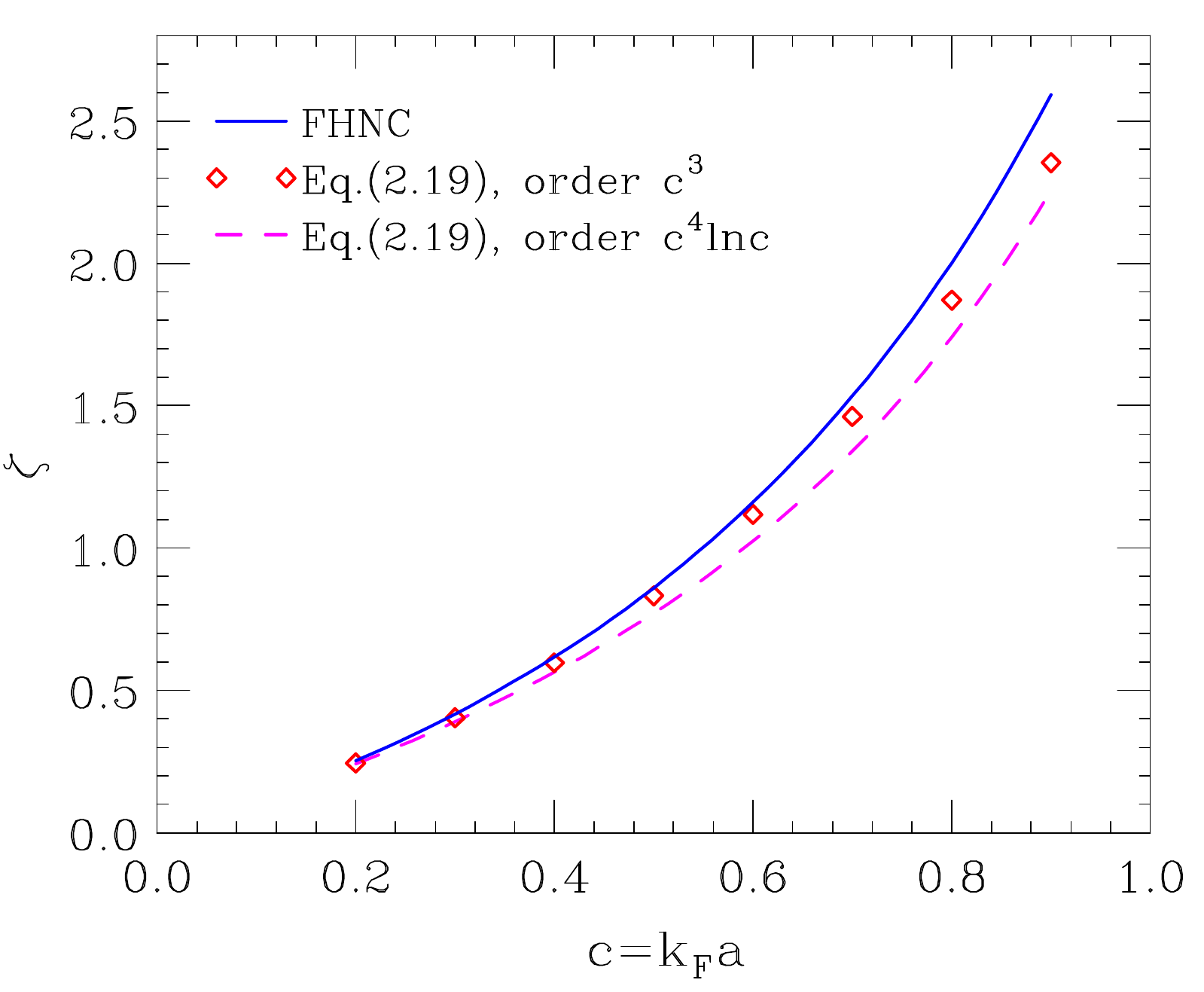}
 \caption[The ground state energy]{The full line shows the $c$-dependence of the dimensionless quantity $\zeta$, defined by Eq. \eqref{def:z},  obtained within the FHNC approach for the system of hard spheres of radius $a=1 \ {\rm fm}$, mass $m = 1 \ {\rm fm}^{-1}$  and degeneracy $\nu=4$. The results obtained from the low-density expansion of 
Eq.~\eqref{zeta_pert} are represented by the dashed line, 
while the diamonds correspond to the perturbative estimates of $\zeta$ computed neglecting terms of order higher than $c^3$.}
\label{fig:E1}
\end{center}
\end{figure}

Figure \ref{fig:E1} reports a comparison between the density dependence of $\zeta$ obtained within the FHNC scheme and  that predicted by the  perturbative low-density expansion of  Eq.~\eqref{eq:E0nu4}
\begin{align}
\label{zeta_pert}
\zeta =  \frac{5}{3} \left[ \frac{2}{\pi}  c \right. & + \left. \frac{12 }{35\pi^2} \left(11 - 2 \ln{2} \right)  c^ 2 + 0.78 c^3  +  \frac{32}{9 \pi^3} \left(4 \pi - 3  \sqrt{3} \right)  c^4 \ln{c} \right] \ .
\end{align}
For reference, we also show, by the diamonds, the perturbative values of $\zeta$ obtained including only contributions up to order $c^3$.

It clearly appears that at low $c$, corresponding to low density, the predictions of the two approaches are very close to one another. At $c=0.2$ (0.3), the difference 
in $\zeta$ turns out to be less than 5\% (7\%), which translates into an energy difference of less than 1\% (2\%). 
The more significant discrepancies observed at higher values of $c$ may be ascribed to a failure of the low-density expansion, although the observation that including the term of order $c^4 \log c$ leads to a decrease of $\zeta$ suggests that the contribution of cluster terms not  
taken into account within the FHNC scheme may also play a role. Note, however, that the full line representing the FHNC results lies 
consistently above the perturbative results. This pattern supports the assumption that the approximations involved in the FHNC calculation of the 
ground state expectation value of the  Hamiltonian do not spoil its upper bound character.

\subsection{Determination  of $v_{\rm eff}$}

As pointed out above, the determination of the CBF effective interaction and the many-body technique  employed to obtain the ground state energy
are conceptually independent. The FHNC summation scheme is , in fact, just one of the possible methods 
that can be employed to estimate the left hand side of Eq. \eqref{H:2body1}. It should be kept in mind that, while it turned out to 
be well suited in the range of densities considered in our study, its accuracy is likely to worsen at higher values of $\rho$. 

Within our scheme, the effective interaction
must reproduce the FHNC ground state energy at first order of perturbation theory in the Fermi gas basis. This goal is achieved by adjusting the range of the correlation function entering the definition of $v_{\rm eff}$, Eq.~\eqref{veff:final2}, 
in such a way that $\langle H \rangle/N$, defined by  Eqs.~\eqref{H:2body1}-\eqref{H:2body2}, coincide with the FHNC result.



\begin{figure}[htbp]
\begin{center}
 \includegraphics[width= 0.7\textwidth]{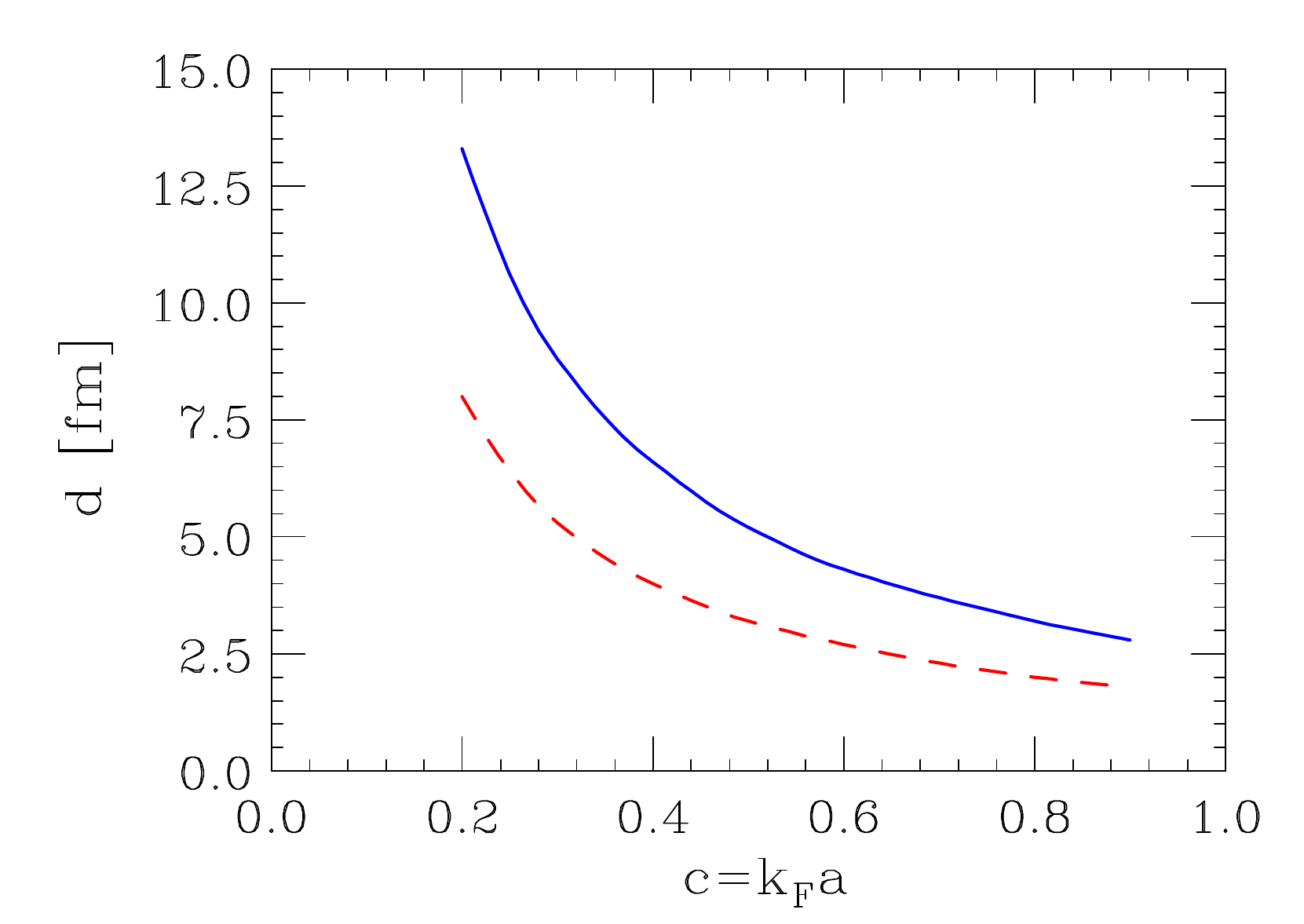}
\caption[The correlation range] { The full line shows the $c$-dependence of the correlation range, $d$, resulting from minimisation of the ground state energy of the 
hard-sphere system computed within the FHNC approach.  The dashed line corresponds to the correlation range employed to obtain the CBF 
effective interaction of Eq.~\eqref{veff:final2}.}

\label{healing}
\end{center}
\end{figure}

In Fig.~\ref{healing}, the correlation range resulting from minimisation of the FHNC ground state energy is compared to that employed to obtain the CBF effective interaction, as a function of the dimensionless variable $c$. The range of the effective interaction turns out to be sizeably smaller than the correlation range 
obtained from the variational calculation for all values of $c$, the difference being $\sim 35 \div 40$\%.
This result is consistent with the observation that the two-body cluster approximation 
underestimates the FHNC energy. Therefore, reproducing the FHNC result at two-body cluster level 
requires a shorter correlation range, leading a steeper correlation function which in turn corresponds to a stronger effective interaction.

\begin{figure}[htbp]
\begin{center}
 \includegraphics[width= 0.7\textwidth]{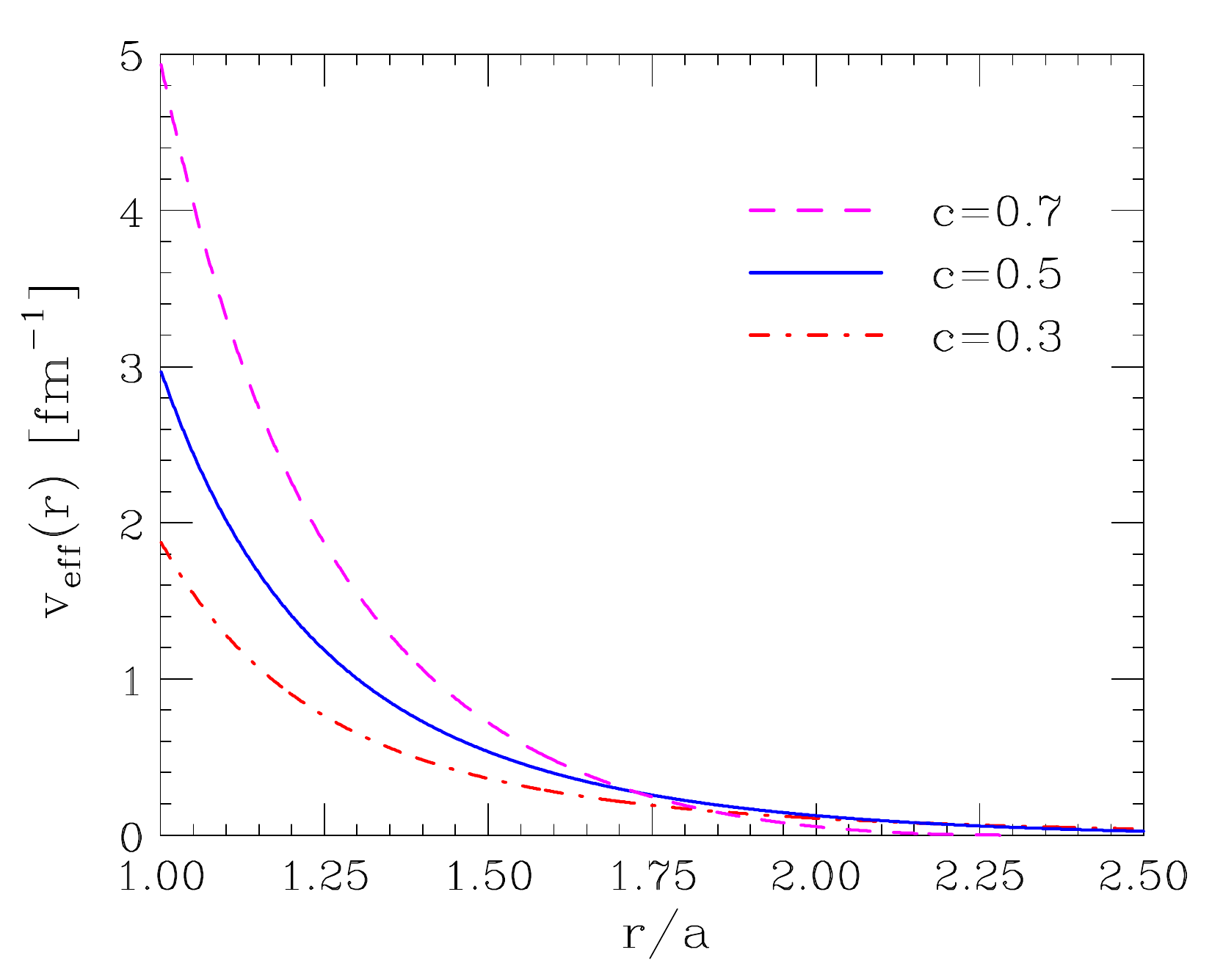}
  \vspace*{-.1in}
\caption[Radial dependence of $v_{\rm eff}$]{ Radial dependence of the effective interaction defined by Eq.~\eqref{veff:final2}.
The dot-dash, solid and dashed lines correspond to $c=k_Fa=$ 0.3, 0.5 and 0.7, respectively. The region 
$(r/a)<1$, where $v_{\rm eff}(r)=0$, is not shown.}
\label{veff:plot}
\end{center}
\end{figure}

The radial dependence of the effective interaction defined by Eq.~\eqref{veff:final2} is illustrated in Fig.~\ref{veff:plot} for 
three different values of the dimensionless variable $c$. Note that the region $(r/a)<1$, where $v_{\rm eff}(r)=0$, is not shown.
The shape of $v_{\rm eff}$ simply reflects the fact that, as the Fermi momentum increases, the correlation range, displayed in Fig.~\ref{healing},  
decreases, and the slope of the correlation function increases. 

\chapter[Equilibrium properties of the fermion hard-sphere systems]{Equilibrium properties of the fermion hard-sphere systems}

\label{chap3}

The effective interaction defined in Chapter \ref{chap2} is designed to reproduce the ground-state expectation value of the  Hamiltonian at first order of perturbation theory in the Fermi gas basis. 
As pointed out above, the   CBF effective interaction approach is based on the assumption  that all properties of the system can be obtained through  perturbative calculations in the Fermi gas basis employing the effective potential $v_{\rm eff}$. This amounts to assuming that, while being defined  
from the expectation value of the  Hamiltonian  in the correlated ground state,  the effective interaction can be also employed  in  calculations of matrix elements involving excited states. 
To gauge the reliability of this scheme,  we have studied a number of equilibrium properties of the fermion hard-sphere system, 
and compared our results to those obtained from different many-body approaches.

 Section \ref{sec:self} will be devoted to the discussion of the two-point Green's function and to the perturbative calculation of
 the real and imaginary parts of the self-energy.
 The numerical results of the calculations of the quasiparticle spectrum, the effective mass and the momentum distribution will be 
 reported and analysed in Sections \ref{sec:spe} and \ref{sec:nk}, respectively.

\section{Self-Energy}
\label{sec:self}

The two-point Green's function $G$, embodying all information on single-particle properties of many-body systems, is obtained from Dyson's 
equation \cite{FW,dickh}
\begin{align}
\label{dyson}
G(k,E) = G_0(k,E) + G_0(k,E) \Sigma(k,E) G(k,E) \ , 
\end{align}
 where
 $G_0$ is the Green's function of the non interacting Fermi gas, the expression of which reads
\begin{align}
\label{green:FG}
G_0(k,E) = \frac{\theta(k-k_F)}{E - e_0(k) + i \eta} +  \frac{\theta(k_F-k)}{E - e_0(k) - i \eta} \  . 
\end{align}
In the above equation, $\eta = 0^+$,  $e_0(k) = k^2/2m$, $\theta(x)$ is the Heaviside step function, and the two terms in the right-hand side 
describe the propagation of particles ($k>k_F$) and holes ($k<k_F$).

The  irreducible,  or proper, self-energy $\Sigma({\bf k},E)$ takes into account the effect of interactions. In the language of the 
diagrammatic representation of the Green's function,  the self-energy corresponds to any part of a diagram which is
connected to the rest  by two $G^0$-lines. The diagrams contributing to the irreducible self-energy, on the other hand,  
are those that cannot be divided into two parts joined by only one $G^0$-line.

Dyson's equation leads to the following expression of the full Green's function
\begin{align}
\label{green:interacting}
G(k,E) = \frac{1}{E - e_0(k) - \Sigma(k,E)} \ .
\end{align}
As we will see, the singularities of Eq. \eqref{green:interacting} fully determine the single-particle spectrum of the system. 
The  proper self-energy, also referred to as  mass operator, can be evaluated in perturbation theory carrying out an expansion in 
powers of the interaction potential, whose terms can be conveniently represented by diagrams. The resulting expression can be written in the
form
\begin{align}
\Sigma(k,E) = \Sigma^{(1)}(k) + \Sigma^{(2)}(k,E) + \ldots   \ .
\end{align}
 Note that the insertion of any finite order approximation to $\Sigma$ in Dyson's integral equation \eqref{dyson} leads 
 to a Green's function including interactions at all orders. 

The diagrammatic representation of  the irreducible self-energy expansion up to second order  terms is reported in Fig.~\ref{self}. For the 
sake of simplicity, we only show the contribution of direct diagrams  (see below). 

\begin{figure}[htbp]
\begin{center}
\subfloat[$\Sigma_{HF}({\bf k})$]{
 \includegraphics[scale= 0.65]{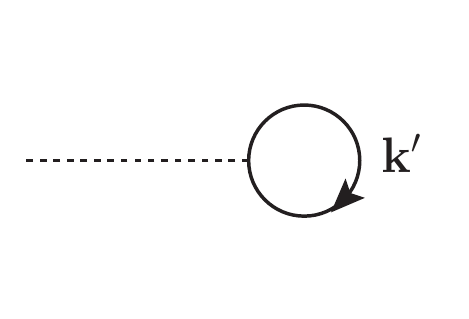}
 \label{fig:hf}
 }
 \\
 \subfloat[ $\Sigma_{2p1h}({\bf k}, E)$]{
 \includegraphics[scale= 0.65]{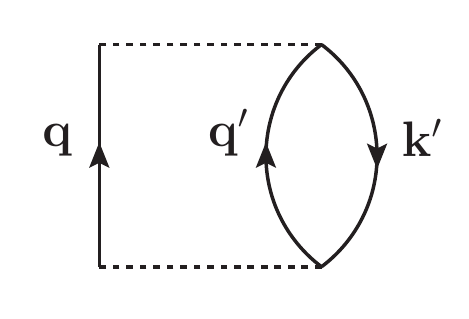}
 \label{fig:pol}
 }
 \subfloat[$\Sigma_{2h1p}({\bf k},E)$]{
 \includegraphics[scale= 0.65]{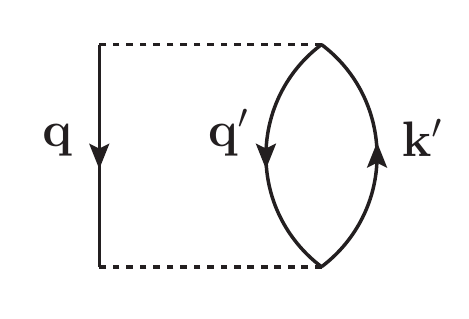}
 \label{fig:correl}
 }
\caption[First and second order irreducible self-energy]{Diagrammatic representation of the direct part of the first and second order contributions to the irreducible self-energy. 
Panels (a), (b) and (c) correspond to the 
Hartree-Fock, polarisation and correlation terms, respectively. Dashed lines represent the CBF effective interaction, while upward and downward oriented 
solid lines depict the  free propagation of  particle and hole states, respectively.}
\label{self}
\end{center}
\end{figure}
\noindent
The first order term, represented by diagram (a),  is the usual Hartree Fock contribution, while the second order terms, corresponding 
to diagrams (b) and (c), involve two-particle\textendash one-hole or two-hole\textendash one-particle intermediate states. 
They are referred to as polarization and correlation contributions, respectively. 

The Hartree-Fock approximation for the self-energy leads to a real and energy independent contribution, whose analytical expression reads
\begin{align}
\label{sigma:HF}
\Sigma_{HF}(k) = \frac{1}{\nu}  \sum_{\sigma, {\bf k}^\prime \sigma^\prime} n^0_<(k^\prime) 
[ {\bf k} \sigma  \ {\bf k}^\prime \sigma^\prime | v_{\rm eff} | {\bf k} \sigma \ {\bf k}^\prime \sigma^\prime ]_a  \ ,
\end{align}
where $n^0_<(k) = \theta(k_F - k)$, the two-particle state is antisymmetrised according to $| \alpha \ \beta \rangle_a = (| \alpha \ \beta \rangle - | \beta \  \alpha \rangle )/\sqrt{2}$, 
and the index $\sigma$ labels the discrete quantum numbers specifying the state of a particle carrying momentum ${\bf k}$.

The explicit expression of the polarisation and correlation contributions are (see  Fig.~\ref{self})
\begin{align}
\label{sigma_p}
\Sigma_{2p1h} \left( k, E \right)   =
 \frac{m}{\nu}    
 \sum_{\sigma, {\bf k}^\prime \sigma^\prime, \mathbf{q}\tau, \mathbf{q}^\prime \tau^\prime}    
\frac{ \left |  [  {\bf q} \tau  \ {\bf q}^\prime \tau^\prime | v_{\rm eff} | {\bf k} \sigma \ {\bf k}^\prime \sigma^\prime ]_a \right | ^2}
{q^2 + {q^\prime}^2- {k^\prime}^2  -2m E  -  i \eta}  \ \ 
n^0_{>} (q) n^0_{>}(q^\prime) n^0_{<} (k^\prime) \ , 
\end{align}
and
\begin{align}
\label{sigma_h}
\Sigma_{2h1p} \left( k, E \right)   =  \frac{m}{\nu}   \sum_{\sigma, {\bf k}^\prime \sigma^\prime, \mathbf{q}\tau, \mathbf{q}^\prime \tau^\prime}    
\frac{   \left | [ {\bf q} \tau  \ {\bf q}^\prime \tau^\prime | v_{\rm eff} | {\bf k} \sigma \ {\bf k}^\prime \sigma^\prime ]_a   \right |^2}
{ {k^\prime}^2 - q^2- {q^\prime}^2  + 2m E  -  i \eta}  \ \ 
 n^0_{<} (q) n^0_{<}(q^\prime) n^0_{>} (k^\prime) \ , 
\end{align}
with $n^0_>(k) = \theta(k-k_F)$.  Equations \eqref{sigma_p} and \eqref{sigma_h} show that,  as the effective interaction is diagonal in the space of the discrete quantum numbers, the self-energy does not depend on $\sigma$.

\subsection{Imaginary part}
The above contributions to the self-energy are complex quantities.   
The corresponding real and imaginary parts can be  easily identified using the relation
\begin{equation}
 \frac{1}{x \pm i \eta}  = \mathcal{P} \left( \frac{1}{x}\right) \mp i \pi \delta(x)  \ , 
 \label{eq:imre}
 \end{equation}
\noindent
The  resulting imaginary  part of  Eq.~\eqref{sigma_p} 
%
\begin{align}
\label{eq:im2p1h}
\IM   \Sigma_{2p1h} \left( k, E \right)  =   \   \pi  \  \frac{   m } {\nu}  
\sum _  { \substack {\sigma, {\bf k}^\prime \sigma^\prime, \\  \mathbf{q}\tau, \mathbf{q}^\prime \tau^\prime}  } &
|  [  {\bf q} \tau  \ {\bf q}^\prime \tau^\prime | v_{\rm eff} | {\bf k} \sigma \ {\bf k}^\prime \sigma^\prime ] _a |^2   \ \delta(   { q}^2 + {{ q} ^{ \prime } }^2-  {{  k} ^{\prime}}^2  -2m E    )      \nonumber \\
   & \ \ \times   n^0_{>} (q) \ n^0_{>}(q^\prime) \  n^0_{<} (k^\prime)    \ ,
\end{align}
is  non vanishing in the energy range  $E >  \epsilon_F$,  with $\epsilon_F = k_F^2/2m$. 

\noindent
On the other hand, the  imaginary  part of  $\Sigma_{2h1p}$ of Eq.~\eqref{sigma_h} 
 \begin{align}
 \label{eq:im2h1p}
\IM  \Sigma_{2h1p} \left( k,  E \right)  =  
\ \pi   \frac{   m } {\nu}   
\sum_{ \substack { \sigma, {\bf k}^\prime \sigma^\prime,  \\ \mathbf{q}\tau, \mathbf{q}^\prime \tau^\prime} }
 & | [ {\bf q} \tau  \ {\bf q}^\prime \tau^\prime | v_{\rm eff} | {\bf k} \sigma \ {\bf k}^\prime \sigma^\prime ]_a |^2  
     \ \delta(      { { k} ^{\prime}}^2   -  {  q} ^2  - { {  q} ^{\prime}} ^2 -2m E    )     \nonumber  \\
&  \ \    \times n^0_{<} (q)  \ n^0_{<}(q^\prime)  \ n^0_{>} (k^\prime) \ , 
\end{align}
is not vanishing for  $E < \epsilon_F$.

The details of the calculation of the matrix element 
 \begin{align}
  [  {\bf q} \tau  \ {\bf q}^\prime \tau^\prime | v_{\rm eff} | {\bf k} \sigma \ {\bf k}^\prime \sigma^\prime ]_a  \ , 
 \end{align}
as well as the explicit expression of the imaginary part of the  polarization and correlation  contributions to the self-energy can be found in Appendix \ref{app_self}.  They have been evaluated numerically, using  the 
 multidimensional integration routines  provided  in the  CUBA library \cite{cuba}.
 We have emptied the VEGAS Monte Carlo  algorithm,  exploiting  a  variance-reduction technique based on importance sampling.

\begin{figure}[htbp]
\begin{center}
 \includegraphics[width= 0.8\textwidth]{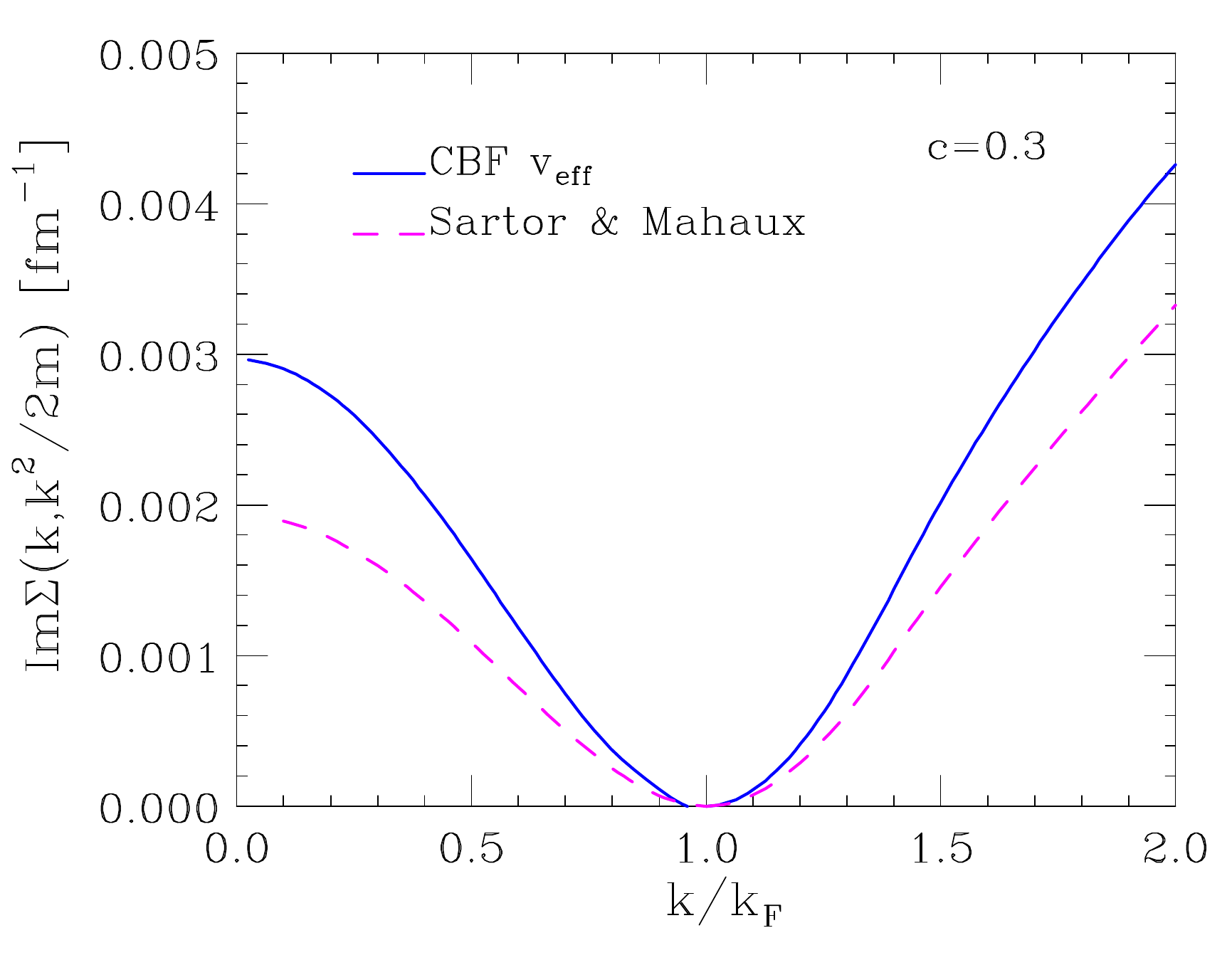}
\caption[On-shell imaginary part of $\Sigma$]{ Imaginary part of the quantities $\Sigma_{2h1p} \left( k<k_F, k^2/2m \right) $ and  $\Sigma_{2p1h} \left( k>k_F, k^2/2m \right)$,
computed at  $c=0.3$ and displayed as a function of the dimensionless variable $k/k_F$. The solid and dashed lines correspond to the 
results obtained from the CBF effective interaction and from the low-density expansion of Ref.~\cite{mahaux}, respectively.}
\label{Im_self}
\end{center}
\end{figure}

Figure~\ref{Im_self} shows the behaviour of the imaginary part of  $\Sigma(k ,E) $ 
corresponding to 
$c=0.3$, computed at $E = k^2/2m$ and displayed as a function of the dimensionless variable $k/k_F$. For comparison, we also show the same 
quantities computed by Sartor and Mahaux using the low-density expansion and including terms up to order $c^2$ \cite{mahaux}.

\begin{figure}[htbp]
\begin{center}
 \includegraphics[width= 0.8\textwidth]{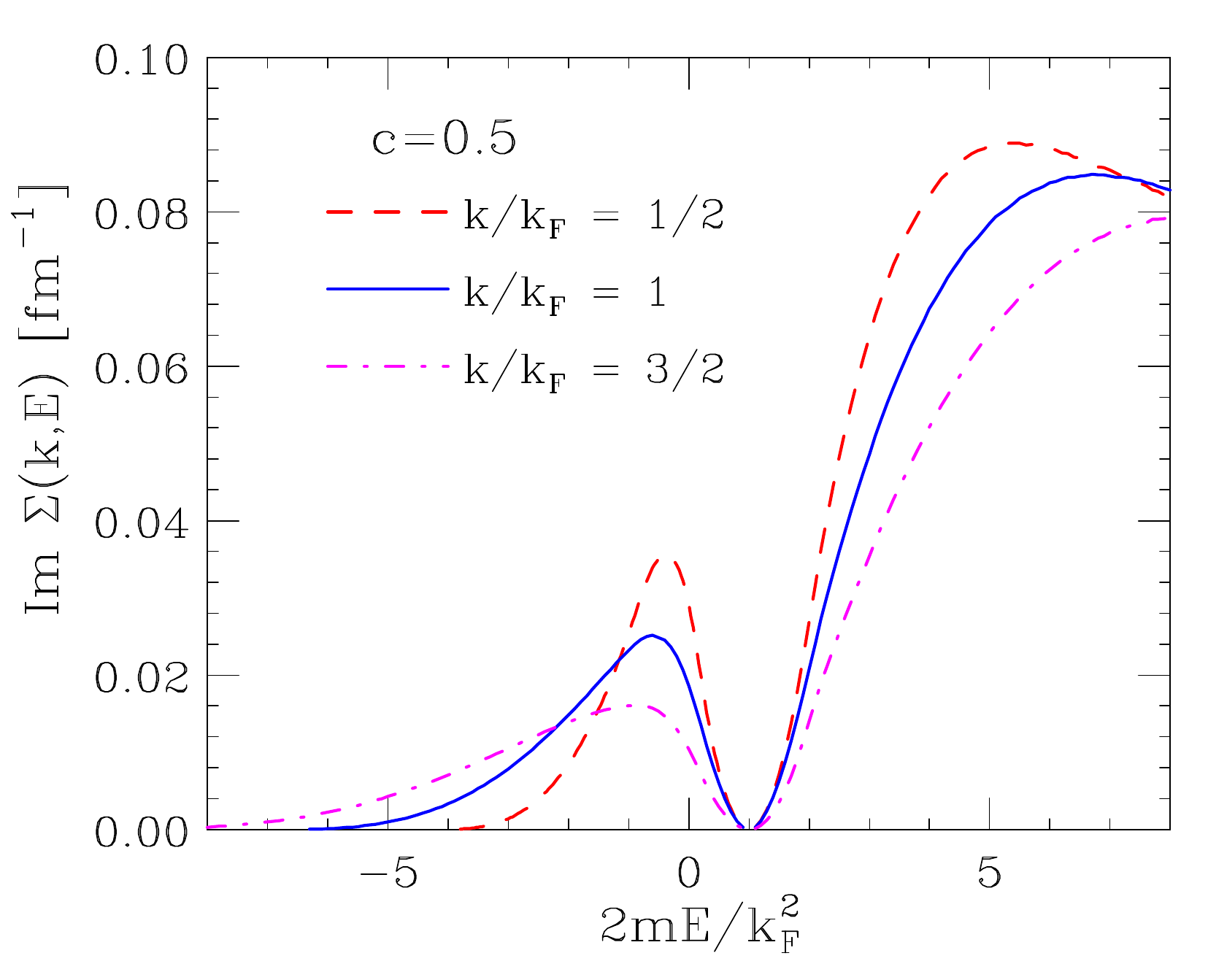}
\caption[Off-shell imaginary part of $\Sigma$]{Energy dependence of the imaginary part of the polarisation ($2mE/k_F^2>1$) and 
correlation ($2mE/k_F^2<1$) contributions to the self-energy of the Fermi hard-sphere system at  $c=0.5$. The dashed, solid 
and dot-dash lines correspond to $k/k_F=$ 0.5, 1 and 1.5, respectively.}
\label{Im_self_off}
\end{center}
\end{figure}

The energy dependence of the imaginary part of the second order contributions to the self-energy is illustrated in Fig.~\ref{Im_self_off}, showing the 
results at $c=0.5$ for three different values of momentum, corresponding to 
$k/k_F=$1/2 (solid line), 1 (dashed line) and 3/2 (dot-dash line).

\section{Quasiparticle spectrum and effective mass}
\label{sec:spe}

The self-energy computed at second order in the CBF effective interaction, discussed in the previous section, has been
used to obtain the single particle spectrum, effective mass and momentum distribution of the 
Fermi hard-sphere system of degeneracy $\nu$=4.

The conceptual framework for the identification of single particle properties in interacting many-body systems
is laid down in Landau's theory of liquid $^3$He (see, e.g. Ref.~\cite{abri}), based on the tenet that there 
is a one-to-one correspondence between the elementary excitations of a Fermi liquid, dubbed quasiparticles, 
and those of the non interacting Fermi gas.

In unpolarized systems, quasiparticle states  of momentum $k$ are specified by their energy, $e(k)$ and lifetime $\tau_k = \Gamma_k^{-1}$. In the limit of small $\Gamma_k$, the 
Green's function describing  the propagation of quasiparticles  can be written in the form
\begin{align}
\label{green:QP}
G(k,E) = \frac{Z_k}{E - e(k) + i \Gamma_k}  \ . 
\end{align}
A comparison between the above expression and Eq.\eqref{green:interacting} clearly shows that 
quasiparticle properties can be readily related to the real and imaginary parts of the self-energy.

\begin{figure}[htbp]
\begin{center}
 \includegraphics[width= 0.8\textwidth]{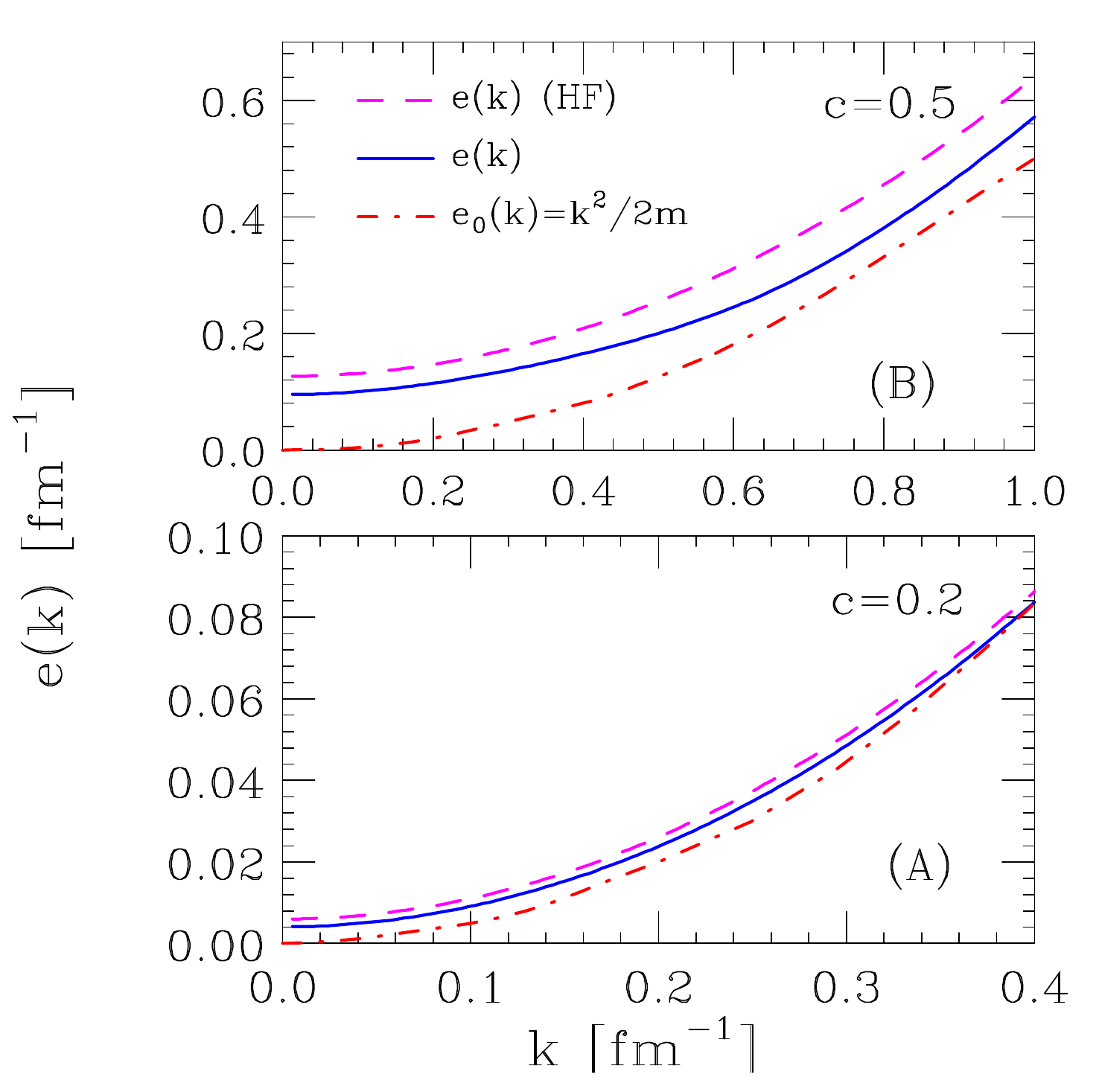}
\caption[Quasiparticle spectrum]{Quasiparticle energy, computed from Eq.~\eqref{def:spectrum} at $c=0.2$ [panel (A)]
and 0.5 [panel(B)]. The dashed and solid lines correspond to the first order (i.e. Hartree-Fock) and second order 
approximations to the self-energy, respectively. For comparison, the dot-dash lines show the kinetic energy spectrum. }
\label{fig:ek}
\end{center}
\end{figure}

The energy of a quasiparticle of momentum $k$, $e(k)$, is obtained solving the equation
\begin{align}
\label{def:spectrum}
e(k) = e_0(k) + {\rm Re} \  \Sigma[k,e(k)] \ .
\end{align}
Substitution of Eq.~\eqref{sigma:HF} in Eq.~\eqref{def:spectrum} yields the Hartee-Fock spectrum, represented by 
the dashed lines of  Fig.~\ref{fig:ek}, while the results obtained including the second order corrections to the self-energy are displayed by
full lines. For comparison, the dot-dash lines show the kinetic energy spectrum.

 From Eqs. \eqref{green:interacting} and \eqref{green:QP} it also follows that the quasiparticle lifetime is
related to the self-energy through
\begin{align}
\label{def:lifetime}
 \tau_k^{-1} =  \Gamma_k = Z_k {\rm Im} \  \Sigma [k,e(k)] \ , 
\end{align}
where
\begin{align}
\label{residue}
Z_k =  \left[ 1 - \frac{\partial}{\partial E} {\rm Re} \ \Sigma(k,E)  \right]^{-1}_{E=e(k)} \ , 
\end{align}
is the residue of the Green's function of Eq.~\eqref{green:QP} at the quasiparticle pole.

Equations \eqref{def:spectrum} and \eqref{def:lifetime} are obtained expanding the energy of the quasiparticle pole
in powers of $\Gamma_k$, and keeping the linear term only. Note that the resulting expressions of $e(k)$ and $\Gamma_k$
obtained using the second order self-energy are {\em not} second order quantities. 

The quasiparticle spectrum is conveniently parametrized in terms of the effective mass $m^\star$, defined by
Eq.~\eqref{def:mstar}. 
The total derivative of $e = e(k)$ is performed using Eq.~\eqref{def:spectrum}, and keeping in mind that, 
since ${\rm Re} \  \Sigma(k,E)$ is evaluated at the quasiparticle pole, 
$k$ and $E$ are not independent of one another. As a consequence, one finds
\begin{align}
\frac{d e}{d k} = \frac{k}{m} + \frac{\partial}{\partial k} {\rm Re} \ \Sigma(k,e) 
+ \frac{\partial}{\partial e} {\rm Re} \ \Sigma(k,e) \frac{d e}{d k} \ ,
\end{align}
implying
\begin{align}
 \frac{d e}{d k} & = \left[ \frac{k}{m} + \frac{\partial}{\partial k} {\rm Re} \ \Sigma(k,E) \right] 
\left[ 1 -  \frac{\partial}{\partial E} {\rm Re} \ \Sigma(k,E)  \right]^{-1}_{E=e(k)} \ .
\end{align}
At first order the self-energy depends on $k$ only, and the above equation reduces to
\begin{align}
\frac{d e}{d k} =   \frac{k}{m} + \frac{\partial  \Sigma_{HF}(k)}{\partial k} \ ,
\end{align}
with $\Sigma_{HF}$ given by Eq.~\eqref{sigma:HF}.

\begin{figure}[htbp]
\begin{center}
\includegraphics[width= 0.6\textwidth]{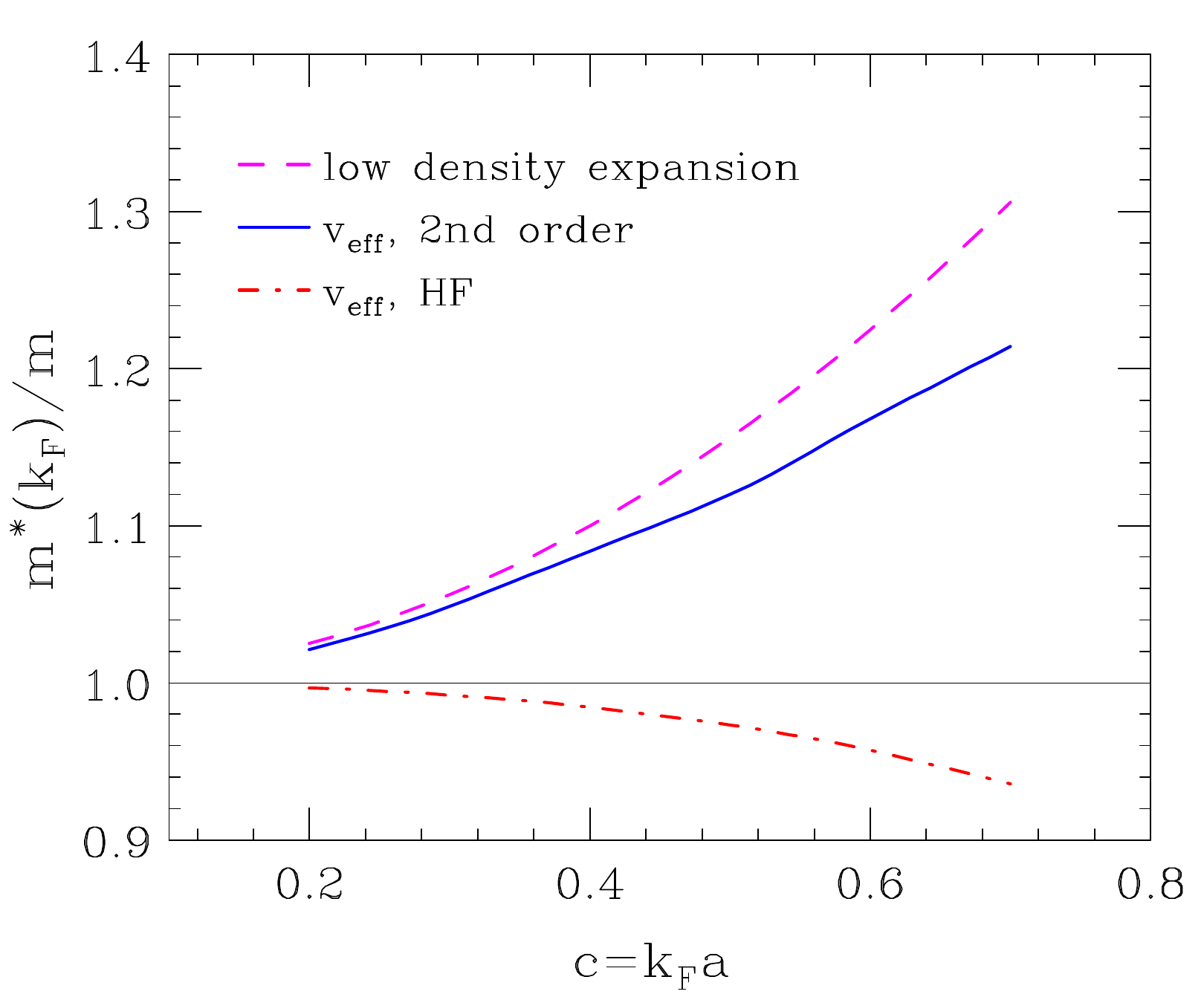}
\caption[The effective mass $m^{\star}$]{$c$-dependence of the ratio $m^\star(k_F)/m$, obtained from Eqs.~\eqref{def:mstar} and \eqref{def:spectrum}. 
The dot-dash and solid lines represent the results of 
calculations carried out using the first and second order approximations to the self-energy. 
For comparison, the dashed line shows the results computed using the low-density expansion of Eq.~\eqref{pert_mstar}.
}
\label{mstar1}
\end{center}
\end{figure}

The dot-dash and solid lines of Fig.~\ref{mstar1} show the $c$-dependence of the ratio $m^\star(k_F)/m$, evaluated using the self 
energy computed at first and second order in the CBF effective interaction, respectively. It is apparent that inclusion of the energy-dependent contributions to the self-energy, resulting in a moderate correction to the spectra of Fig.~\ref{fig:ek}, 
leads instead to a drastic change in the behaviour of the effective mass. While in the Hartee-Fock approximation the ratio $m^\star(k_F)/m$ is less than one and 
monotonically decreasing with $c$, the full result turns out to be larger than one and monotonically increasing. 

The dashed line of Fig.~\ref{mstar1}, representing the ratio obtained from the low-density expansion at order $c^2$, Eq.~\eqref{pert_mstar}, 
exhibits the same features as the solid line. The low-density expansion appears to provide quite accurate results for $c \lsim 0.3$. 
A comparison with Fig.~\ref{fig:E1} suggests that in the case of the ground state energy the inclusion of higher order contributions extends 
the range of applicability of the expansion to $c \lsim 0.4$.

It is worth pointing out that the striking differences between the effective masses computed using the first and second order expressions of
the self-energy are a consequence of their different functional dependences.  While the former is a function of momentum only, 
the latter depends on both momentum and energy. Because the enhancement of the effective mass, as well as the modification of its behaviour as
a function of density, arise from the appearance of the energy dependence, it is arguable that the inclusion of higher order terms  would result 
    in small corrections.

\section{Momentum distributions}
\label{sec:nk}

In translationally invariant systems, the momentum distribution, $n(k)$, describes the occupation probability
of the single-particle state of momentum $k$.  

The connection between $n(k)$ and the Green's function, or the self-energy, can be best understood introducing the 
spectral functions appearing in the Lehmann representation of the two-point
Green's function (see, e.g., Refs.~\cite{dickh,czyz})
\begin{align}
\label{KL}
G(k,E)  = \int_0^\infty dE^\prime \ & \left[  \frac{ P_p(k,E)}{E-E^\prime - \mu +i \eta}
+ \frac{ P_h(k,E)}{E+E^\prime - \mu - i \eta} \right] \ ,
\end{align}
where $\mu=e(k_F)$ denotes the chemical potential. 

The particle (hole) spectral function $P_p(k,E)$ [$P_h(k,E)$] yields 
the probability of adding to (removing from) the ground state a particle of momentum $k$, leaving the 
resulting $(N+1)$- [$(N-1)$-] particle system with energy $E$. It follows that  
\begin{align}
\label{E:int}
n(k) = \int_0^\infty dE P_h(k,E)  = 1 -  \int_0^\infty dE P_p(k,E) \ .
\end{align}

The momentum distribution obtained from Eq.~\eqref{E:int}, with 
\begin{align}
\label{def:Ph}
P_h(k,E) = &
\frac{1}{\pi} {\rm Im} \ G(k,\mu - E)   \nonumber \\ 
 = & \frac{1}{\pi} \frac{ {\rm Im} \Sigma(k,\mu - E) }{ [ \mu - E - e_0(k) - {\rm Re} \Sigma(k,\mu - E) ]^2 + [ {\rm Im} \Sigma(k,\mu - E) ]^2 } \ ,
\end{align}
and
\begin{align}
\label{def:Pp}
P_p(k,E) = &
- \frac{1}{\pi} {\rm Im} \ G(k,\mu + E)    \nonumber \\ 
  = & - \frac{1}{\pi} \frac{ {\rm Im} \Sigma(k,\mu + E) }{ [ \mu + E - e_0(k) - {\rm Re} \Sigma(k,\mu + E) ]^2 + [ {\rm Im} \Sigma(k,\mu + E) ]^2 } \ ,
\end{align}
can be cast in the form \cite{benfabrofanto1990}
\begin{align}
\label{nkpole}
n(k) = Z_k \theta(k_F - k) + \delta n(k) \ .
\end{align}
The first term in the right-hand side of the above equation, with $Z_k$ defined by Eq.~\eqref{residue}, originates from 
the quasiparticle pole in Eq.~\eqref{green:QP}, while $\delta n(k)$ is a smooth contribution,  extending to momenta both below and above $k_F$, 
arising from more complex excitations of the system.  
Equation~\eqref{nkpole} shows that the discontinuity 
of $n(k)$ at $k=k_F$ is given by 
\begin{align}
n(k_F - \eta) - n(k_F + \eta) = Z_{k_F} = Z \ .
\end{align}

At second order in the effective interaction, the momentum distribution obtained from Eqs.\eqref{KL}-\eqref{def:Pp} can be 
conveniently written in the form
\begin{align}
n(k) = n_<(k) + n_>(k) \ ,
\end{align}
where $n_<(k>k_F) = n_>(k<k_F) = 0$, and  
\begin{align}
\label{eq:nk<}
n_<(k<k_F) & = 1 + \left[ \frac{ \partial}{\partial E} {\rm Re} \Sigma_{2p1h}(k,E) \right]_{E=e_0(k)} \ , \\
\label{eq:nk>}
n_>(k>k_F) & = - \left[ \frac{ \partial}{\partial E} {\rm Re} \Sigma_{2h1p}(k,E) \right]_{E=e_0(k)} \ .
\end{align}
Note that the above equations imply that within the Hartree-Fock approximation $n(k) = \theta(k_F-k)$, and $Z~=~1$.

\begin{figure}[htbp]
\begin{center}
 \includegraphics[width= 0.8\textwidth]{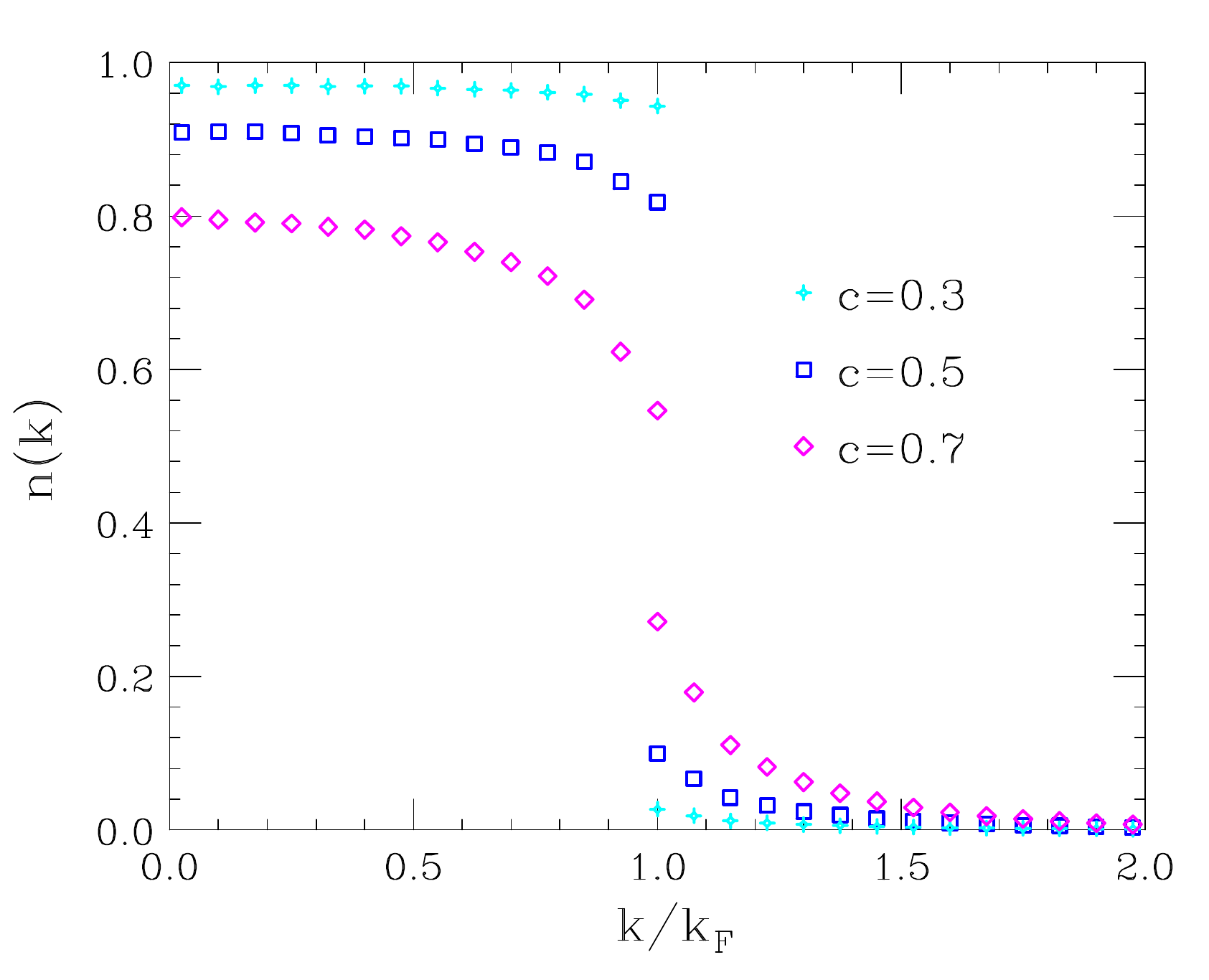}
\caption[Momentum distributions]{ Momentum distributions computed at second order in the CBF effective interactions, for
three different values of $c=k_F a$. The values of the discontinuity corresponding  $c=$ 0.3, 0.5 and 0.7 are 0.92, 
0.72 and 0.28, respectively.}
\label{nk}
\end{center}
\end{figure}

\begin{figure}[htbp]
\begin{center}
 \includegraphics[width= 0.8\textwidth]{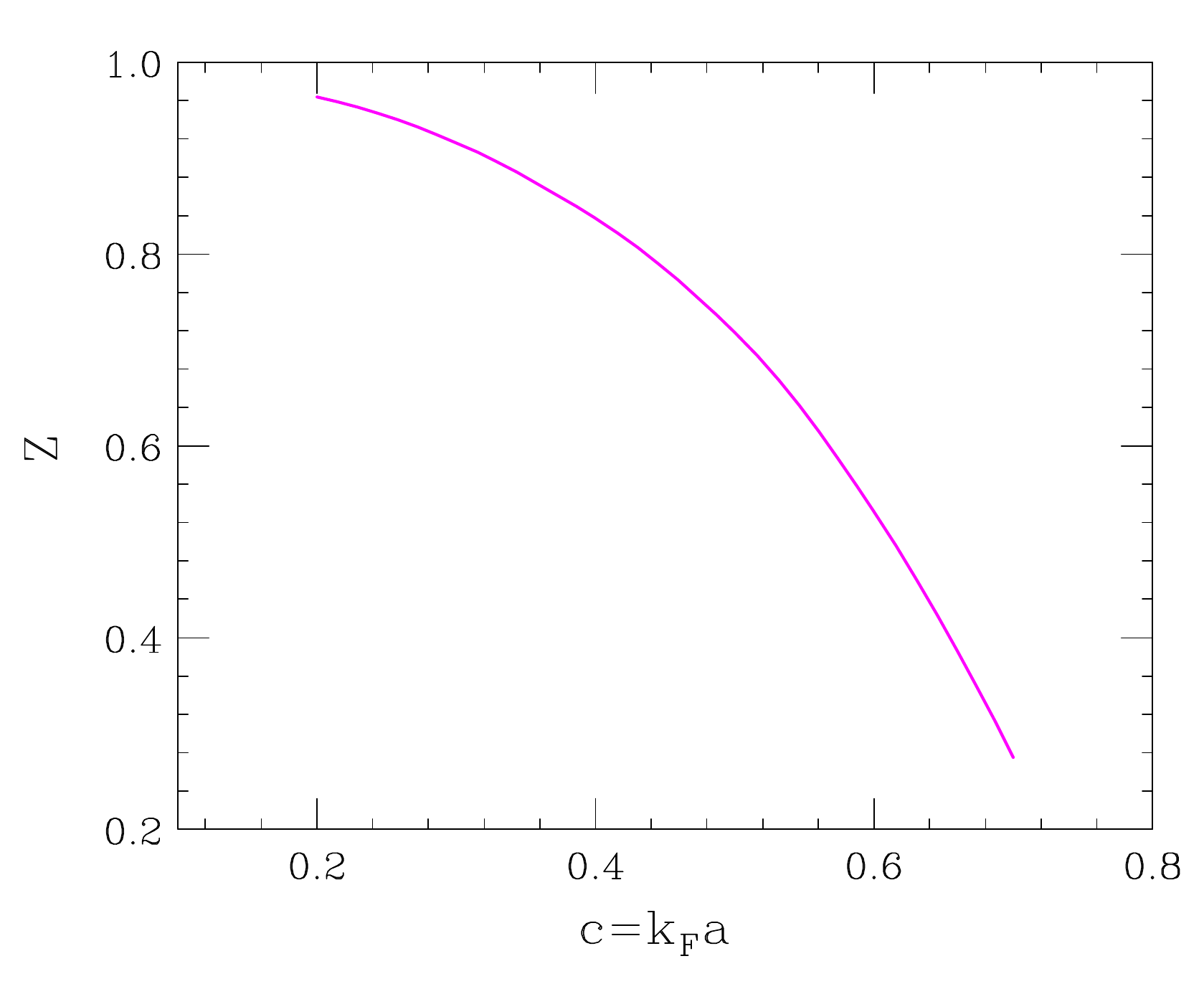}
\caption[Discontinuity  of $n(k) $]{ Discontinuity of the momentum distribution of the Fermi hard-sphere system, as a function 
of $c=k_Fa$.}
\label{z_k}
\end{center}
\end{figure}

Figure \ref{nk} shows the momentum distributions obtained including contributions up to second order in the CBF effective interaction, 
for three different values of the dimensionless parameter $c$. It clearly appears that the deviation 
from the Fermi gas result rapidly increases with density. 
A measure of interaction effects is provided by
the discontinuity $Z$, shown in  Fig.~\ref{z_k} as a function of $c$.

\begin{figure}[htbp]
\begin{center}
 \includegraphics[width= 0.8\textwidth]{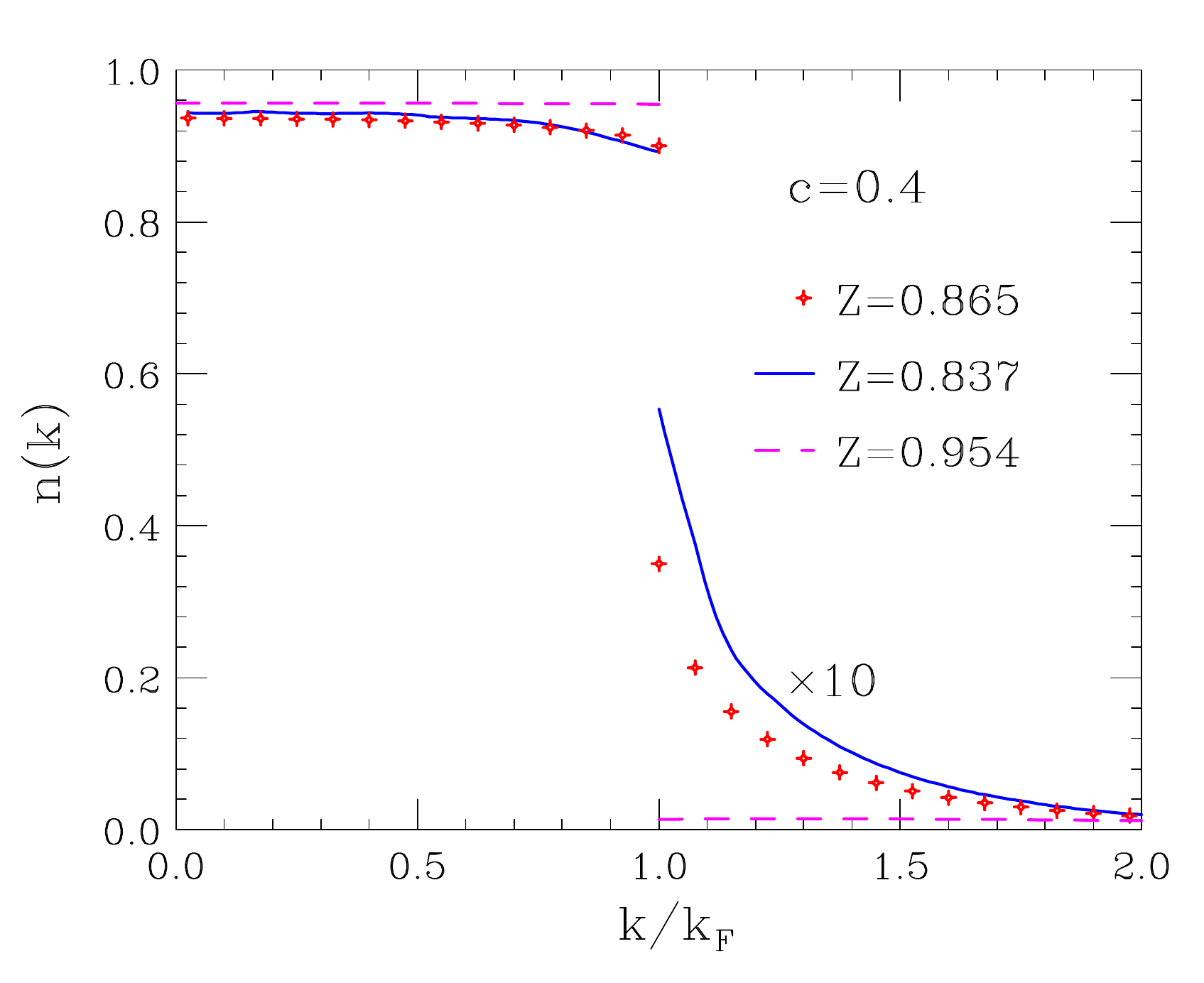}
\caption[$v_{\rm  eff } $ and CBF $n(k)$]{ Momentum distribution of the Fermi hard-sphere system at $c=0.4$. Solid line: results 
obtained at second order in the CBF effective interaction; dashed line: variational results of Ref. \cite{FFPR}; 
crosses: results of the low-density expansion at order $c^2$. All values of $n(k>k_F)$ are multiplied by a factor 10.}
\label{nk_comp1}
\end{center}
\end{figure}

In  Fig.~\ref{nk_comp1} we compare the momentum distribution resulting from our calculation, represented by the solid line,  
to those reported in Ref. \cite{FFPR} for $c=0.4$. The dashed line shows the results computed using the variational wave function 
obtained from minimisation of the ground state energy within the FHNC scheme, while the crosses correspond to the 
predictions of the  the low-density expansion discussed in 
Refs.\cite{Galitskii1958,mahaux,mahaux_erratum,belyakov} (see Appendix \ref{nkApp}),  including contributions up to order $c^2$. Note that the values of $n(k>k_F)$ are multiplied 
by a factor 10.


It clearly appears that the variational approach sizeably underestimates interaction effects, and fails to provide the 
correct logarithmic behaviour at $k$ close to the Fermi momentum. On the other hand, the momentum distributions obtained 
from the CBF effective interaction and from the low-density expansion are in close agreement at $k<k_F$ and exhibit 
discontinuities that turn out to be within $\sim3$\% of one another.

The kinetic energy computed using the variational $n(k)$ exactly agrees with the variational energy.
On the other hand,  the result obtained from the perturbative momentum distribution 
does not necessarily reproduce the kinetic energy calculated using the effective interaction, Eq.~\eqref{def:veff}, which  
coincides with the variational energy by definition.  

\begin{figure}[htbp]
\begin{center}
\includegraphics[scale= 0.7]{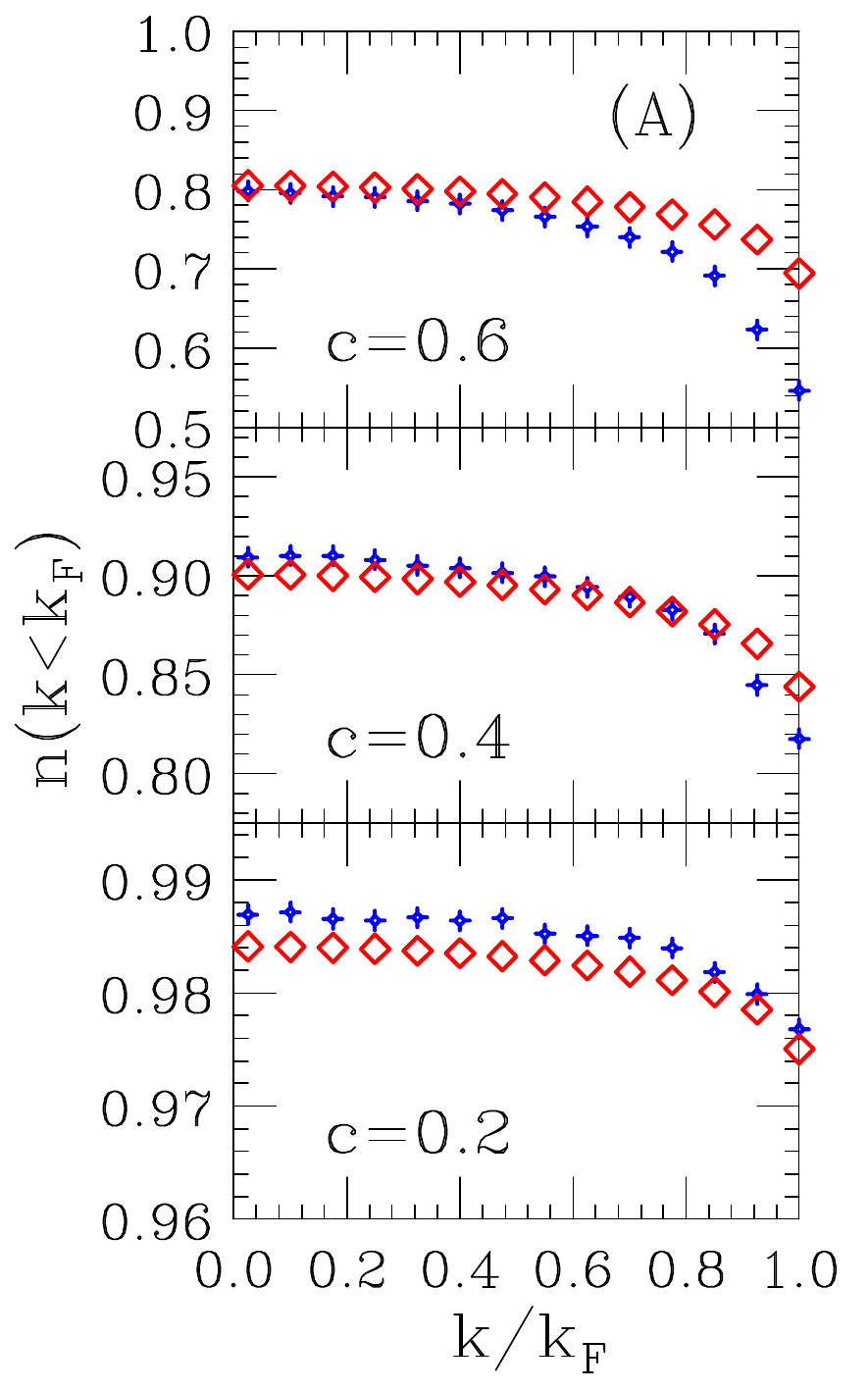}
 \includegraphics[scale= 0.7]{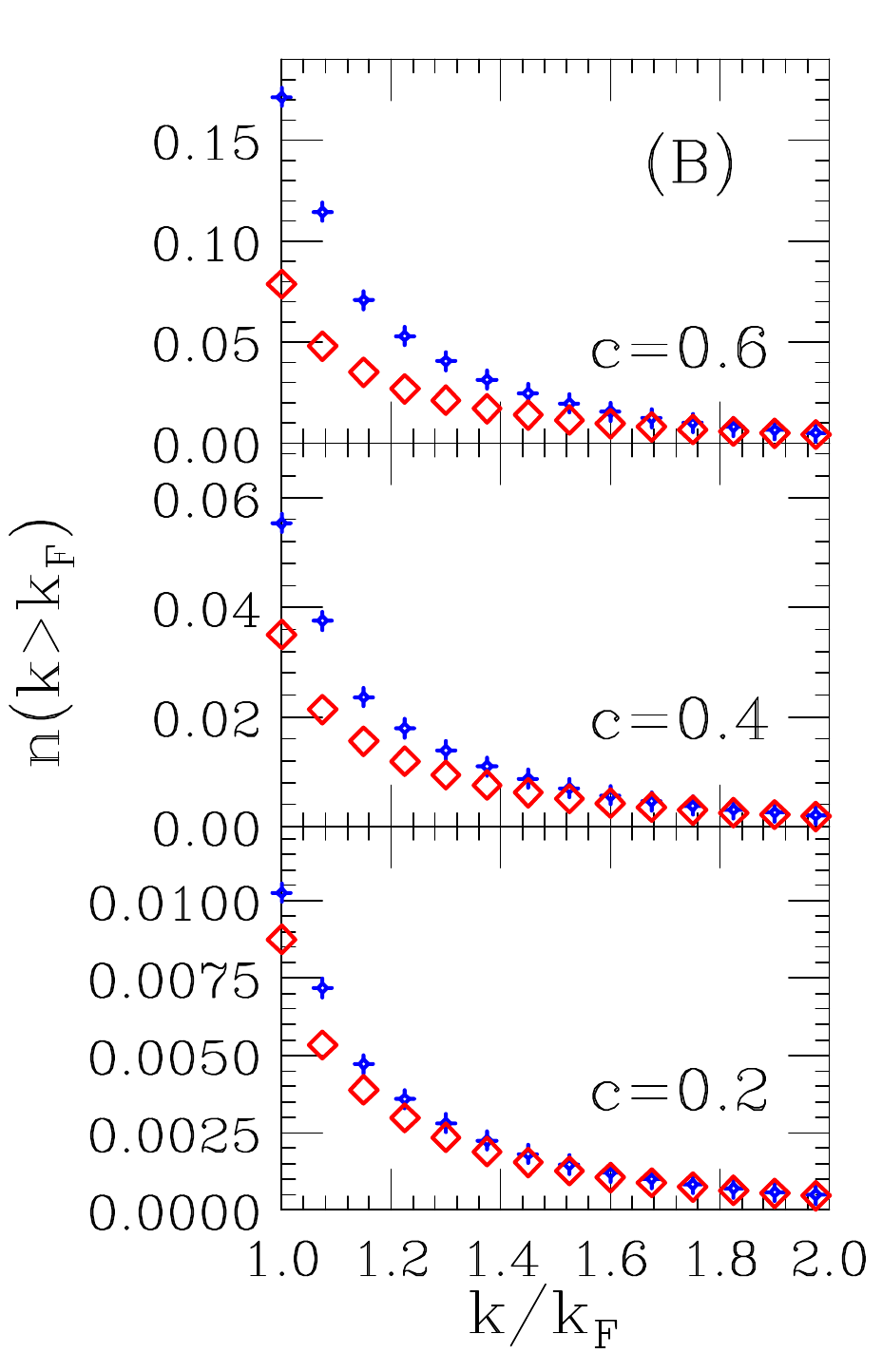}
\caption [CBF  and low-density expansion $n(k)$] {Comparison between the momentum distribution obtained from the CBF effective
interactions (crosses) and the low-density expansion discussed in Refs.~\cite{Galitskii1958,mahaux} (diamonds), for different
values of the dimensionless parameter $c=k_F a$.  Panels (A) and (B) correspond to the regions $k<k_F$
and $k>k_F$, respectively.}
\label{nk_comp2}
\end{center}
\end{figure}

In  Fig.~\ref{nk_comp2}, the difference between the momentum distribution computed using the effective interaction 
and that obtained from the low-density expansion is illustrated for different values of $c$, ranging from 0.2 to 0.6.
The emerging picture is consistent with that observed in Figs. \ref{fig:E1} and \ref{mstar1}, and suggests that the low density 
expansion provides accurate predictions for $c\lsim0.3$. Sizable discrepancies appear at larger values of $c$, most notably 
in the vicinity of the Fermi surface.


\begin{figure}[htbp]
\begin{center}
 \includegraphics[width= 0.8\textwidth]{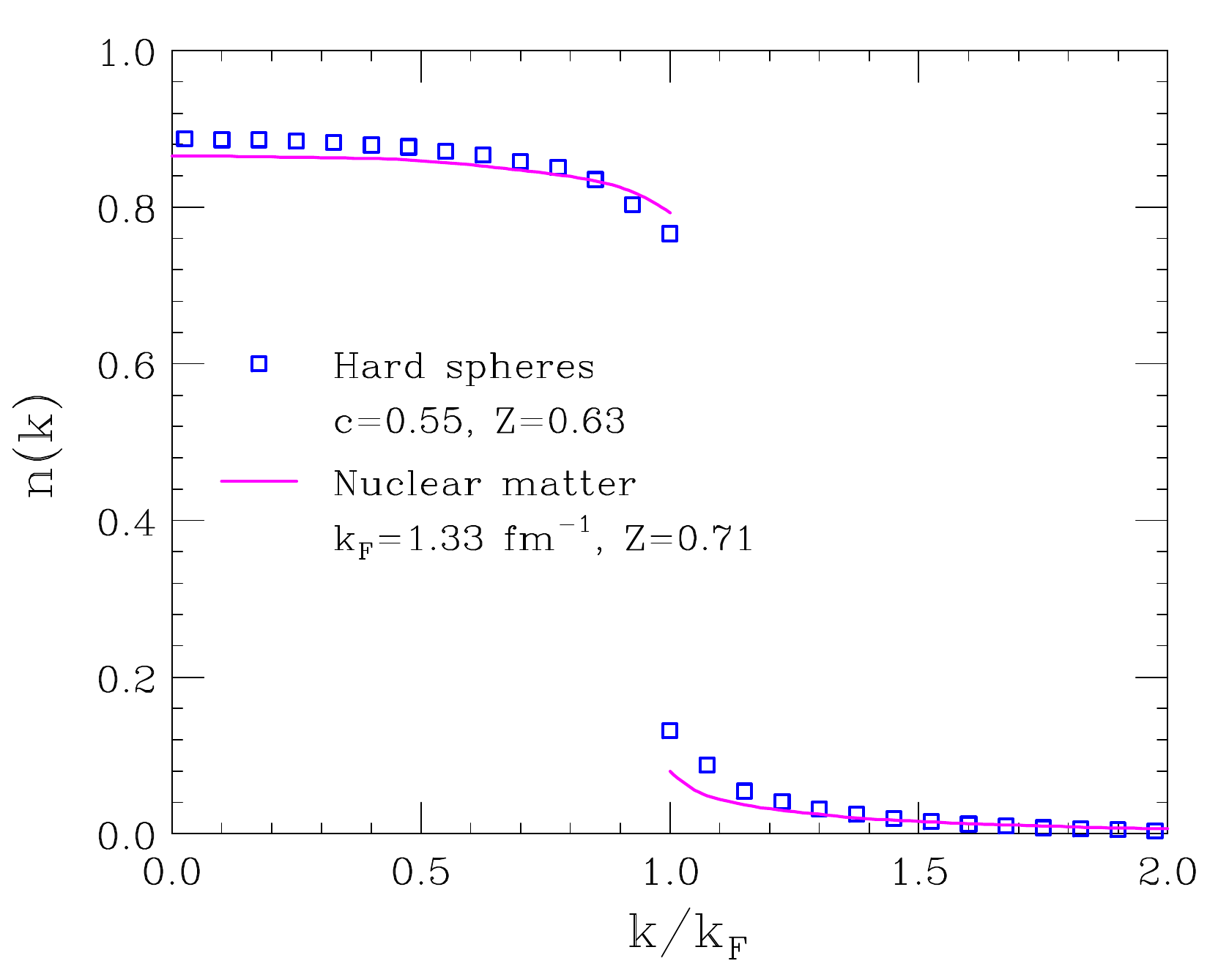}
\caption[HS and nuclear matter $n(k)$]{  Comparison between the momentum distribution of the Fermi hard-sphere system obtained from the effective interaction 
approach discussed in this article (squares) and that of isospin symmetric nuclear matter at equilibrium density reported in Ref.~\cite{fantonipandha} (solid line), computed using correlated wave functions and second order CBF perturbation theory.}
\label{nk_comp_nm}
\end{center}
\end{figure}

In order to establish a correspondence between the hard-sphere system and isospin symmetric nuclear matter at equilibrium density, 
we have analysed the corresponding momentum distributions. In  Fig.~\ref{nk_comp_nm} the results of our calculations at $c=0.55$ are compared
to the results of the the calculation of Fantoni and Pandharipande \cite{fantonipandha}, carried out using a correlated wave function 
and including second order contributions in CBF perturbation theory. Note that the approach of Ref. \cite{fantonipandha} is conceptually 
very similar to ours, although the effects of correlations are taken into account modifying the basis states, instead of 
replacing the bare potential with an effective interaction.

It appears that, as far as the momentum distribution is concerned,  the system of hard spheres of radius $a=1\ {\rm fm}$ and  
$k_F = 0.55 \ {\rm fm}^{-1}$ corresponds to nuclear matter at density  $\rho_{\rm NM} = 0.16 \ {\rm fm}^{-3}$, or
Fermi momentum $k_F~=~1.33 \ {\rm fm}^{-1}$. Because $n(k)$ is mainly determined by the dimensionless parameter 
$c=k_Fa$,  the results of Fig.~\ref{nk_comp_nm} suggest that nucleons in nuclear matter behave like hard 
spheres of radius $a = 0.55/1.33 \approx 0.4 \ {\rm fm}$. A comparison with nuclear matter momentum distributions 
obtained from other methods \cite{RPD} leads to the same conclusion. 

Note that, because the momentum distribution provides a measure of the occupation probability of single 
particle levels,  the deviations  of $n(k)$ from the prediction of the Fermi gas model reflect the occurrence of 
virtual scattering processes involving pairs of strongly correlated particles, leading to their excitation to states 
outside the  Fermi sea. Therefore, our results suggest that these processes are mainly driven by the short-range repulsive core 
of the nucleon-nucleon interaction. On the other hand, the crude description in terms of hard spheres is not 
expected to explain nuclear matter properties driven by low momentum, i.e. long distance, physics. 
In this context, it is worth mentioning that the discussion of the hard-core model of nuclear matter of Ref.~\cite{FW}, based 
on the solution of the Bethe-Goldstone equation, also assumes a hard-core radius $a=0.4 \ {\rm fm}$.

\chapter{Transport properties of the fermion hard-sphere systems}
 \label{chap4}
 
 In this chapter we will discuss the application of the approach based on the CBF effective interaction to the calculation of 
 transport coefficients, focusing on shear viscosity and thermal conductivity. Because these properties are of paramount importance in 
 astrophysical applications, we will concentrate on the hard-sphere system of degeneracy $\nu=2$, which can be seen as a model
 of pure neutron matter. 
 It should be kept in mind,  however, that this analogy is limited to densities below the solidification point. The analysis of Ref.~\cite{solidification} indicates that for $\nu$=2 solidification occurs at a density $\rho_s$ such that ${\rho}_s a$=0.23.  Moreover, at subnuclear densities nuclear matter is known to undergo transitions to superfluid and/or superconducting phases, which are not allowed by the purely repulsive interaction.   Theoretical studies suggest that neutron matter becomes superfluid at density $\lsim$0.08 fm$^{-3}$ \cite{superfluidphase}. 

 As in the previous chapters, the mass of the particles and the hard-core radius will be set 
 to $m = 1\ {\rm fm}^{-1}$ and $a =$ 1 fm, respectively.
 

\section{Landau\textendash Abrikosov\textendash Khalatnikov formalism}
\label{sec:LAK}

We follow the approach based on Landau's theory of normal Fermi liquids (see, e.g., Ref.~\cite{Baym-Pethick}), originally 
developed by Abrikosov and Khalatnikov \cite{kinetic_abrikal, abrikala_rep}. Within this scheme, the shear viscosity and thermal conductivity 
coefficients---denoted $\eta$ and $\kappa$, respectively---are determined from the momentum and energy fluxes
obtained from the kinetic equation for the distribution function, $n_{\bf k}$, which can be written in the form
\begin{align}
\label{boltzmann}
\frac{ \partial n_{\bf k} }{ \partial t } +  \frac{ \partial n_{\bf k} }{ \partial {\bf r} } \cdot \frac{ \partial \epsilon_{\bf k} }{ \partial {\bf k} } 
- \frac{ \partial n_{\bf k} }{ \partial {\bf k } } \cdot \frac{ \partial \epsilon_{\bf k} }{ \partial {\bf r}  } = I[n_{\bf k}] \ ,
\end{align}
In the above equation,  $\epsilon_{\bf k}$ is the energy of a quasiparticle carrying  momentum ${\bf k}$, while $I[n_{\bf k}]$ is the collision integral, 
the definition of which involves the in medium scattering probability~$W$.

In general, the scattering probability depends on the initial and final momenta of the particles participating in the process. In the low-temperature limit, however, 
the system is strongly degenerate, and only quasiparticles occupying states in the vicinity of the Fermi surface can be involved in interactions.  
As a consequence, the magnitudes of their momenta can be all set equal to the Fermi momentum, and $W$ reduces to a function of only two angular 
variables, $\theta$ and $\phi$. The former is the angle between the initial momenta,  whereas the latter is the angle between the planes
specified by the initial and final momenta, respectively. 

The above procedure leads to the expressions   \cite{kinetic_abrikal,abrikala_rep}
\begin{align}
\label{eta_AK}
\eta_{AK} = \frac{16}{15}\frac{1}{T^2}\frac{k_F^5}{{m^\star}^4}  \ \frac{1}{ \langle W \rangle (1-\lambda_\eta)} \ ,
\end{align}
and
\begin{align}
\label{cappa_AK}
\kappa_{AK} = \frac{16}{3} \frac{1}{T} \frac{ \pi^2 k_F^3}{{m^\star}^4} \  \frac{1}{\langle W \rangle (3 - \lambda_\kappa)}
\end{align}
where $T$ is the temperature,  
\begin{align}
\label{lambda:eta}
\lambda_\eta = \frac{ \langle W [ 1 - 3 \sin^4 (\theta/2) \sin^2 \phi ] \rangle}{\langle W \rangle} \ \ \ , \ \ \ 
\lambda_\kappa = \frac{\langle W (1+2 \cos \theta) \rangle}{\langle W \rangle} \ ,
\end{align}
and  the angular average is defined according to
\begin{align}
\label{Wavg}
\langle W \rangle = \int \frac{d\Omega}{2\pi} \ \frac{ W(\theta,\phi) }{\cos \theta/2} \ ,
\end{align}
with $d \Omega = \sin \theta d \theta d \phi$. Note that, as we are considering a system of identical particles, the angular integration is normalised to $2 \pi$.

In the above equations, corresponding to the leading terms of low-temperature expansions,  $m^\star$ denotes
the quasiparticle effective mass evaluated at momentum such that $|{\bf k}| = k_F$. 
Note that the shear viscosity and thermal conductivity coefficients exhibit different $T$-dependence.

The quasiparticle lifetime $\tau$ can also be written in terms of the  angular average  of the scattering probability, Eq.~\eqref{Wavg}, 
according to
\begin{align}
\label{tau_AK}
\tau = \frac{1}{T^2} \ \frac{8\pi^4}{{m^\star}^3}\    \frac{ 1 }{\langle W \rangle} \ .
\end{align}

Corrections to the Abrikosov-Khalatnikov results were derived by Brooker and Sykes in the late 1980 \cite{BS_PRL}. 
Their final results can be cast in the form
\begin{align}
\eta = \eta_{AK} \frac{1-\lambda_\eta}{4} 
 \sum_{k=0}^\infty \frac{4k+3}{(k+1)(2k+1)[(k+1)(2k+1)-\lambda_\eta]} \ ,
\label{eta_sb}
\end{align}
and 
\begin{align}
\label{cappa_sb}
\kappa = \kappa_{AK} \frac{3 - \lambda_\kappa}{4} \sum_{k=0}^\infty \frac{4k+5}{(k+1)(2k+3)[(k+1)(2k+3)-\lambda_\kappa]} \ .
\end{align}
The effect of the corrections, measured by the ratio between the results of Ref.~\cite{BS_PRL} and those of Ref.~\cite{kinetic_abrikal,abrikala_rep}, 
while being small to moderate on viscosity, turns out to be large on thermal conductivity.  One finds $0.750 < (\eta/\eta_{AK}) < 0.925$, 
 and $0.417 < (\kappa/\kappa_{AK}) < 0.561$.
 
 Equations \eqref{eta_sb} and \eqref{cappa_sb} show that the input required to obtain $\eta$ and $\kappa$ includes the effective mass,  the calculation of which has been discussed in Chapter \ref{chap3}, and the in medium scattering probability, which 
 can be readily obtained in Born approximation using the  CBF effective interaction.


\section{Calculation of the transport coefficients}
\label{sec:calctransp}

Before focusing on the effective masses and scattering probabilities employed in the calculation of the transport coefficients,  we 
briefly discuss the derivation of the CBF effective interaction for the 
hard-sphere system of degeneracy $\nu=2$, which turned out to exhibit a significant 
distinctive feature, with respect to the case $\nu=4$.

\subsection{Effective interaction for the hard-sphere system with $ \nu =2$}

The CBF effective interaction is fully determined from the correlation function $f(r)$, which is in turn
obtained solving the Euler\textendash Lagrange equation \eqref{ELeq}. Equations \eqref{eq:grel} and \eqref{eq:phiel} 
clearly show  that $f(r)$ depends on the degeneracy of the system through the  coefficient of the Slater function describing  
statistical correlations. The lower the degeneracy of the system the larger the effect of these correlations, the range
of which monotonically increases as the density of the system decreases. As a consequence, in the low-density region the 
determination of $f(r)$ from the numerical solution of Eq. \eqref{ELeq} with $\nu=2$ is hindered by the presence of long-range  
statistical correlations, whose effect is much larger than in the case $\nu=4$.

Owing to the above difficulty, at $c < 0.5$, the ground-state energy computed within the FHNC scheme does not develop a clear minimum as
a function of the variational parameter $d$. To overcome this problem, and obtain the accurate estimate of $\langle H \rangle$
needed to determine the CBF effective interaction at all densities, we have replaced the FHNC variational estimate with 
the ground state expectation value of the Hamiltonian resulting from Diffusion Monte Carlo (DMC) calculations. In addition, 
we have used the results of the Variational Monte Carlo (VMC) approach to gauge the accuracy of the FHNC approach at $c \geq0.5$. 


Within VMC, the multidimensional integrations involved in the calculation of the expectation value of the Hamiltonian in the correlated ground-state  is evaluated using Metropolis Monte Carlo quadrature\cite{metropolis:1953}. The trial wave function, chosen to be the same  as in the FHNC calculation, is defined in terms of the Jastrow-type correlation functions of 
Eq. (\ref{eq:jastrow_central}) and the Fermi gas ground state 
\begin{equation}
\label{trialwf}
\Psi_T(\mathbf{R})=\langle \mathbf{R} | \prod_{j<i}f(r_{ij})| 0_{FG} ] \ .
\end{equation}
In the above equation, $\mathbf{R} \equiv \{{\bf r}_1, \ldots , {\bf r}_N \}$ denotes the set of coordinates specifying the system in configuration space, 
and $| \mathbf{R} \rangle$ is the corresponding eigenstate.

The infinite system is modeled by considering a finite number of particles in a box and imposing periodic boundary conditions. As a consequence, 
the spectrum of eigenvalues of  the wave vector $\mathbf{k}$ is discretized. For a cubic box of side $L$, one finds the familiar result
\begin{equation}
k_i=\frac{2\pi}{L} n_i \ \ , \ \  i=x,y,z  \ \ , \ \  n_i=0,\pm 1, \pm 2, \dots \ .
\end{equation}
In order for the  wave function to describe a system with vanishing total momentum and angular momentum, all shells corresponding to momenta 
such that $|{\bf k}| < k_F$ must be filled. This requirement determines a set of ``magic numbers'', which are commonly employed in simulations 
of periodic systems. For example, the VMC\textemdash as well as the DMC\textemdash calculation whose results  are used in this Thesis have been obtained 
with 132 particles, corresponding to 66 and 33 momentum states for degeneracy $\nu=2$ and 4, respectively.

The expectation value of the Hamiltonian in the state described by the the trial wave function of Eq. \eqref{trialwf} can be cast in the form
\begin{equation}
\langle \Psi_T | H | \Psi_T\rangle = \int d\mathbf{R} \  E_L(\mathbf{R}) P(\mathbf{R}) \ ,
\end{equation}
where the local energy $E_L(\mathbf{R}) $ is defined as
\begin{equation}
E_L(\mathbf{R}) = \frac{H\Psi_T(\mathbf{R})}{\Psi_T(\mathbf{R})} \ ,
\end{equation}
and we have introduced the probability density $P(\mathbf{R})~\equiv~\Psi_{T}^*(\mathbf{R}) \Psi_T(\mathbf{R})$. Within VMC, the above integral 
is estimated by a sum over the set $ \{{\bf R} \} $, consisting of $N_c$ configurations  sampled from the distribution $P(\mathbf{R})$ using the Metropolis algorithm
\begin{equation}
\langle \Psi_T |  H | \Psi_T \rangle \approx \frac{1}{N_c} \  \sum_{{\bf R}_i \in \{\mathbf{R}\}} \frac{\Psi_{T}^*(\mathbf{R}_i) H \Psi_T(\mathbf{R}_i)}{P(\mathbf{R}_i)} \ . 
\end{equation}

The VMC approach can be seen as an alternative to the cluster expansion technique underlying the FHNC approach, allowing for a stringent test of the approximation implied by the  neglect of  cluster contributions associated with the so-called elementary diagrams \cite{clarkvariational}.

The main drawback of VMC, obviously shared by FHNC, 
is that the accuracy of the result entirely depends on the quality of the trial wave function. The DMC method\cite{kalos:2012,grimm:1971} overcomes the limitations of the variational approach by using a projection technique to enhance the true ground-state component of the trial wave function.
This
result is achieved expanding $|\Psi_T\rangle$  in eigenstates of the Hamiltonian according to
\begin{equation}
|\Psi_T\rangle=\sum_n c_n |n\rangle \qquad ,\qquad {H}|n\rangle = E_n |n\rangle\ ,
\end{equation}
which implies
\begin{equation}
\lim_{\tau\to\infty}e^{-({H}-E_0)\tau} |\Psi_T\rangle=c_0 |0\rangle\, ,
\end{equation}
with $\tau$ being the imaginary time. Provided  $|\Psi_T \rangle$ it is not
orthogonal to the true ground state, i.e. for $c_0\neq 0$, in the limit of large $\tau$ the above procedure projects out the exact lowest-energy state.

Beacuse the direct calculation of $\exp[-({H} -E_0)\tau]$ involves prohibitive difficulties, the imaginary-time evolution is broken into $N$ small imaginary-time steps,  and complete sets of position  eigenstates are inserted, in such a way that only the calculation of the short-time propagator is required. This procedure yields the expression
\begin{align}
\langle \mathbf{R}_{N+1}   | e^{-({H}-E_0)\tau}  | \mathbf{R}_1 \rangle
 & =   \int     d{\bf R}_2 \ldots d{\bf R}_N 
 \langle \mathbf{R}_{N+1} | e^{-({H} -E_0)\Delta\tau} | \mathbf{R}_{N}\rangle  \nonumber \\
&  \times \langle \mathbf{R}_{N} | e^{-({H} -E_0)\Delta\tau} | \mathbf{R}_{N-1} \rangle \dots
 \langle \mathbf{R}_2| e^{-({H} -E_0) \Delta \tau} | \mathbf{R}_1 \rangle \ ,
\label{eq:propagation}
\end{align}
where, for the sake of simplicity, the dependence on the discrete  degrees of freedom has been omitted. Monte Carlo techniques are used to sample the
paths $\mathbf{R}_i$ in the propagation.
Note that, although Eq.~\eqref{eq:propagation} is only exact in the  $\Delta\tau\to 0$ limit, its accuracy 
can be tested performing several simulations with smaller and smaller time step and extrapolating to zero.

\begin{figure}[ht]
\begin{center}
 \includegraphics[width= 0.47\textwidth]{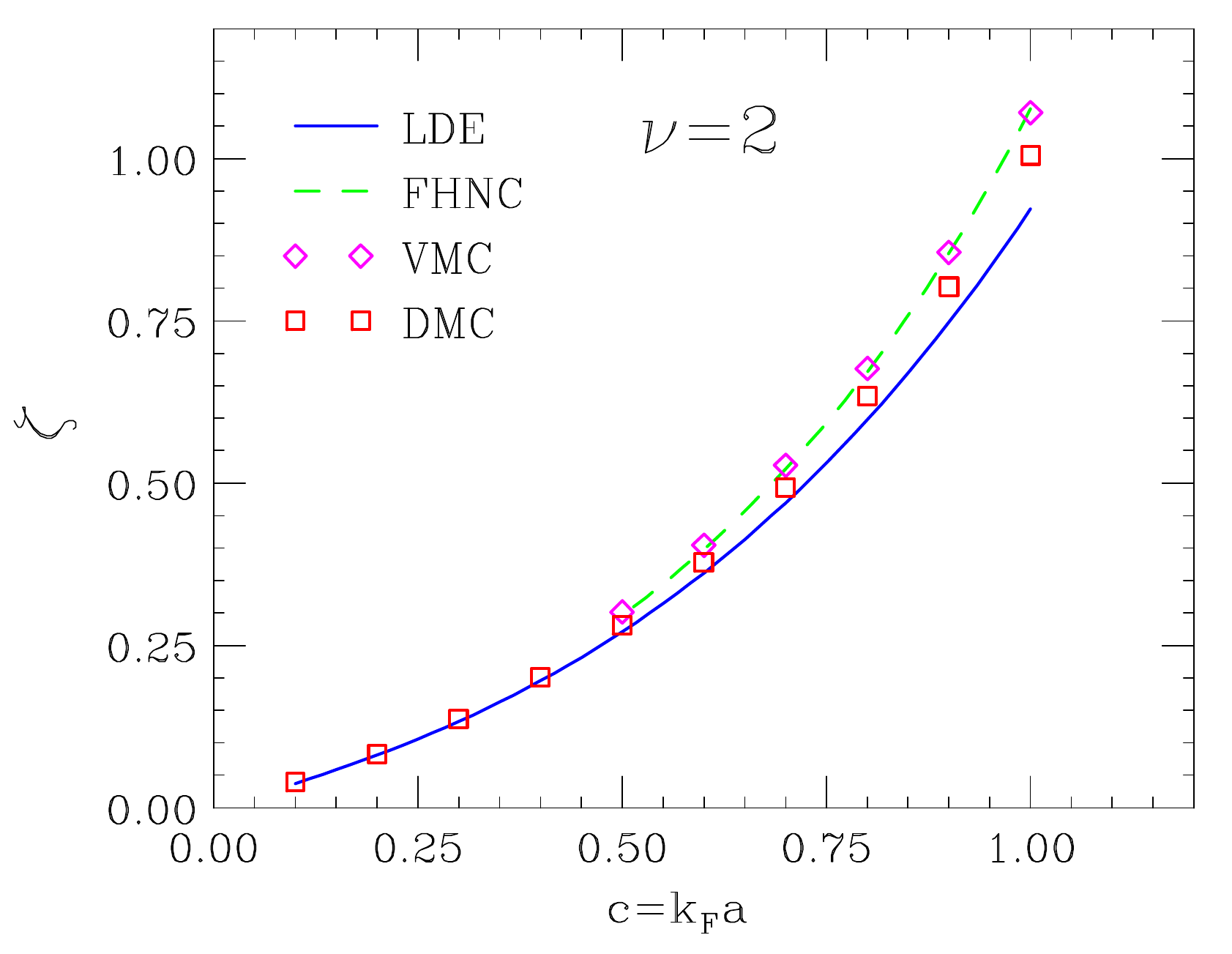}
  \includegraphics[width= 0.45\textwidth]{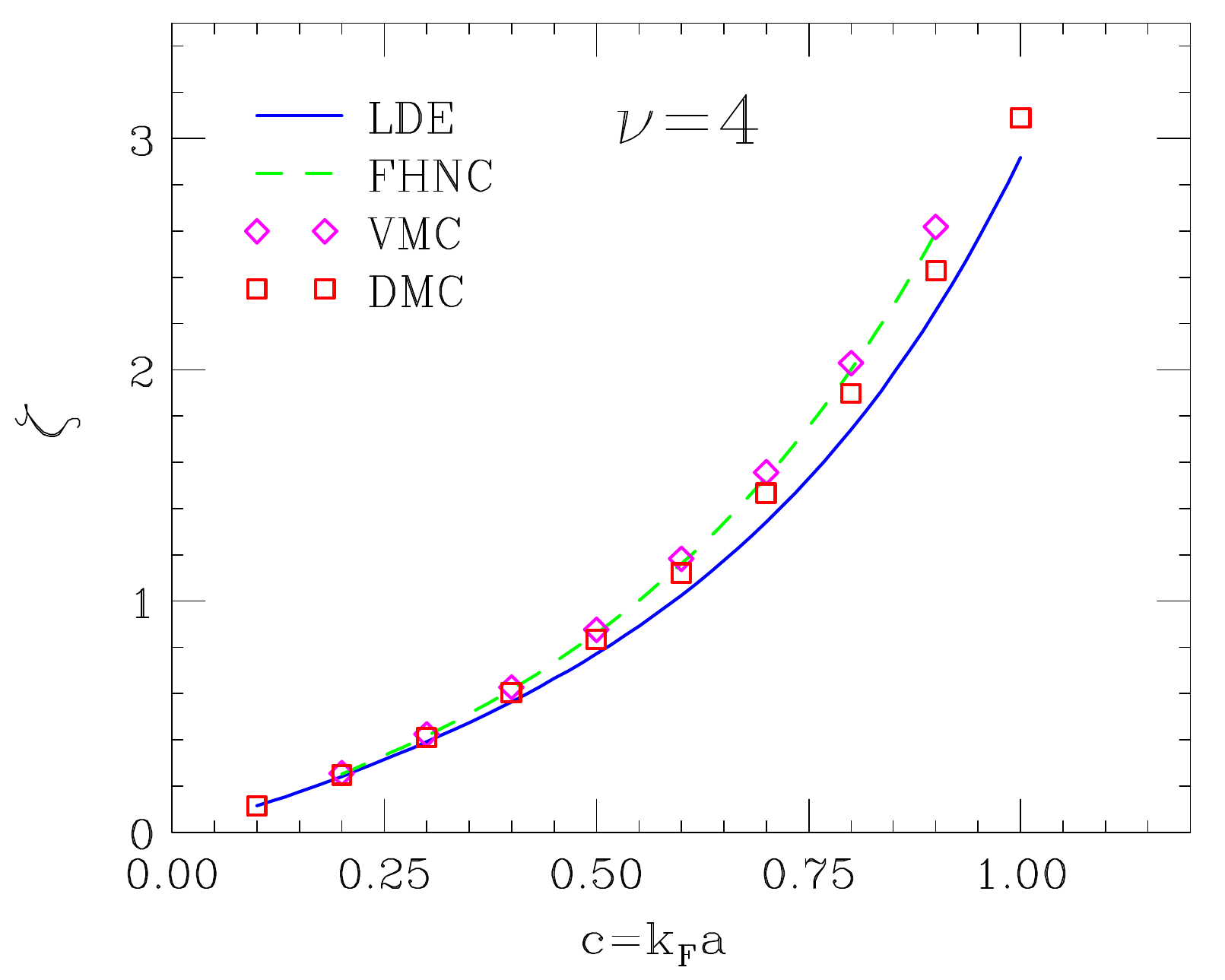}
  \caption[HS ground state energy for $\nu = $2 and $\nu = $4]{c-dependence of the quantity $\zeta$, defined by Eq. \eqref{def:z}, for degeneracy $\nu=$2 (left panel) and 4 (right panel). The solid and dashed lines show the results obtained from 
  the low-density expansions \eqref{eq:E0nu2} and \eqref{eq:E0nu4} and the variational FHNC approach, respectively. The VMC and DMC results 
  are represented by diamonds and squares. Monte Carlo error bars are not visible on the scale of the figure.}
\label{fig:MC24}
\end{center}
\end{figure}

In Fig. \ref{fig:MC24} the results of DMC calculations of the quantity $\zeta$ of Eq. \eqref{def:z}, yielding the deviation of the ground state energy from the Fermi gas result, are compared to the values obtained from the VMC and FHNC approaches\textemdash which are only available at $c \geq 0.5$\textemdash as well as to the predictions of the low-energy expansions \eqref{eq:E0nu2} and \eqref{eq:E0nu4}. 
It clearly appears that the VMC and FHNC 
results are very close to one another,  thus showing that at $c\geq 0.5$ the FHNC approximation does provide an upper bound to the ground state energy. 
The accuracy of the variational result is measured by the difference between the VMC\textemdash or, equivalently, FHNC\textemdash values of $\zeta$ and those obtained from DMC. In the case of degeneracy $\nu =2$, illustrated in the left panel, this difference ranges between  2\% and 9\% at $0.2 < c < 1$.  Note that a 9\% difference in $\zeta$ translates in a difference of less that 3\% in the ground 
state energy $E_0$.  The low-density expansion turns out to be quite accurate, its predictions being within 
5\% of the DMC results at $c < 0.5$. The right panel of Fig. \ref{fig:MC24} shows the results corresponding to $\nu=4$, which exhibit    the same 
pattern.

The CBF effective interaction has been computed from Eq. \eqref{veff:final2} choosing the 
correlation range $d$ in such a way as to to reproduce the ground-state expectation value of the Hamiltonian obtained using the  
DMC technique.

\subsection{Quasiparticle spectrum and effective mass}
\label{masseff}

Once the effective interaction has been determined, the calculation of the quasiparticle energy and effective mass can be performed following the 
procedure described in Chapter \ref{chap3} for the case of degeneracy 4.

\begin{figure}[htbp]
\begin{center}
\includegraphics[width= 0.6\textwidth]{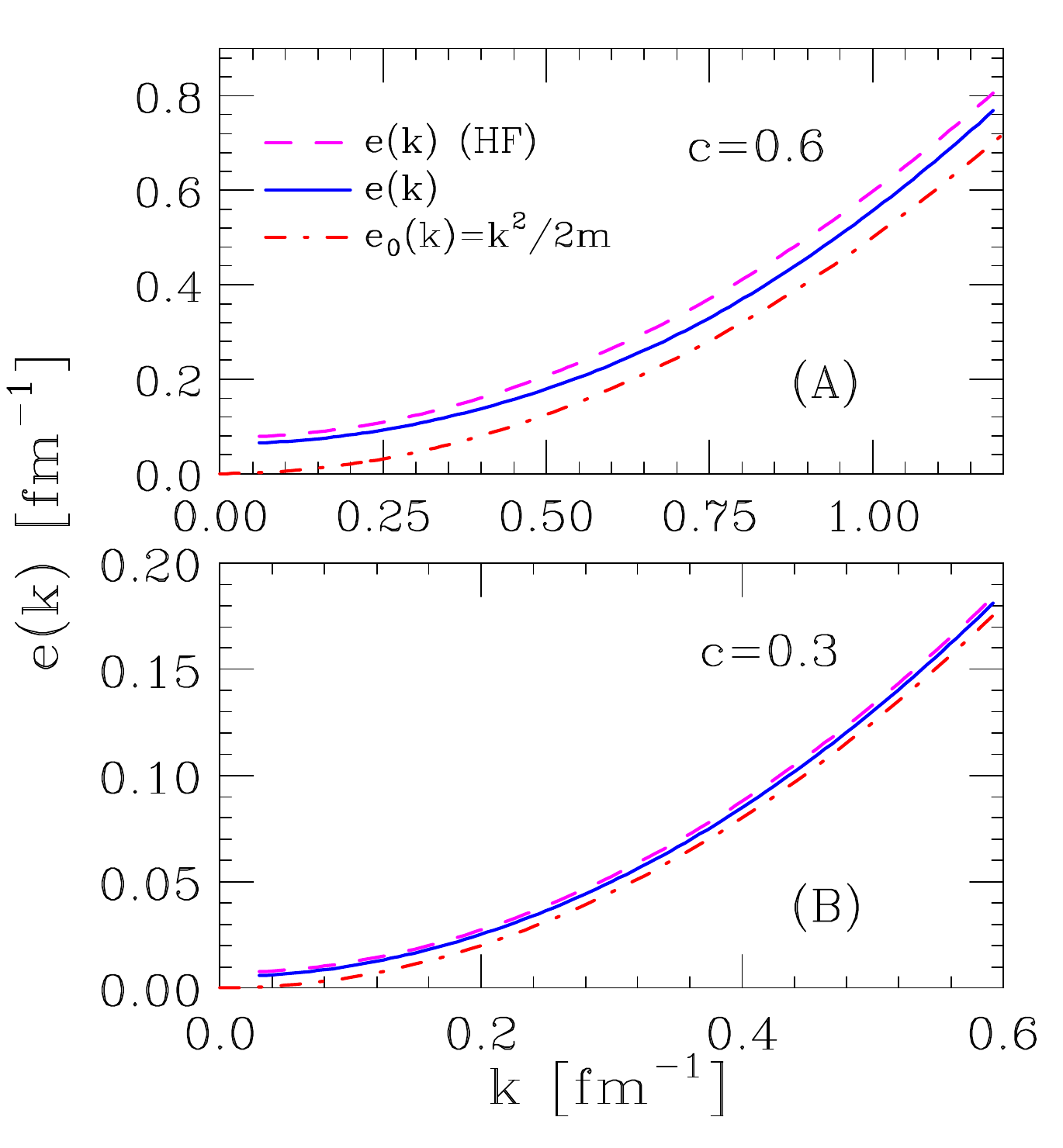}
\caption[Quasiparticle spectrum for  $\nu = 2$]{Quasiparticle energy of the fermion hard-sphere system of degeneracy $\nu = 2$ at $c=0.3$ [panel (B)]
and 0.6 [panel(A)]. The meaning of the lines is the same as in Fig.~\ref{fig:ek}:
the dashed and solid lines correspond to the first order (\emph{i.e.} Hartree-Fock) and second order
approximations to the self-energy, respectively. For comparison, the dot-dash lines show the kinetic energy spectrum. }
\label{fig:spe_nu2}
\end{center}
\end{figure}

\begin{figure}[htbp]
\begin{center}
\includegraphics[width= 0.6\textwidth]{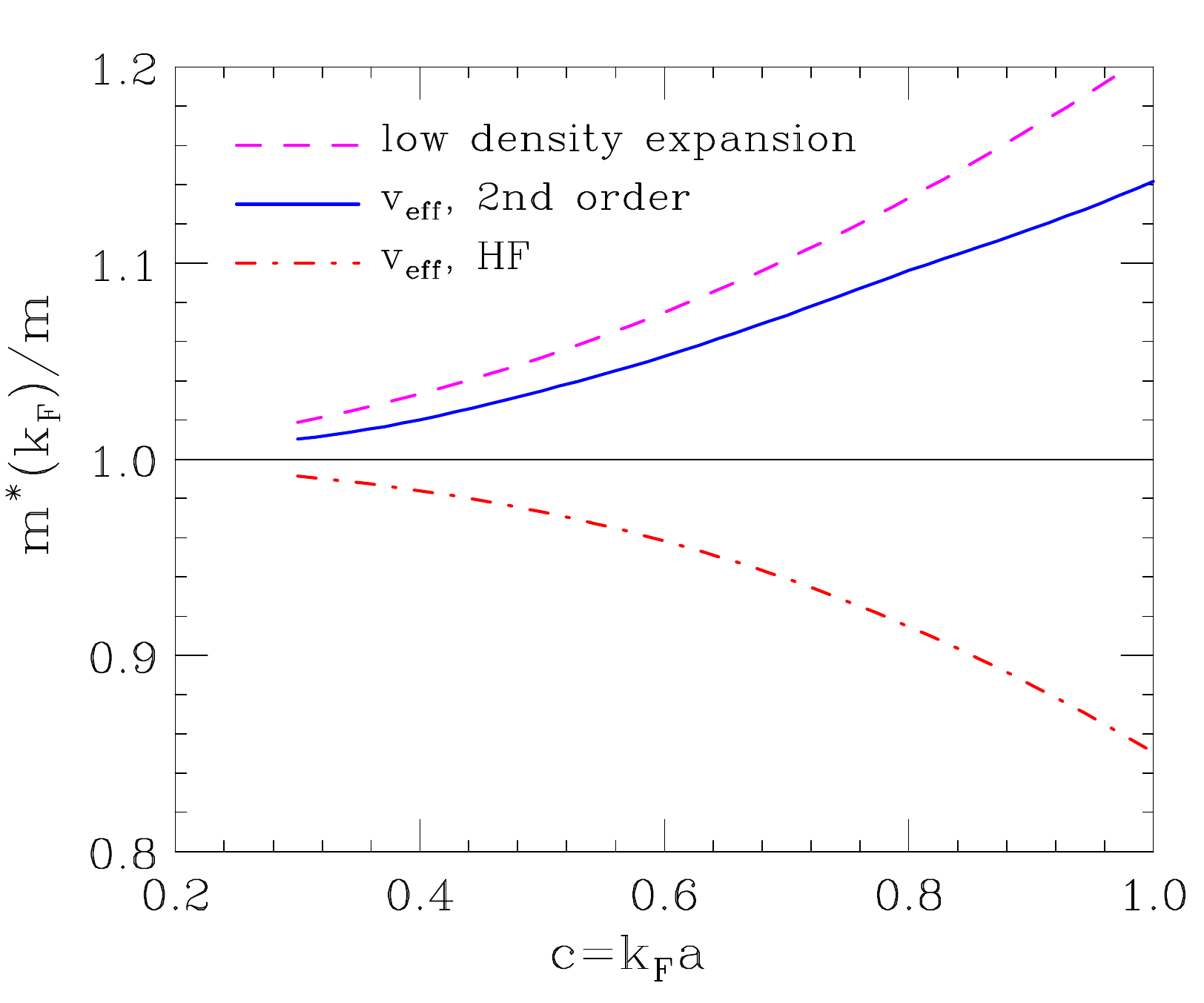}
\caption[The effective mass $m^{\star}$ for  $\nu = 2$]{$c$-dependence of the ratio $m^\star(k_F)/m$ of the hard-sphere system of
degeneracy  $\nu = 2$. The meaning of the lines is the same as in Fig. \ref{mstar1}: the dot-dash and solid lines represent the results of calculations carried out using the first and second order approximations to the self-energy.  For comparison, the dashed line shows the results computed using the low-density expansion of Eq.~\eqref{pert_mstar}.
}
\label{fig:mstar_nu2}
\end{center}
\end{figure}

The resulting spectrum,  defined by  Eq.~\eqref{def:spectrum}, is displayed in Fig.~\ref{fig:spe_nu2}, while Fig.~\ref{fig:mstar_nu2} shows the 
corresponding effective mass, obtained  from of Eq.~\eqref{def:mstar}.
Second order corrections to the self-energy have the same effects  observed in the case $\nu=4$. 
The inclusion of the energy-dependent contributions results in small modifications of the quasiparticle spectrum, but  dramatically  affects both the magnitude and the  density  dependence of the effective mass at $|{\bf k}| = k_F$. 

For reference, Fig ~\ref{fig:mstar_nu2} also shows the effective mass computed using the low-density expansions,  Eq. \eqref{pert_mstar}. 
The difference between the value of  $m^{\star}(k_F)$ evaluated using the  CBF effective interaction and the one obtained  
from Eq. \eqref{pert_mstar} turns out to be $\leq 2\%$ for $c\leq 0.6$, and grows up to $5.5\%$ as the value of $c$ increases up to $c=1.0$.

\subsection{Scattering probability}
\label{sec:scatprob}

The scattering probability $W(\theta, \phi)$ appearing in the collision integral of Eq. \eqref{boltzmann} is  trivially related to  the scattering 
amplitude $f(\theta, \phi)$ through the relation
\begin{align}
\label{eq:W}
W(\theta, \phi) = \frac{2 \pi}{2} \left |  \frac{4 \pi}{ m }  f(\theta, \phi) \right |^2   \ .
\end{align}
The scattering amplitude is in turn related to  the differential cross section according to
\begin{align}
\label{eq:scat_ampl}
\frac{d \sigma}{ d \Omega } = |f(\theta,\phi) |^2 \ . 
\end{align}
Combining the above equations one finds\footnote{
Note that the relation between scattering probability and cross section reported in Ref. \cite{BV}, in which the 
factor $16 \pi^3/m^2$  is replaced by $16 \pi^2/m^2$,  is incorrect.
}
\begin{align}
\label{dsigma:W}
W(\theta, \phi) = \frac{16      \pi^3}{m^2}  \left( \frac{d\sigma}{d\Omega} \right) \ .
\end{align}

The scattering cross section is usually expressed in either the laboratory (L) or the center-of-mass (CM) frame. However, the 
Abrikosov\textendash Khalatnikov formalism is derived in a different  frame, referred to as AK frame, in which the 
Fermi sphere is at rest.

To clarify the connection betweel AK and CM frame, let us consider the process in which two particles carrying momenta ${\bf k}_1$ and
${\bf k}_2$ scatter to final states of momenta ${\bf k}_1^\prime$ and ${\bf k}_2^\prime$.
The total energy of the initial state
\begin{align}
E = \frac{{\bf k}_1^2}{2m} + \frac{{\bf k}_2^2}{2m}
\end{align}
can be conveniently rewritten in terms of the center of mass and relative momenta,
${\bf K}~= {\bf k}_1~+~{\bf k}_2$ and ${\bf k}~=~({\bf k}_1~-~{\bf k}_2)/2$, as
\begin{align}
E = \frac{{\bf K}^2}{2M} + \frac{{\bf k}^2}{2\mu} = {\cal E}+ {\cal E}_{{\rm rel}} \ ,
\end{align}
with $M = 2m$ and $\mu = m/2$. In the CM reference frame, in which the center of mass of the system is at rest, 
$E=E_{{\rm cm}}={\cal E}_{{\rm rel}}$, while in the L frame, in which ${\bf k}_2=0$,
$E=E_{{\rm L}}= 2{\cal E}_{{\rm rel}}$.

In strongly degenerate systems, the magnitude of all momenta
playing a a role in the determination of the transport coefficients is equal to the Fermi momentum,
 and conservation of energy requires that the angle between the 
momenta of the particles participating in the scattering process be the same before
and after the collision. In general, however, the angle $\phi$ between the initial and
final relative momenta, ${\bf k}$ and ${\bf k}^\prime = ({\bf k}_1^\prime - {\bf k}_2^\prime)/2$, defined
through
\begin{align}
\cos \phi = \frac{ ({\bf k} \cdot {\bf k}^\prime) }{ |{\bf k}||{\bf k}^\prime| } \ ,
\label{def:phi}
\end{align}
does not vanish. Hence, for any given Fermi momentum, i.e. for any given matter density, the
scattering process in the AK frame is specified by the center of mass energy
\begin{align}
{\cal E}_{{\rm AK}} = \frac{k_F^2}{2m} (1 + \cos \theta ) \ ,
\end{align}
and the two angles $\theta$ and $\phi$.


The AK-frame variables can be easily connected to those of the CM reference frame. Since the relative kinetic energy, \emph{i.e.} the energy in the CM reference frame $E_{\rm cm}$,  is the same in any  frame, we have
\begin{align}
\label{ecm}
E_{\rm  cm} = E^{ AK}_{  rel}  = \frac{k_F^2}{2m} (1-\cos \theta)  \ , 
\end{align}
where we have used again the condition that  scattering processes involve particles in momentum states close to the Fermi surface.
Moreover, the  angle between the planes containing ingoing and outgoing momenta, $\phi$ , is nothing but the angle  between the  initial and final relative momenta, and can therefore be identified with the scattering angle in the CM frame, setting  
\begin{align}
\label{thetacm}
\Theta_{\rm cm} = \phi \ .
\end{align}
Through the above relations, the differential cross section in the CM frame, written as a function of the two variables $E_{\rm  cm}$ and 
$\Theta_{\rm cm}$, can be transformed into the corresponding quantity in the AK frame, depending on the two angular variables 
$\theta$ and $\phi$, needed for the calculation of the transport coefficients. We can write
\begin{equation}
\frac{d \sigma}{ d \Omega} [ E_{\rm cm}(\theta),  \Theta_{\rm cm}(\phi)] =  \frac{d \sigma}{ d \Omega} ( \theta,  \phi ) \ , 
\end{equation}
with $E_{\rm cm}(\theta)$ and $\Theta_{\rm cm}(\phi)$ given by Eqs. \eqref{ecm} and \eqref{thetacm}.

In the pioneering works of Refs. \cite{tranpsI,tranpsII}, the scattering probability in neutron star matter was computed from Eq.\eqref{dsigma:W} replacing the 
bare nucleon mass with an effective mass and using the nucleon-nucleon scattering cross section in free space, obtained from the measured phase shifts.
This procedure accounts for the fact that both the incoming flux and the phase space available to the final state particles are affected by the presence of the medium. However, it neglects possible medium modifications of the scattering probability.

The authors of Ref. \cite{BV} improved upon the approximation of Refs. \cite{tranpsI,tranpsII}, using the CBF effective interaction to obtain both the effective mass and the in medium scattering cross section of pure neutron matter within a consistent framework.

In this Thesis, we have applied the approach of Ref. \cite{BV} to the fermion hard sphere system. The in medium scattering probability  
has been computed in Born approximation using  the CBF effective interaction and the definition
\begin{align}
W(\theta, \phi ) = \pi \left | [ {\bf k}_1^{\prime},{\bf k}_2^{\prime} |  \veff  |   {\bf k}_1 ,  {\bf k}_2  ] _a^{} \right|^2  \ ,
\end{align}
where ${\bf k}_i$ and ${\bf k}_i^{\prime}$ are the  initial and final momenta, respectively.
The calculation of the matrix element is essentially the same as that performed to obtain the second order contributions to the self-energy, described in 
Appendix \ref{app_self}. The only differences stem from the fact that, because here we are considering a scattering process, 
in Eq. \eqref{eq:msquared} we need to average over the  two spins of the initial state particles. The result can be written in the form
\begin{align}
 \frac{1}{\nu^2}\sum_{\sigma, \sigma^{\prime}  } 
\left | [ {\bf k}_1^{\prime},{\bf k}_2^{\prime} |  \veff  |   {\bf k}_1 ,  {\bf k}_2  ]_a ^{} \right|^2  
=
   M^2 (u)     +  M^2(v)          -  \frac{2}{\nu}  M(u) M(v)  \  .
\end{align}
Following the  notation of Appendix \ref{app_self},  in the above equation $ M(u)$ and $M(v)$ denote the Fourier transforms of the effective potential for arguments given by the  following   combination of initial  and final relative momenta, ${\bf k}$ and $ {\bf k}^{\prime}$
\begin{align}
 u &= | {\bf k} - {\bf k}^{\prime} |  =    k_F\sqrt{ (1 - \cos \theta) (1 - \cos \phi )  } \nonumber \\
 v &= | {\bf k} + {\bf k}^{\prime} | =    k_F\sqrt{ (1 - \cos \theta) (1 + \cos \phi )  }   \ . 
 \end{align}
 
\begin{figure}[htbp]
\begin{center}
\includegraphics[width= 0.6\textwidth]{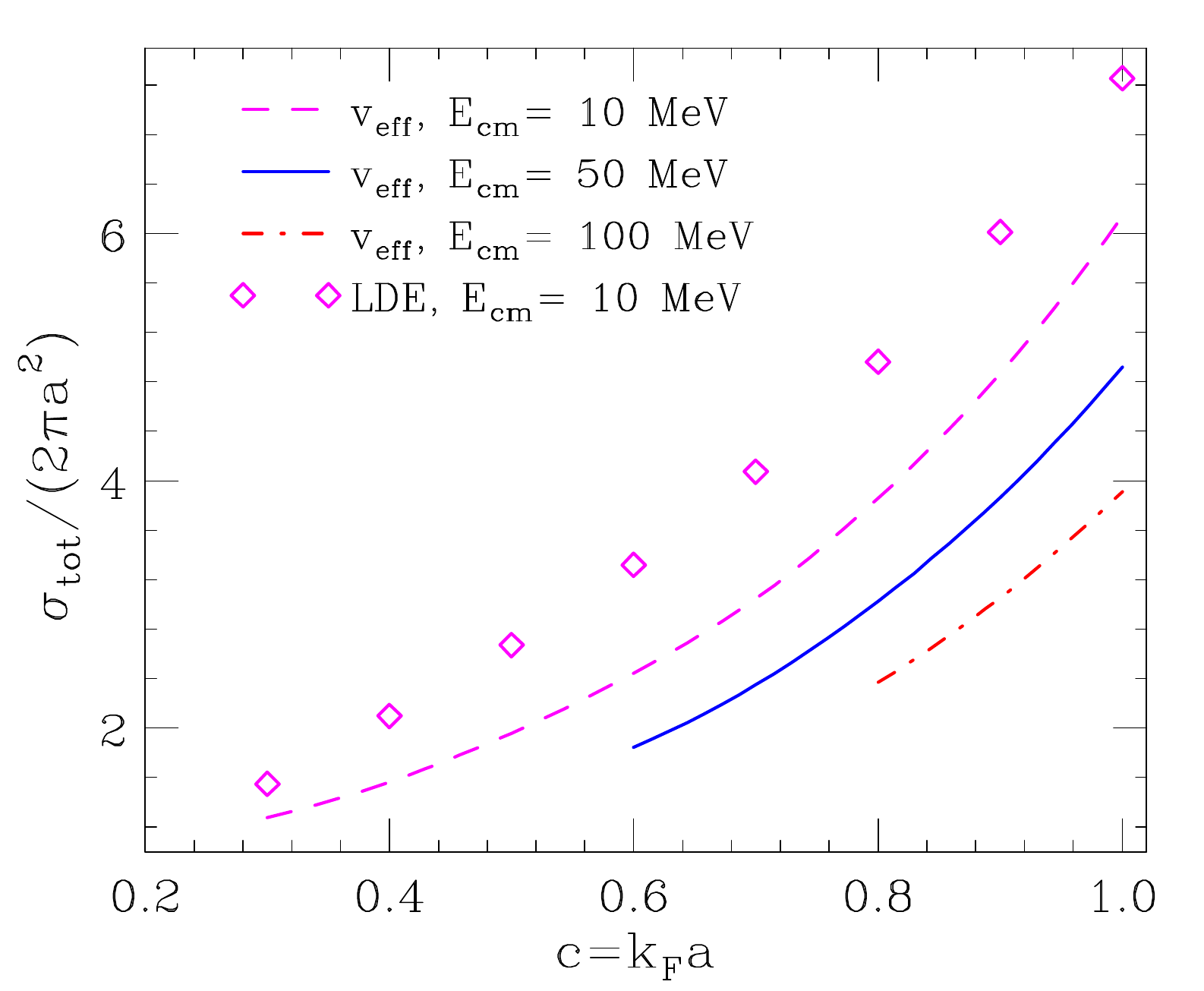}
\caption[$c$-dependence of in-medium total cross section]{$c$-dependence of the in-medium total cross section of the
fermion hard-sphere system with $\nu$=2\textemdash normalized to the low-energy limit in vacuum\textemdash computed using the CBF effective interaction for different values of the CM energy $E_{\rm cm}$. For comparison, the diamonds show the results of the low-density expansion derived in Ref.~\cite{JLTP1}.
}
\label{fig:sigma_E}
\end{center}
\end{figure}

\begin{figure}[htbp]
\begin{center}
\includegraphics[width= 0.6\textwidth]{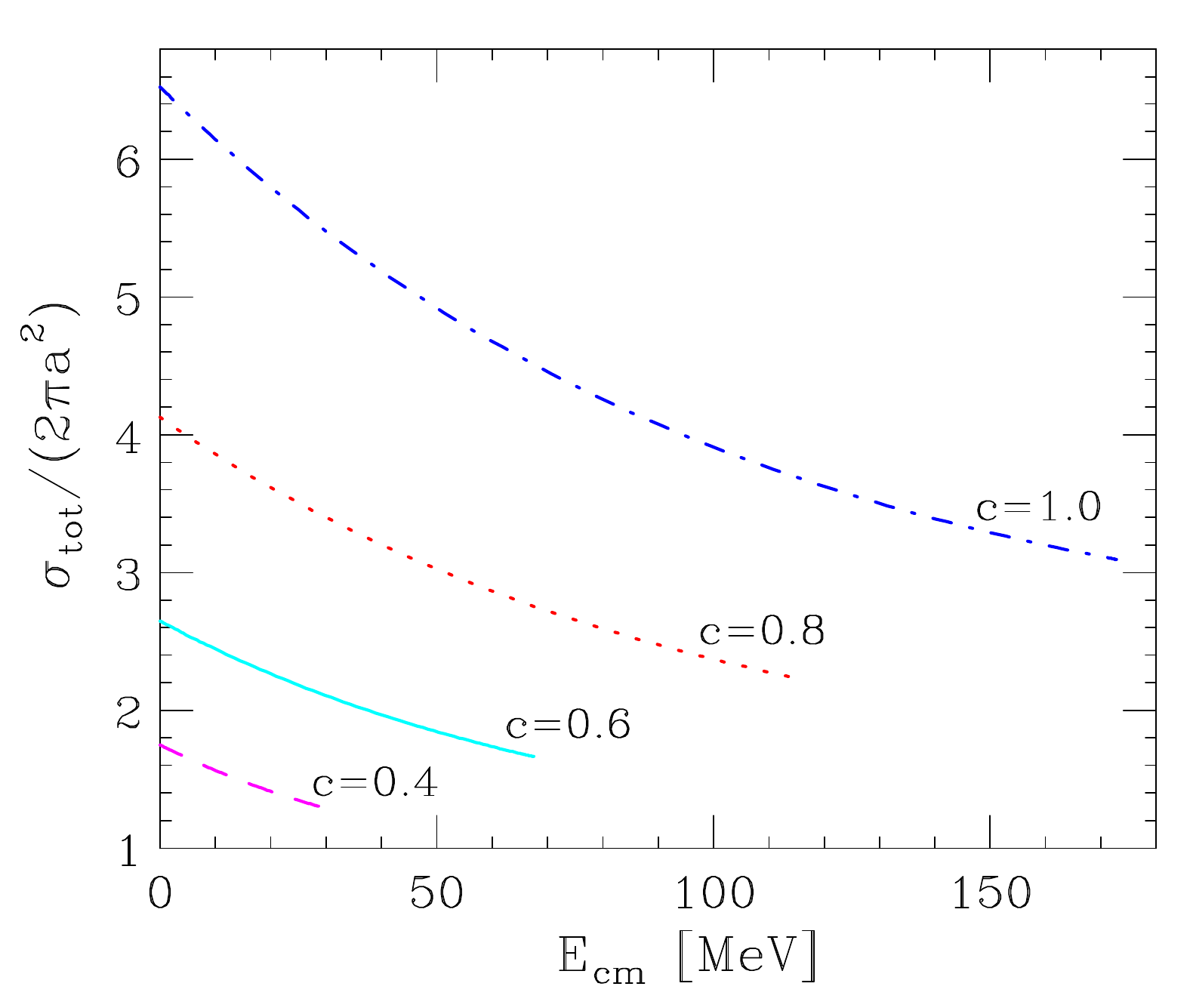}
\caption[$E_{\rm cm}$-dependence of in-medium total cross section]{
$E_{\rm cm}$-dependence of the in-medium total cross sections  of the
fermion hard-sphere system with $\nu$=2\textemdash normalized to the low-energy limit in vacuum\textemdash computed using the CBF effective interaction for different values of $c = k_F a$.
}
\label{fig:sigmatot}
\end{center}
\end{figure}

 The density dependence of the total cross section
 \begin{align}
 \label{sigmatot}
 \sigma_{\rm tot} = \int d \Omega \ \left(  \frac{d\sigma}{d\Omega} \right) , 
 \end{align}
 resulting from our calculations is shown in Fig.~\ref{fig:sigma_E} for center of mass energies $0 \leq E_{\rm cm} \leq 140$ MeV. 
 For any given value 
 of $E_{\rm cm}$, Eq. ~\eqref{ecm} implies that the Fermi momentum must satisfy  the constraint $k_F > \sqrt{ m E_{\rm cm} }$.
 Note that 
 $\sigma_{\rm tot}$ is normalized to the low-energy limit obtained from the partial wave expansion of the cross section in vacuum, $\sigma_{\rm tot} = 2 \pi a^2$.
 In Fig.~\ref{fig:sigmatot},  the same quantity is  shown as a function of CM energy, with $E_{\rm cm} < k_F^2/m$,  for different  densities 
 in the range  $0.4 \leq c \leq 1$.

The in medium scattering probability, defined as in  Eq. \eqref{eq:W},  
has been also studied within the framework of 
``standard'' perturbation theory  \cite{JLTP1,JLTP2}.  The authors of Ref. \cite{JLTP1} were able to obtain  the expression of $W(\theta, \phi)$  by solving the  generalised  Bethe\textendash Salpeter  equation for the scattering amplitude of a dilute gas of Fermi hard spheres, including terms up to order $c$, corresponding to order $c^2$ for the scattering probability. 
A more detailed description of the work of Ref. \cite{JLTP1} can be found in Appendix \ref{app_ampl}. 
For comparison, in Fig.~\ref{fig:sigma_E} the corresponding results at $E_{\rm cm}=10$ MeV are shown by the diamonds. 

The results displayed in Figs.~\ref{fig:sigma_E} and \ref{fig:sigmatot}, showing that $\sigma_{\rm tot}$ increases with density,  can be 
explained considering that the range of  the effective interaction\textemdash that takes into account screening arising form dynamical correlations\textemdash  is larger than the hard-sphere radius $a$, and grows with $c$ (see Fig. \ref{veff:plot}).

\section{Numerical results}

\begin{figure}[htbp]
\begin{center}
\includegraphics[width= 0.6\textwidth]{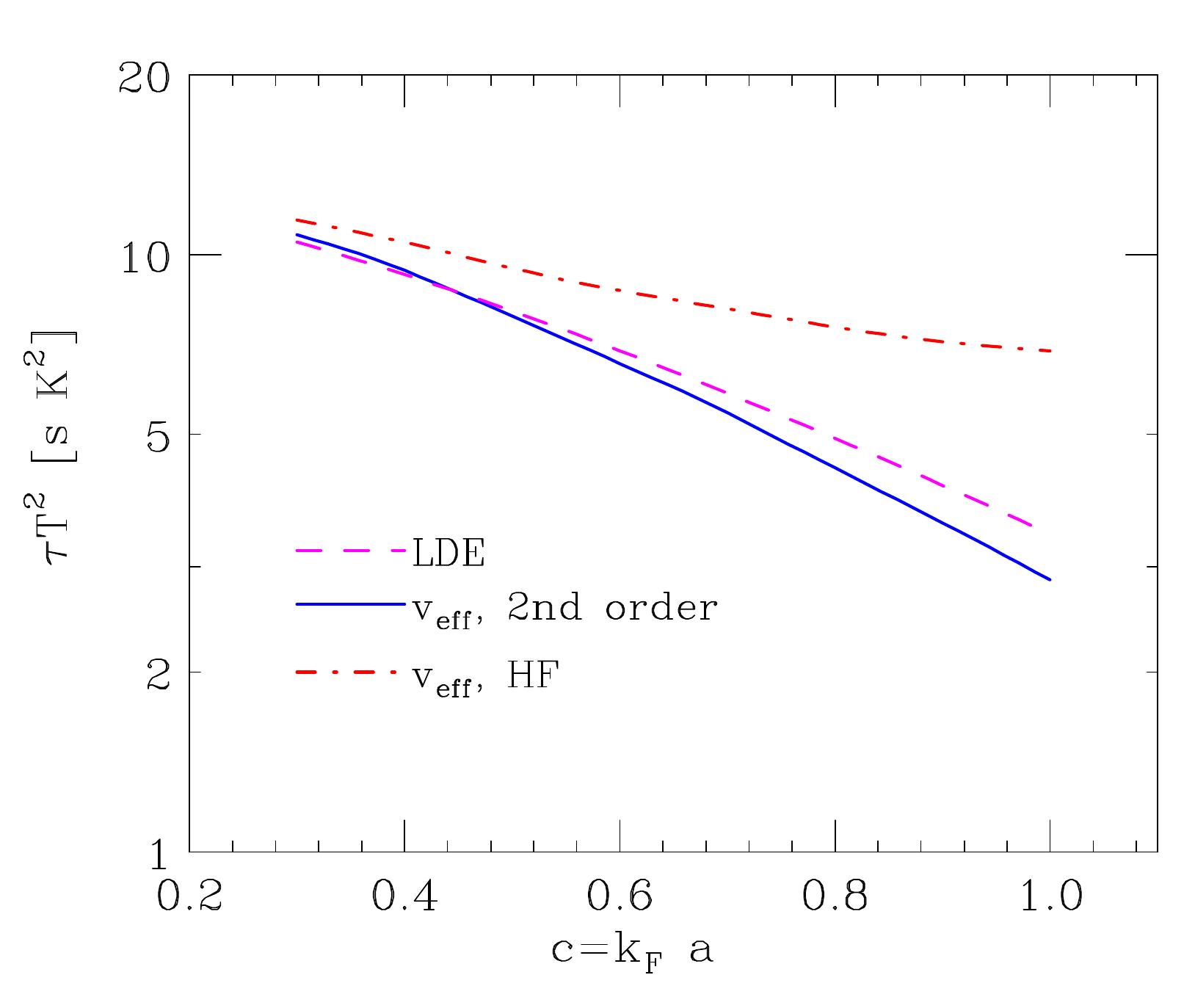}
\caption[$c$-dependence of  the quasiparticle lifetime]{
$c$-dependence of the temperature-independent quantity $\tau T^2$, where
$\tau$ is the quasiparticle lifetime of the fermion hard-sphere system of degeneracy  $\nu$=2. The dot-dash and solid lines represent the results of calculations carried out using the first and second order approximations for the effective mass, respectively, while the dashed line shows the results of the low-density expansion of Ref. \cite{JLTP2}, including terms of order up to $c$.
}
\label{lifetime}
\end{center}
\end{figure}

The shear viscosity coefficient of the fermion hard-sphere system of degeneracy $\nu$=2, $\eta$,  has been obtained
from Eqs. \eqref{eta_AK}, \eqref{lambda:eta} and \eqref{eta_sb} with the effective mass
and the in-medium scattering probability computed using  the CBF interaction.
Before analyzing  the shear viscosity and thermal conductivity, in Fig. \ref{lifetime} we illustrate the $c$-dependence of the time independent quantity 
$\tau T^2$, where $\tau$  is the  quasiparticle lifetime of Eq. \eqref{tau_AK},  computed using the CBF effective interaction. 
For comparison the prediction of the low-density expansion of Ref. \cite{JLTP2} is also shown. Overall, the emerging pattern reflects the one observed 
in Fig.~\ref{fig:mstar_nu2}. As  expected, the large corrections to the Hartee-Fock estimate of the effective mass translate into large corrections to the quasiparticle lifetime.

\subsection{Shear viscosity and thermal conductivity}

The shear viscosity coefficient of the fermion hard-sphere system of degeneracy $\nu=2$, $\eta$,  has been obtained 
from Eqs.~\eqref{eta_AK}, \eqref{lambda:eta} and \eqref{eta_sb} with the effective mass
and the in medium scattering probability computed using  the CBF interaction. 

\begin{figure}[htbp]
\begin{center}
\includegraphics[width= 0.7\textwidth]{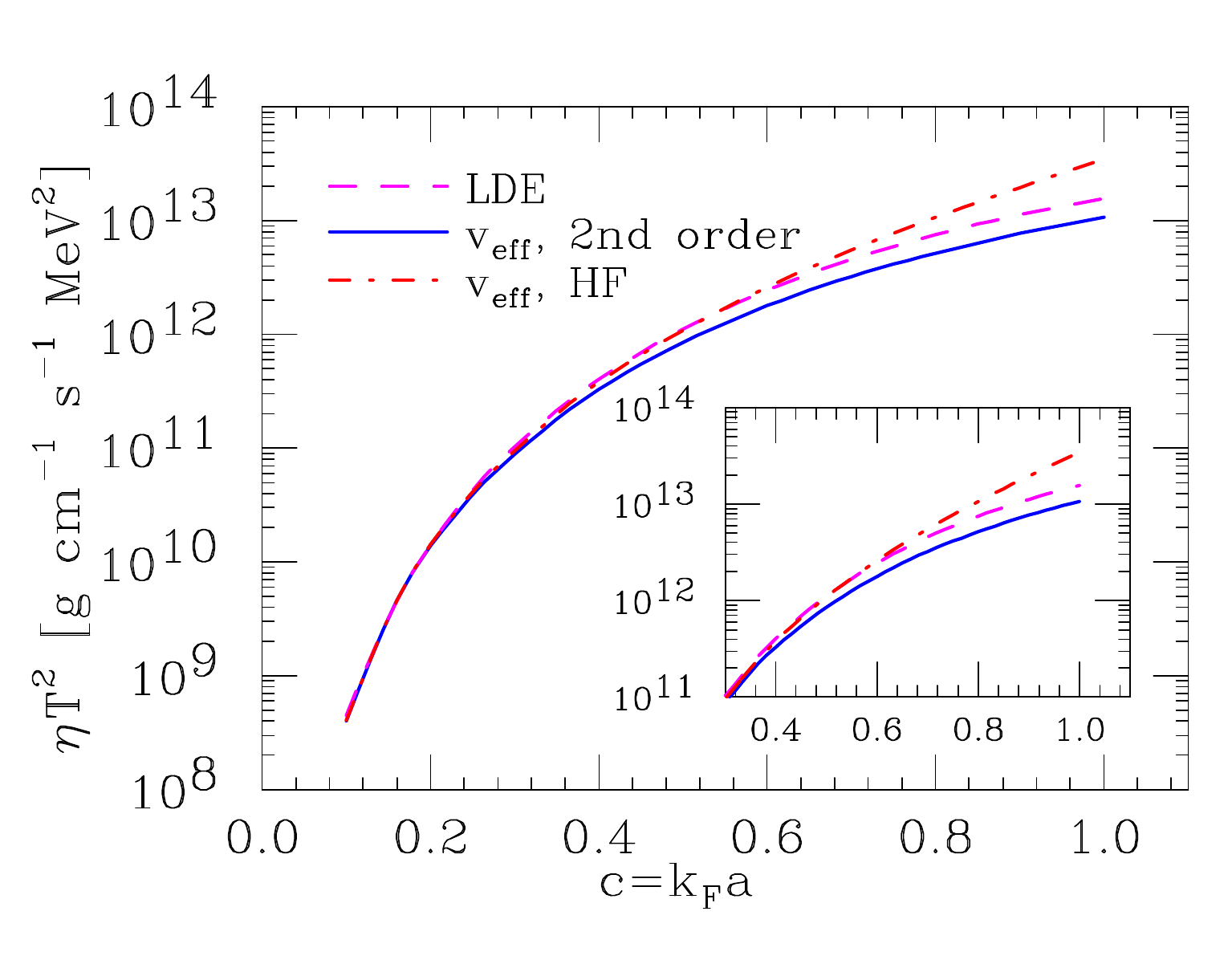}
\caption[$c$-dependence of the shear viscosity]{$c$-dependence of the temperature-independent quantity $\eta T^2$, where 
$\eta$ is the shear viscosity coefficient of the  fermion hard-sphere system of degeneracy  $\nu = 2$. The dot-dash and solid lines represent the results of calculations carried out using the first and second order approximations for the effective mass, respectively, while the dashed line shows the results of the low-density expansion of Ref. \cite{JLTP2}, including terms of order up to $c$.}
\label{eta1}
\end{center}
\end{figure}

Figure~\ref{eta1} shows the $c$-dependence of the
$T$-independent quantity   $\eta T^2$. The most relevant feature of the results displayed in the figure is, again,  the sizable effect
of second order contributions to the effective mass. As shown in Fig \ref{fig:mstar_nu2}, these corrections lead to sharp increase 
of $m^\star$, which in turn implies  a decrease 
of the shear viscosity coefficient $\eta$.   

\begin{figure}[htbp]
\begin{center}
\includegraphics[width= 0.7\textwidth]{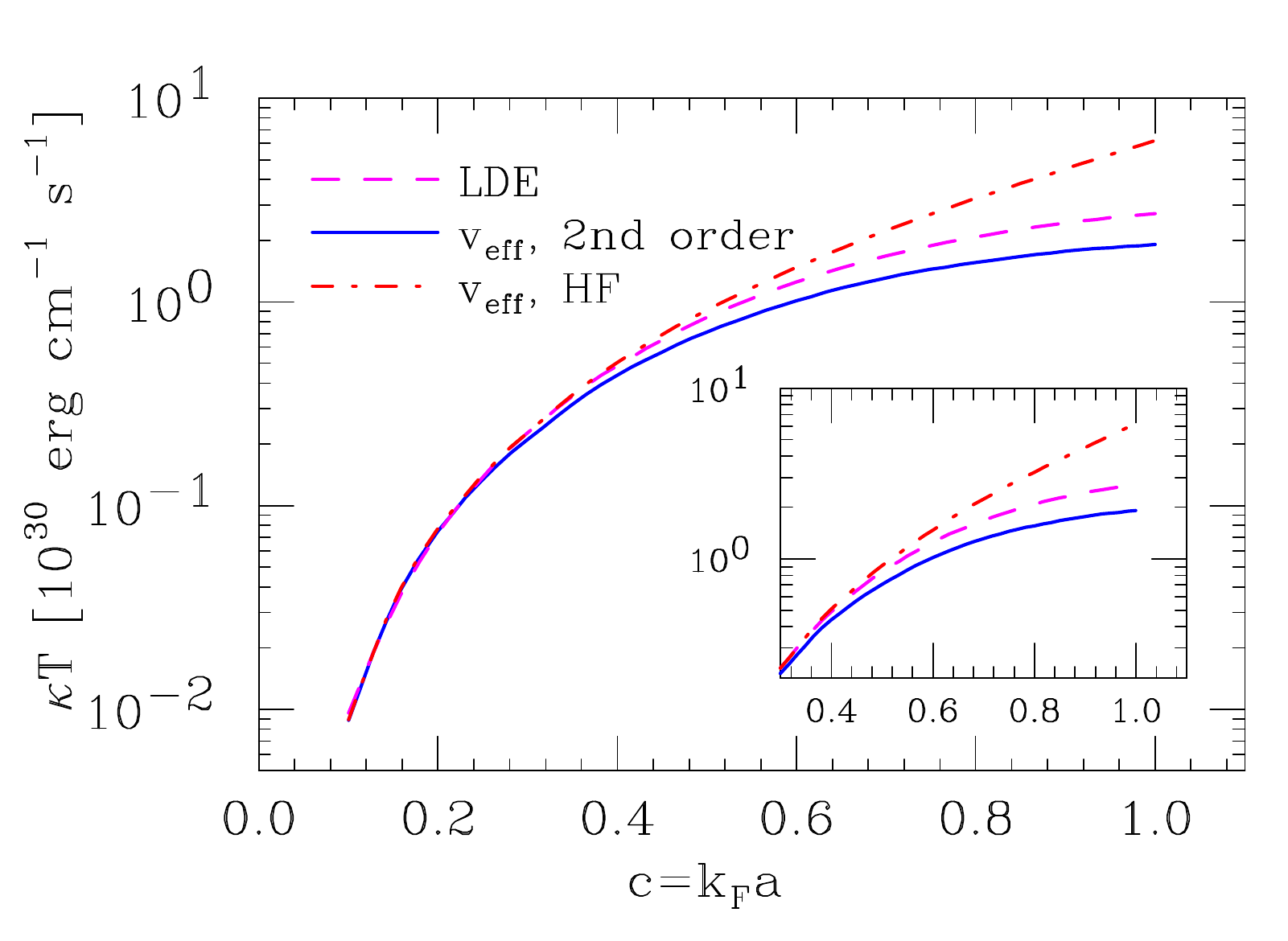}
\caption[$c$-dependence of the thermal conductivity]{$c$-dependence of the temperature-independent quantity $\kappa T$, where 
$\kappa$ is the thermal conductivity of the fermion hard-sphere system of degeneracy  $\nu = 2$. The dot-dash and solid lines represent the results of calculations carried out using the first and second order approximations for the effective mass, respectively, while the dashed line shows the results of the low-density expansion of Ref. \cite{JLTP2}, including terms of order up to $c$.}
\label{cappa1}
\end{center}
\end{figure}

The $T$-independent quantity $\kappa T$,  where $\kappa$ is the  
thermal conductivity defined by Eqs.~\eqref{cappa_AK}, \eqref{lambda:eta} and \eqref{cappa_sb}, is shown in Fig. \ref{cappa1} as a 
function of the dimensionless parameter $c$. Overall, the pattern is close to the one observed in Fig.~\ref{eta1}.

In order to establish a connection between our results and those corresponding to neutron matter, in Fig.~\ref{nk} we compare 
the momentum distribution of the fermion hard-sphere system at density corresponding to $c=0.4$\textemdash computed with the CBF effective interaction following the procedure described in  Sec.~\ref{sec:nk}\textemdash to those reported in Ref.\cite{nk_MC}, 
obtained using a quantum Monte Carlo technique. The shaded region illustrates the variation of the momentum distribution 
of Ref.~\cite{nk_MC} in the density range  0.08$\leq \rho \leq$0.24 fm$^{-3}$. The corresponding values of the renormalisation constant 
are $Z$=0.9566 for the hard-sphere system and $0.9579 \leq Z \leq 0.9378$ for neutron matter. The appreciably higher values of $n(0)$
obtained from the Monte Carlo approach reflect the softness  of the chiral neutron-neutron potential employed by the authors 
of Ref.~\cite{nk_MC}.  The results of Fig.~\ref{nk} suggest that neutrons in pure neutron matter behave similarly to hard spheres of radius
$\lsim$ 0.3 fm. The same analysis for isoscalar nucleons, performed in Sec.~\ref{sec:nk}, leads to a radius of $\sim$0.4 fm.

\begin{figure}[htbp]
\begin{center}
\includegraphics[width= 0.5\textwidth]{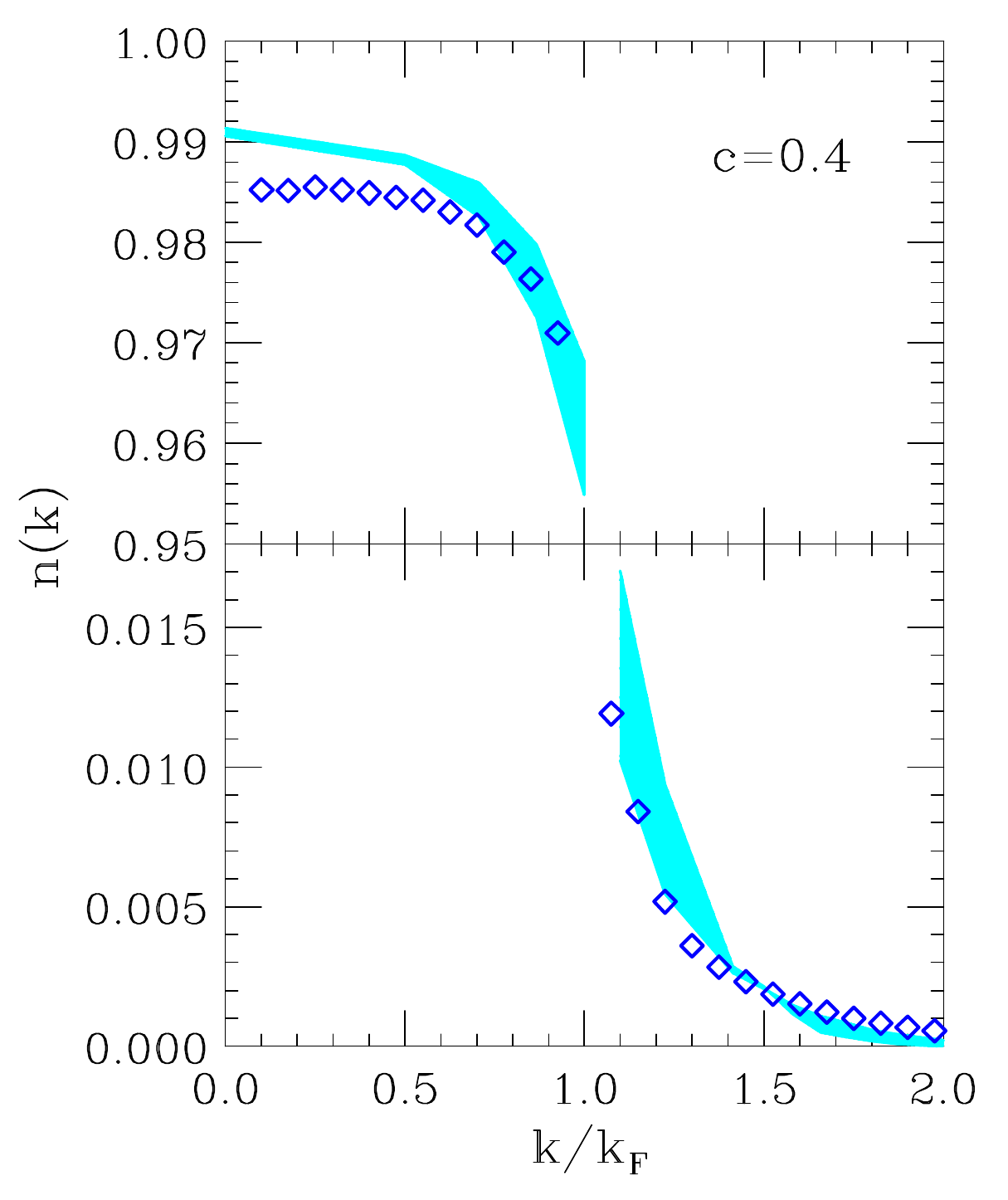}
\caption[HS and neutron matter $n(k)$ ]{Comparison between the momentum of the fermion hard-sphere system of degeneracy $\nu$=2 at $c=0.4$ (diamonds) and those
reported in Ref.~\cite{nk_MC}, corresponding to the density range $0.08  \leq  \rho \leq 0.24$ fm$^{-3}$  (shaded area). }
\label{nk}
\end{center}
\end{figure}

\cleardoublepage 
\addcontentsline{toc}{chapter}{Summary and Prospects}
\chapter*{Summary and prospects}
\label{summary}

This Thesis reports the results  of  perturbative calculations of a variety of properties of the Fermi hard-sphere system, carried out  using an effective
interaction derived within the CBF formalism and the cluster expansion technique.
Our study, while being very interesting in its own right, is mainly aimed at establishing  the accuracy of the proposed approach in view of future applications to neutron star matter. To achieve this goal, we have performed a systematic 
comparison between its results and those obtained from perturbative calculations,  providing exact predictions in the low-density limit.

The main advantage of our scheme is the ability to combine the effectiveness of including correlation effects 
through a modification of the basis states with the flexibility of perturbation theory in the Fermi gas basis. This feature is 
fully apparent in the calculated momentum distribution, which, unlike the one obtained using correlated wave functions in 
the context of the variational method, exhibits the correct logarithmic behavior  in the vicinity of the Fermi surface.  
Attaining the same result using the bare interaction and a correlated basis involves
non-trivial  difficulties, arising from the use of non-orthogonal perturbation theory \cite{fantonipandha}.

The quasiparticle properties obtained from the self-energy computed using the CBF effective interaction turn out to be significantly affected
by second order contributions to $\Sigma(k,E)$, exhibiting an explicit energy dependence. In the case of the effective mass at momentum $k=k_F$, 
the inclusion of these
contributions leads to a dramatic change of both the magnitude and the density-dependence of the ratio $m^\star(k_F)/m$, with respect to the
predictions of the Hartree-Fock approximation. Similar results had been previously found in nuclear matter calculations, carried out within
G-matrix \cite{Gmatrix,Gmatrix2}, Self~Consistent Green's Function \cite{SCGF}  and non orthogonal CBF perturbation theory \cite{fantonifrimanpandha_eCBF}.

Overall, the comparison between the results reported in this Thesis and those obtained from low-density expansions suggests that the
latter  provide accurate predictions in the density range corresponding to $k_F \lsim 0.3 - 0.4\ {\rm fm^{-1}}$.
Based on the argument made in Section \ref{sec:nk}, these values of $k_F$ correspond to densities of
isospin symmetric nuclear matte in the range $0.2  \lsim (\rho /  \rho_{NM}) \lsim 0.4$, $\rho_{NM}$ being the equilibrium density.


In order to gauge the  accuracy of the effective interaction approach in a different context, 
we have extended our study to the calculation of  quasiparticle scattering, the description of which is needed to 
obtain the transport coefficients from Boltzmann's equation.  At low density,  the resulting shear viscosity and thermal conductivity coefficients  
turn out to agree with those obtained from Landau's kinetic theory \cite{JLTP1,JLTP2}. However, in this case the range of applicability of the 
perturbative expansion is somewhat limited, since it only includes terms linear in the parameter $c = \myc$.

The enhancement of the effective mass  resulting from the inclusion of second order contributions to the self-energy has important implications for the calculation of the in medium scattering cross section, and consequently of  the transport coefficients,  since the value of the effective mass affects both the incoming flux and the phase-space available to the particles in
the final state.  The enhanced ratio $m^\star(k_F)/m >1$ brings about an increase of the cross section, resulting in turn in a decrease of the
shear viscosity and thermal conductivity coefficients  (see, e.g., Ref.~ \cite{BV}). This feature is potentially very important, because {\em all} existing calculations of the transport 
coefficients of neutron star matter use effective masses obtained within the Hartee\textendash Fock approximation.

As pointed out above, the ultimate goal of our study is the application of the approach based on the CBF effective interaction to the description of dense matter of astrophysical interest.  
The formalism discussed in this Thesis can be readily generalised, along the line discussed in Ref. \cite{response, response2}, to obtain a number of nuclear matter properties, such as the spectral functions defined by  Eq. \eqref{KL} and the density and spin-density responses \cite{response,response2}.
In this context, the CBF effective interaction is a unique tool, allowing to describe short- and long-range correlations, associated with the appearance
of collective excitations, in a fully consistent fashion. 
Comparison between the results obtained from the CBF effective
interaction and those derived from different many-body techniques and using different nuclear
Hamiltonians \cite{GMSF,spectral,green,response,response2,baldo1,carbone1} will allow to firmly assess
the potential of this new approach.

\cleardoublepage 

\appendix

\chapter {Low-density expansion of the momentum distribution}
\label{nkApp}

For the sake of completeness, we report the explicit expression of the momentum distribution
obtained from the low-density expansion including terms of order up to $c^2=(k_Fa)^2$.  As pointed out in Section \ref{sec:nk}, 
$n(k)$ can be conveniently written in terms  of two contributions, associated with hole and particle states, according to
\begin{equation}
n(k) = n_<(k) + n_>(k) \ ,
\end{equation}
where
\begin{align}
 n_<(k) = 0 \ \ \mbox{ for } \ \ k > k_F \ ,  \nonumber  \\
 n_> (k) = 0 \ \ \mbox{ for } \ \ k < k_F \ .
\end{align}

At $x=k/k_F<1$, the algebraic expression of $n_<(k)$,  derived by the authors of Refs. \cite{mahaux, belyakov}, reads  
\begin{alignat}{2}
\label{n<}
n_< (k)  =   1  -&  \frac{\nu - 1}{ 3 \pi^2 x } c^2    \left [    \right. && \left.  (7 \ln 2 - 8 ) x^3 + (10-3 \ln 2) x  
     \right . 	\nonumber \\ 
& && \left .    + 2 \ln {\frac{1+x}{1-x}}   -    2 (2-x^2) ^{3/2} \ln{ \frac{(2-x^2 )^{1/2} + x  } { (2-x) ^{1/2}  - x} }  \right ]  \ ,
\end{alignat}
where $\nu$ denotes the degeneracy of the momentum eigenstates. The form of $n_>(k) $, reported in Refs. \cite{mahaux,mahaux_erratum}, depends on the range of $x$.
For $1 < x  < \sqrt {2}$
\begin{alignat}{2}
\label{n1>}
 n_> (k)    =&   
   \frac{\nu- 1 }{6 \pi ^2 x } c^2 
   \left \{  \right.     &&  \left .
(7 x^3 - 3 x - 6) \ln { \frac{x-1}{x+1} } + (7 x^3 - 3x + 2) \ln 2  -  8 x^3   + 22 x^2 \right.  \nonumber \\ 
 &  &&\left.   + 6 x - 24  +   2 (2 - x ^2 )^{3/2}  
     \left [
     \ln\frac{ 2 + x + (2-x^2) ^{1/2 } } { 2 + x - (2-x^ 2 ) ^{1/2 } }  \right.   \right. \nonumber \\ 
     & &&\left. \left.
+ \ln    \frac{ 1 +  (2 - x^2  )^{1/2 } }{1 - (2-x^2 )^{1/2 } } 
- 2 \ln \frac{ x  + (2 - x^2 )^{1/2 } } { x - (2-x^2) ^{1/2} }
  \right]
  \right \}   \  ,
\end{alignat}
while for  $ \sqrt{2} < x < 3$
\begin{alignat}{2}
\label{n2>}
n_>(k) =   \frac{\nu - 1 }{6 \pi ^2 x} c^2
& \left \{ \right. && \left. 
(7 x^3 - 3 x - 6) \ln { \frac{x-1}{x+1} } + (7 x^3 - 3x + 2) \ln 2  - 8 x^3  + 22 x^2
 \right.  \nonumber \\ 
 &\left . \right. && \left. 
  + 6 x - 24   -  4 ( x ^2 - 2 )^{3/2}    
    \left [
     \tan^{-1} \frac{ (x + 2) } { (x^2 - 2) ^{1/2}}       \right.  \right.  \nonumber \\
&  &&   \left. \left .
+   \tan ^{-1}  (x^2 - 2)^{-1/2}    - 2 \tan^{-1}  x(x^2 - 2 )^{-1/2}  
\right]
\right \}   \  .
\end{alignat}
Finally, in the domain $ x > 3$
\begin{align}
\label{n3>}
n_>(k) =  2 \frac{\nu - 1}{ 3 \pi ^2 x } c^2
& \left \{   2 \ln \frac{x+1}{x-1} - 2 x + (x^2 - 2 )^{3/2}  \left [
2 \tan^{-1} x ( x^2 -2)^{-1/2}  \right. \right. \nonumber \\
 &\left. \left.
- \tan ^{-1} (x-2)(x^2 - 2 )^{-1/2} - \tan^{-1} (x+2)(x^2 -2)^{-1/2}
\right]
\right \} \ .
\end{align}


\chapter{Cluster expansions}
\label{app:2bc}

\section{Energy at two-body cluster level}

The factorized cluster expansion of the ground state expectation value of the Hamiltonian, discussed in Section~\ref{sec:cluster}, reads
 \begin{align}
E_0 = T _ F  +     \sum_{n = 1 } ^ {N } \left(  \Delta E  \right)_n \nonumber      \tag{\ref{eq:expansionE}}    \ , 
\end{align}
with
\begin{align}
\label{fact:clust}
\left( \Delta E \right) _ n =  \sum_{i_1<\ldots <  i_n}  \left [    \frac{1}{I_{i_1 \ldots i_n} } \frac{\partial}{\partial \beta} I _{i_1 \ldots i_n}  - \sum_{k= 1}^{n}  \frac{1}{I_{i_1 \ldots i_k }  }    \frac{\partial}{\partial \beta} I _{i_1 \ldots i_k}       \right ]_{\beta = 0} \ .  
\nonumber \tag{\ref{eq:deltaEnbis}}
\end{align}
Equation~\eqref{fact:clust} can be rewritten in terms of the $n$-body  operator $ {w}_n$  in the form 
\begin{align}
 \left( \Delta E \right) _ n    =&  \frac{1}{n !} \sum_{i_1\ldots i_n} [ i_1 \ldots i_n |  {w}_n  | i_1 \ldots i_n ] _ a  \ .
\end{align}
The ${w}_n$ are obtained from Eq.~\eqref{eq:deltaEnbis}, exploiting the property that the normalization factors 
 $   \left . I_{i_1 \ldots i_n} \right |^{-1} _{\beta =  0} $
  differ  from by quantities $\mathcal{O} \left( 1/ N \right)$ at most, which can be neglected in the limit $N\rightarrow \infty$.

The hermiticity  of   the  Jastrow-type  correlation function
\begin{align}
F = \prod_{j,i=1}^N f(r_{ij}) \  ,  \nonumber  \tag{\ref{eq:jastrow_central}}
\end{align}
makes the two-body  operator $w_{2}$ reducible to the simple form 
 to
\begin{equation}
\label{eq:w12}
  {w}_{2}  =   \frac{1}{2}  \Big [   F_2({12})  ,   \left [ t(1) + t(2) , F_2({12})  \right]   \Big]  +    {F_2}^2({12}) v({12})      \ ,
 \end{equation}
 with $F_2(12) = f(r_{12})$,  yielding  the explicit formula for  two-body cluster contribution
\begin{align}
\label{eq:d2}
 \left( \Delta E \right) _ 2     =  &    \frac{1}{2} \sum_{n_1 n_2   }   \int d x_1 dx_2  \ 
  \Phi^{\star}_{n_1 n_2} \frac{1}{2}    \Big [   F_2({12})  ,   \left [ t(1) + t(2) , F_2({12})  \right]   \Big]     \Phi_{n_1 n_2}     \nonumber \\ 
   & +  \frac{1}{2} \sum_{n_1 n_2 }     \int  d x_1 dx_2  \  
 \Phi^{\star}_{n_1 n_2}    v({12}) {F_2}^2({12})      \Phi_{n_1 n_2}  \ .
\end{align}
In the above equation,   $dx_i$ denotes both integration over the spatial coordinates ${\bf r}_i $ and  trace over the unices associated with the discrete degrees of freedom.
The two-particle Fermi gas state $\Phi_{n_1 n_2} $ is  obtained  as Slater determinant of two  
  single particle wave-functions $\phi_{n_i}(x_i)$ 
\begin{equation}
\label{eq:planewave}
\phi_{n_i} (x_i) =  \frac{1}{ \sqrt {V} } e^{i {\bf k}_{n_i} {\bf r }_i }  \chi_{\sigma_i} \chi_{\tau_i} \ ,
\end{equation}
where $n_i \equiv \{ {\bf k}_i, \sigma_i, \tau_i \}$  and the $\chi$'s are Pauli spinors in spin($\sigma_i$)-isospin($\tau_i$) space. 

The antisymmetric product of  any number $p$ of  sinlge-particle wave functions  can be written in a compact way introducing the antisymmetrisation   operator~$\mathcal{A}$
\begin{equation}
\Phi_{n_1 	\ldots n_ p} =   \mathcal{A}   \left[  \phi_{n_1}(x_1)  \ldots  \phi_{n_p}(x_p) \right]  \ ,
\end{equation}
whose expression, involving the the pair exchange operator  ${P}_{ij}$,  is
\begin{align}
\label{eq;antisymOperator}
\mathcal{A} = 1 -  \sum_{i < j} {P}_{ij}  + \sum_{i< j < k}   ( P_{ij} P_{jk} + P_{ik} P_{jk} ) + \ldots    \ \ \ . 
\end{align}
It follows that for a system of two particles 
$\mathcal{A} = 1 - P_{12} $.

The pair exchange operator
 $  {P}_{12} $ can in turn by defined by the result of its action on the two-particle state  
\begin{equation}
  {P}_{12}   \left [  \phi_{n_1} (x_1)  \phi_{n_2} (x_2 )  \right ]  =   \phi_{n_1} (x_2) \phi_{n_2} (x_1 ) \ ,
\end{equation}
and can be factorized into the product of  the operator that exchanges the coordinates, 
 $ {P}^{r}_{12}$,  and the operator acting on discrete degrees of freedom, $ {P}^{\sigma  \tau}_{12}$. The resulting expression is
\begin{align}
\label{eq:P12}
 {P}_{12} =    \    {P}^{r}_{12} \times  {P}^{\sigma \tau}_{12}  \ , 
\end{align}
with
\begin{align}
\label{eq:pr}
  {P}^{r}_{12}  =   \  \exp \left [  - i \left(  {\bf k}_1 - {\bf k }_2   \right  )   \cdot     ({\bf r }_ {1} -{\bf r }_ {2} )         \right ] \ ,   
\end{align}
and
\begin{align}
\label{eq:pst}
 {P}^{\sigma  \tau}_{12} =     \   \frac{1}{4} \left (  1+\sigma_{12}   \right ) \left ( 1+\tau_{12}  \right )  \ ,
\end{align}
where $\sigma_{12}$ and $\tau_{12}$ denote the products of Pauli matrices
\begin{equation}
\sigma_{12} = \vec {\sigma} _1 	\cdot  \vec {\sigma} _2 \ ,  \ \tau_{12} = \vec {\tau} _1 	\cdot  \vec {\tau} _2 \ .
\end{equation}
Since the Pauli matrices  are traceless, and   the potential  and the correlation  functions are both diagonal in spin-isospin space,  after summation over the discrete degrees of freedom,  the  operator $ {P}_{12}$ implied in  Eq.~\eqref{eq:d2}  reduces to the the expression
\begin{equation}
  {P}^{r}_{12}   \rightarrow \frac{1}{\nu}    \exp \left [  - i \left(  {\bf k}_1- {\bf k }_2   \right  )   \cdot     ({\bf r }_ {1} -{\bf r }_ {2} )     \right ] \ .
 \end{equation} 
In addition, because of translation invariance, in  Eq.~\eqref{eq:d2} we can conveniently use  the center of mass and relative  coordinates   
\begin{equation}
\label{eq:transform}
{\bf r} = {\bf r}_1 -   {\bf r}_2  \ \ , \ \
{\bf R } = \frac{1}{2} \left (  {\bf r } _ 1 +  {\bf r} _ 2  \right )  \ .
\end{equation}

Using the above variable transformation, the trace operation and the summation over all  possible occupied states the  term  $ \mathop {Tr }  \sum_{n_1 n_ 2 } \Phi^{\star}_{n_1 n_2}    \Phi_{n_1 n_2}$ in Eq.~\eqref{eq:d2}  can be cast in the form
\begin{equation}
\label{eq:tmp}
 \mathop {Tr } \sum_{n_1 n_ 2 } \Phi^{\star}_{n_1 n_2}    \Phi_{n_1 n_2} = \sum_{k_{n_1} k_{n_ 2} \leq k_F } \frac{  \nu^ 2} {V^2} 
 \left [ 1 - \frac{1}{	\nu} e ^  {-i (  {\bf k}_1 -  {\bf k}_2 )  \cdot {\bf r}  }    \right]    \ . 
\end{equation}
In addition, from the relation between the  Slater function  $\ell(k_F r)$ and the non diagonal density matrix of  Eq.~\eqref{eq:onebodydens}
\begin{align}
\rho(x_1, x_2)  \equiv \rho \ell(k_F r) \ , 
\end{align}
one obtains
\begin{align}
\label{eq:ell}
  \ell (k_F r)  =   \frac{1} {\rho} \sum_{ n \in \{F\} }  \phi^{\star}_n (x_1) \phi^{}_n(x_2) =  \frac{\nu} {N} \sum_{ k _n \leq k_F }  e^{i {\bf k } _n  \, \cdot {\bf r} }   \ , 
\end{align}
with the  explicit expression of $\ell(x)$ reported in Eq.~\eqref{eq:slater}
\begin{align}
\ell(x)~=~\frac{3}{x^3} \   \left( \sin x - x \cos x \right) \ .   \nonumber \tag{\ref{eq:slater}} 
\end{align}
The final result for  Eq.~\eqref{eq:tmp}  reads
\begin{equation}
\label{eq:p2fg}
 \mathop {Tr } \sum_{n_1 n_ 2 } \Phi^{\star}_{n_1 n_2}    \Phi_{n_1 n_2} =  \rho^2   \left [ 1 - \frac{1}{	\nu} \ell^2(k_F r)    \right]   \ .
\end{equation}
Finally, the potential energy  contribution to the two-body cluster approximation to the ground state energy can be evaluated from
\begin{align}
 \langle V   \rangle _ {2b}  
=   \frac{1}{2} N \rho   \int d{\bf r} f^2(r) v(r)  \left [ 1 - 	\frac{1}{\nu} \ell^2 \left (k_F r  \right) \right]   \ .
\end{align}
As for the kinetic term of the Hamiltonian, the sum of the  single particle  operators
$$  t(i)   = -  \frac{1} {2 m}  \boldsymbol { \nabla}^2_{ r_i}  \ , $$
 can be written, after the transformation into the variables of Eq. \eqref {eq:transform},   as a sum of  center of mass and relative kinetic energies
\begin{align}
t(1) + t(2)  =  -  \frac{1} {4 m}  \boldsymbol { \nabla}^2_{  R}  -  \frac{1} {m} \boldsymbol { \nabla}^2_{ r} \equiv T(R) + t(r)    \ . 
\end{align}

\noindent
Because $\left [ T(R) , f(r)  \right] = 0 $, it follows that 
\begin{equation}
\left [ t(1) + t(2) , F_2(12)  \right]  = \left [  t(r) , f(r)  \right]   \ .
\end{equation}
The calculation of the kinetic energy contribution at two-body cluster level amounts to evaluating the expectation value of the commutator in Eq.~\eqref{eq:w12} in the 
two-body wave function of the non interacting system, $\Phi_{n_1 n_2}$, involving the quantity
\begin{equation}
\frac{1}{2}   \left [ f(r), \left[ T(r), f(r) \right] \right ]  =
- \frac{1}{2m}\Phi^{\star}_{n_1 n_2} \left \{  f(r) \left [ \boldsymbol { \nabla}^2_r, f(r)   \right ]      -     \left [ \boldsymbol { \nabla}_r^2, f(r)  \right ]     f(r)    \right \} \Phi_{n_1 n_2}   \ . 
\end{equation}
After simple algebraic manipulations, the above expression  can be reduced to the form
\begin{equation}
\frac{1}{m}  \Phi^{\star}_{ij}   \left [  \boldsymbol { \nabla}_r f \right]^2 \Phi_{ij} \ , 
\end{equation}
yielding the two-body cluster approximation to the kinetic energy
\begin{align}
   \langle T \rangle  _{2b} =  
   \frac{1}{2}     N \rho  \int   d{\bf r}  \  \frac{1}{m}   \left [  \boldsymbol { \nabla}_r f \right]^2
 \left [ 1 - \frac{1}{\nu}  \ell^2(k_F r )\right ]   \ .
\end{align}
Collecting the above results we can finally write the corresponding approximation to the ground-state energy in the form
\begin{equation}
\label{eq:e2bc}
\frac{1}{N} \left ( \Delta E \right ) _ 2 = 
   \frac{1}{2}  \rho  \int   d{\bf r} \      \left [    \frac{1}{m}   \left (  \boldsymbol { \nabla}_r f \right)^2   + v^2(r) f^2(r) \right]
  \left [ 1 - \frac{1}{\nu}  \ell^2(k_F r )\right ]   \ .
\end{equation}

\section{Expectation value of the effective interaction}

The expectation value of the  effective interaction, or of a any two-body operator of the form    
\begin{equation}
{\Omega}= \sum_{i<j} {\omega}(r_{ij})  \ , 
\end{equation}
in the Fermi gas ground state can be evaluated from its explicit expression
\begin{align}
\label{eq:Op2b}
[ 0_{FG} |  \Omega  | 0 _ {FG} ]   = &  \frac{1 }{2}  \sum_{ij}  \int dx_1 \ldots dx_N \Phi^{\star}(x_1\ldots x_N)  \omega(r_{ij}) \Phi(x_1\ldots x_N) \nonumber \\
= &   \frac{1}{2} N(N-1)  \int dx_1 \ldots dx_N \Phi^{\star}(x_1\ldots x_N)  \omega(r_{12}) \Phi(x_1\ldots x_N)   \ ,
\end{align}
where $\Phi(x_1\ldots x_N)$ is the  $N$-particle  ground state  wave-function, represented by a Slater determinant of single particle states defined as in Eq.~\eqref{eq:planewave}.
Extraction of two particles of coordinates $x_1$, $ x_2  $ from the  Slater determinant allows one to isolate the two-particle state by writing the  $N$-particle wave function as 
\begin{align}
\Phi (x_1 	\ldots x_N)=&   \frac{1}{\sqrt{N (N-1)  }} \sum_{n_1 n_2} (-)^{n_1+n_2+1} \mathcal{A} \left [ \phi_{n_1}(x_1) \phi_{n_2} (x_2) \right]  \tilde{\Phi}_{n_1 n_2} (x_3 \ldots x_N)  \nonumber \\
=&  \frac{1}{\sqrt{N (N-1)  }} \sum_{n_1 n_2} (-)^{n_1+n_2+1}  \Phi_{n_1 n_2}(x_1, x_2)   \tilde{\Phi}_{n_1 n_2} (x_3 \ldots x_N) \ .
\end{align} 
In the above equation,  ${\Phi}_{n_1 n_2}(x_1, x_2) $ is the two-particle state obtained through the antisymmetric combination of  the orbitals $n_1$ and $n_2$, while  $\tilde{\Phi}_{n_1 n_2} (x_3 \ldots x_N)$ is the Slater determinant describing $(N-2)$ particles, which does not include the states $n_1$ and $n_2$.
The orthonormality relations obeyed  by the states of $\tilde{\Phi}_{n_1 n_2} (x_3 \ldots x_N)$ 
\begin{equation}
\int dx_3  \ldots dx_N  \tilde{\Phi}^{\star}_{n_1 n_2} (x_3 \ldots x_N) \tilde{\Phi}_{p_1 p_2} (x_3 \ldots x_N) = \delta_{n_1 p_1} \delta_{n_2 p_2} \ , 
\end{equation}
implies
\begin{equation}
\int dx_3  \ldots dx_N \left | \Phi (x_1 	\ldots x_N) \right|^2 =   \frac{1}{N(N-1)}  \sum_{n_1 n_2 } \left |  \Phi_{n_1 n_2}(x_1, x_2)   \right |^2  \ . 
\end{equation}

Substitution of the above result and  Eq.~\eqref{eq:p2fg} in Eq.~\eqref{eq:Op2b} leads to the final expression of  the expectation value of the two-body operator $\Omega$
in the Fermi gas ground state
\begin{align}
\frac{1}{N} [ 0_{FG} |   \Omega  | 0 _ {FG} ]  = & 
 \frac{1}{2N} \sum_{n_1, n_2} \mathop{Tr}  \int d{\bf r}_1d{\bf r}_2 \ \omega(r_{12}) \left |  \Phi_{n_1 n_2}  \right |^2  \nonumber \\
= & \frac{1}{2} \rho  	\int d{\bf r} \ \omega(r) \left [ 1 - 	\frac{1}{\nu} \ell^2(k_F r) \right]   \ . 
\end{align}

\chapter{Euler-Lagrange equation}
\label{app:euler}

The  functional minimization of the two-body cluster approximation to the ground state energy, discussed in Appendix \ref{app:2bc}, 
leads to a Euler-Lagrange equation determining the shape of the correlation function $f(r)$. 

The satrting point of this procedure is the functional defined by  the Eq.~\eqref{H:2body2}
\begin{align}
\frac{1}{N}(\Delta E)_2  =  & \ 2 \pi \rho     \int dr F  \left[ f, f^{\prime} \right ]    \ ,
\end{align} 
where $f^\prime = (df/dr)$ and 
\begin{align}
F \left[ f , f^{\prime}  \right]  \equiv  \ &   	\left [ \frac{1}{m} {f^{\prime} (r)}^2  + f^2(r) v(r)  \right ]   \Phi^2(r) , 
\end{align}
with $\Phi(r)$ defined as
\begin{align}
\Phi (r) \equiv r \ &  \sqrt{   1 - \frac{1}{\nu} \ell^2(k_F r)  }   \ .  \nonumber \tag{\ref{eq:phiel}}
\end{align}
In order to  obtain the  Euler-Lagrange equation
\begin{equation}
\frac{d }{ d r }   \left ( \frac{ \partial  F }{  \partial f ^\prime }  \right ) =   \left ( \frac{ \partial  F }{ \partial f}  \right )
\end{equation}	
one has to
 compute the derivatives  of the function $F$
\begin{align} 
\frac{ \partial  F }{ \partial f } = &    \   2 f(r) v(r) \Phi^2(r) \\ 
\nonumber \\
\frac{ \partial  F }{ \partial f^{\prime}} =   &    \   \frac{2}{m} f^{\prime }(r)  \Phi^2(r)   \\
\nonumber \\
\frac{d }{ d r }   \left ( \frac{ \partial  F }{ \partial f^{\prime}}  \right ) =  &  \  \frac{2}{m} \left [   f^{\prime \prime }(r)  \Phi^2(r)  + 2  f^{\prime} (r) \Phi(r) 
\Phi^{\prime}(r) \right ]
\end{align}
yielding the differential equation
\begin{align}
f^{\prime \prime }(r)  \Phi(r)  + 2  f^{\prime} \Phi(r) \frac{ \Phi^{\prime}(r)}{\Phi(r) } =   m v(r) f(r)  \Phi(r) \ .
\end{align}

In terms of the function  $g(r)$, defined as 
\begin{align}
g(r) \equiv  f(r) \Phi(r) \nonumber \tag{\ref{eq:grel}} \ ,
\end{align}
the  Euler-Lagrange equation can be conveniently rewritten in the form
\begin{align}
g^{\prime \prime} (r) - \left[   \frac{\Phi^{\prime \prime} (r)  }{\Phi(r)  }   + m v(r)    \right]  g(r) = 0 \ .
\end{align}
In the case of  the   hard-sphere system,  the interaction  potential vanishes in the region $r \geq a $, while the correlation function is zero inside the core.  Therefore, the boundary conditions
to be fulfilled by the correlation function are
\begin{align}
f(a) = 0 \ , \ f( d) = 1  \ ,       
\end{align}
with the additional requirement that the derivative of $f$ be continuous at $r=d$
\begin{align}
\left.  f^{\prime} \right | _ {r = d} = 0  \ , 
\end{align}
The above constraints translate into a set of boundary conditions  to be satisfied by the the function $g(r)$
\begin{align}
g(a ) = 0 \ , \
g ( d) = \Phi(d)  \  . 
\end{align}
The additional requirement 
\begin{align}
  \left.  g ^{\prime} \right | _ {r = d} =      \left.    \phi^{\prime}(r)  \right | _ {r = d} \ , 
\end{align}
is fulfilled introducing a Lagrange multiplier, denoted  $\lambda$. The resulting Euler-Lagrange equation reads
\begin{align}
g^{\prime \prime}(r) - g(r) \left[ \frac{\Phi^{\prime \prime}(r)}{\Phi(r)}  + m \lambda \right] = 0 \ .    \nonumber   \tag{\ref{ELeq}}
\end{align}

\chapter{Calculation of the self-energy}
\label{app_self}

\section{Direct and exchange diagrams}
The second order  contributions to the self-energy of  Fig.~\ref{self} can be represented in two different ways, depending on whether the ingoing  momentum  is $k> k_F $  or $k < k_F$.
 
\begin{figure}
\begin{center}
 \subfloat[]{
 \includegraphics[scale= 0.7]{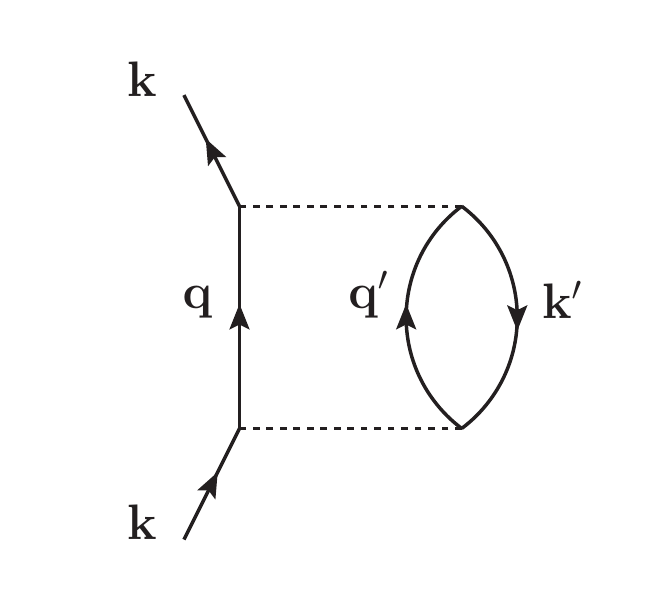}
 \label{fig:pol1}
 }
 \subfloat[]{
 \includegraphics[scale= 0.7]{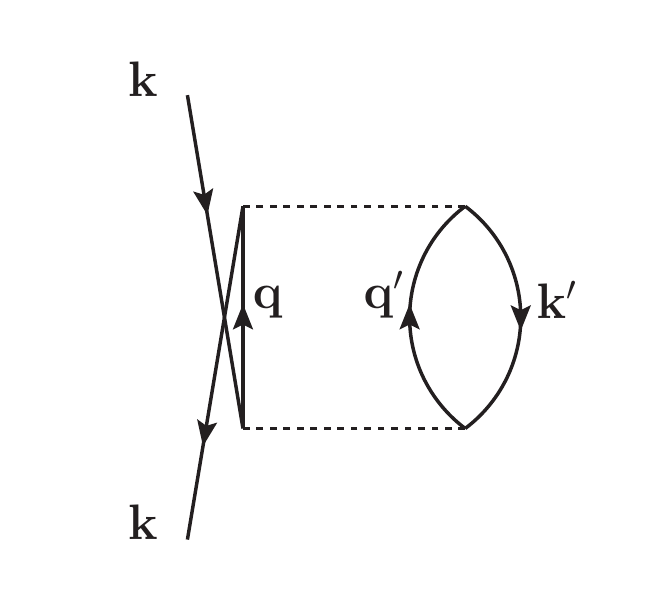}
 \label{fig:pol2}
 }
 \caption[The \emph{polarisation} term]{The self-energy contribution of $\Sigma_{2p1h}$, Eq.~ \eqref{sigma_p},  for particle states ($k > k_F$)  in the left panel  and for  hole states ($ k<k_F$) in the right panel.}
\end{center}
\end{figure}

The diagram associated with  the term $\Sigma_{2p1h}$  of Fig.~\ref{fig:pol1} has a simple physical interpretation. It represents the interaction between a particle of momentum $\bf k$ and energy $E$ 
and a hole of momentum  ${\bf k}^{\prime}$,  belonging to the Fermi sea, which leads to the excitation of a two-particle\textendash one-hole state of the $(N+1)$-particle system.
In other words, the particle carrying momentum $\bf k $ polarises the  $N$-body system through the excitation of a  one particle-one hole state. 
Because of energy conservation, these  excitations can only take place for  $E > \epsilon_F$, where $\epsilon_F$ denotes the Fermi energy.


On the other hand, for hole states ($k < k_F $) the classification as  polarisation contribution is misleading. The diagram of Fig.~\ref{fig:pol2} is in fact associated with  
ground-state correlations, taking place before the creation of the hole of momentum $\bf k$ and energy $E$, leaving the $(N-1)$-particle system in a 
two-particle\textendash one-hole state.

\begin{figure}
\begin{center}
 \subfloat[]{
 \includegraphics[scale= 0.7]{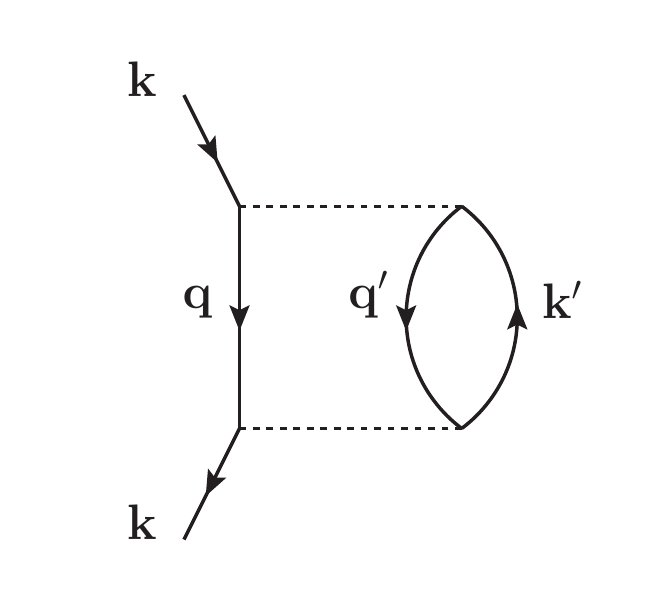}
 \label{fig:cor1}
 }
 \subfloat[]{
 \includegraphics[scale= 0.7]{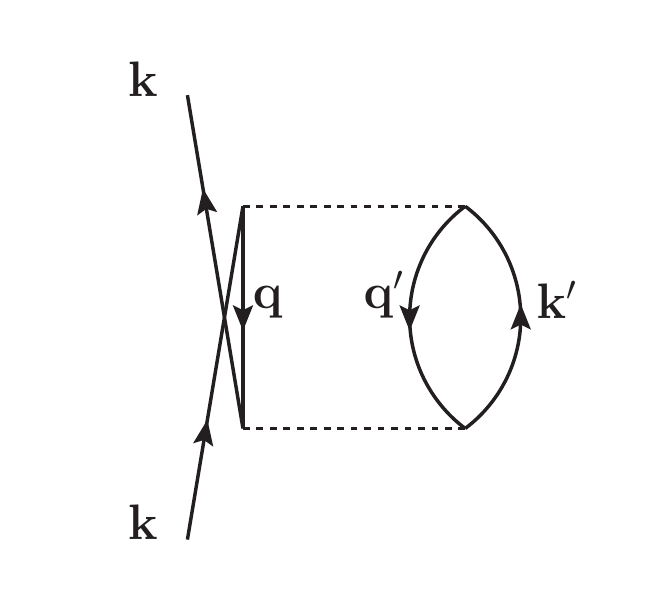}
 \label{fig:cor2}
 }
 \caption[The \emph{correlation} term]{The self-energy contribution of $\Sigma_{2h1p}$, Eq.~ \eqref{sigma_h},  for particle states ($k > k_F$)  in the left panel  and for  hole states ($ k< k_F$) in the right panel.}
\end{center}
\end{figure}

The physical polarisation effect for hole states is described by the the diagram in Fig.~\ref{fig:cor2}, which  is the 
second order term of the self-energy  of Fig.~\ref{fig:correl}   for $k<k_F$.
Finally, Fig.~\ref{fig:cor1} describes the modification of  ground-state correlations due to the introduction of an additional particle of momentum $k> k_F$.

In spite of the above considerations, $\Sigma_{2p1h}$ and $\Sigma_{2h1p}$ are referred to as \emph{polarisation} and \emph{correlation} terms, respectively. Owing to their  
different analytical properties, the corresponding contributions to  the momentum distribution $n(k)$ are determined by $\Sigma_{2p1h}$, for $k< k_F$, and $\Sigma_{2h1p}$, for $k> k_F$, as 
shown by Eqs.~\eqref{eq:nk<} and \eqref{eq:nk>} \cite{mahaux1985}.

\section{The matrix element}

The calculation of the second order contribution to the self-energy involves the evaluation of the matrix element 
\begin{equation}
\label{eq:M}
 M =  [   {\bf q} \tau, {\bf q}^{\prime} \tau^{\prime}  | \veff |  {\bf k} \sigma ,  {\bf k}^{\prime} \sigma^{\prime} ]_ a  \ .
\end{equation}
The antisymmetric combination of single particle states specified by the quantum numbers $ {\bf k} \sigma, {\bf k} \sigma^{\prime}$
can be easily obtained using the operators $\mathcal{A}$ and $ {P}_{12}$ defined in Appendix \ref{app:2bc} 
\begin{equation}
| {\bf k} \sigma, {\bf k} \sigma^{\prime} ] _ a  =  \mathcal{A } | {\bf k} \sigma, {\bf k} \sigma^{\prime} ]  = ( 1-  {P}_{12} )  | {\bf k} \sigma, {\bf k} \sigma^{\prime} ] \, .
 \end{equation}

Let us first expand the squared matrix element  of the effective interaction
\begin{align}
|M|^2&=    [   {\bf k} \sigma, {\bf k}^{\prime} \sigma^{\prime}  |   \veff (1- {P}_{12})       | {\bf q} \tau, {\bf q} \tau^{\prime} ] 
    [   {\bf q} \tau, {\bf q}^{\prime} \tau^{\prime}  |   \veff (1- {P}_{12})   | {\bf k} \sigma, {\bf k} \sigma^{\prime} ]  \ . 
\end{align}
Exploiting completeness of the spin-isospin basis states, we obtain
\begin{equation}
\label{eq:msquared}
 \frac{1}{\nu}\sum_{\sigma, \sigma^{\prime} \tau, \tau^{\prime} } 
| M| ^ 2=  \nu |M_d|^2 - M_d^{\star} M_e - M_e^{\star} M_d + \nu |M_e|^2 	 \ ,
\end{equation}
where the direct $M_d$ and exchange $M_e$ matrix elements are defined as
\begin{align}
&M_d=[ {\mathbf{k},\mathbf{k}^\prime} |   \veff                      | {\mathbf{q},\mathbf{q}^\prime} ]   \ , \\ 	\nonumber
&M_e=[ {\mathbf{k},\mathbf{k}^\prime} |   \veff P^{r}_{12}   | {\mathbf{q},\mathbf{q}^\prime} ]  \  .
\end{align}
The explicit expressions  of $M_d$ and $M_e$ are 
\begin{align}
M_d  =  
 \frac{ 1 }{V^2} 
 \int   d {\bf r}_1 d {\bf r}_2  \ 
 e ^ {-i  { \bf k \cdot r}_1 } \    e^ { -i    { \bf k} ^{\prime}  \cdot { \bf  r}_2  } \ 
 \veff (r_{12}) 
 \  e^{i  { \bf q \cdot r}_1 } \    e^ { i    { \bf q} ^{\prime}  \cdot { \bf  r}_2 }    \ , 
\end{align}
\begin{align}
M_e  =  
 \frac{ 1 }{V^2} 
 \int   d {\bf r}_1 d {\bf r}_2  \ 
 e ^ {-i  { \bf k \cdot r}_1 } \    e^ { -i    { \bf k} ^{\prime}  \cdot { \bf  r}_2  } \
 \veff (r_{12}) 
 \    e^ { i    { \bf q} ^{\prime}  \cdot { \bf  r}_1 } \  e^{i  { \bf q \cdot r}_2 }    \ .
\end{align}

In terms of the center of mass [${\bf R }=  ( {\bf r}_1 + {\bf r }_ 2 ) / 2$] and relative  [$ {\bf r } = {\bf r}_1 - {\bf r }_2$] coordinates , the direct term $M_d$ 
can be rewritten in the form
\begin{align}
\label{eq:Md2}
M_d = \frac{1}{V} \delta_{ {\bf q } +{ \bf q}^{\prime} - { \bf k} -  { \bf k } ^{\prime} }
 \int d{ \bf r} \  e^ {i { \bf u  \cdot  r  }}  \veff  (r) =
\frac{1}{V} \delta_{\mathbf{q}+\mathbf{q}^\prime-\mathbf{k}-\mathbf{k}^\prime} M(u)\,,
\end{align}
where the Kronecker $\delta$  accounts for conservation of the total momentum, and  $\bf u$ is defined as
\begin{align}
{\bf u}\equiv  \frac{1}{2}  ( {\bf q}- {\bf q}^{\prime} - { \bf k} +{ \bf k}^{\prime})   \ . 
\end{align}
The same procedure can be followed for the exchange term, wit the result
\begin{equation}
\label{eq:Me2}
M_e =\frac{1}{V} \delta_{\mathbf{q}+\mathbf{q}^\prime-\mathbf{k}-\mathbf{k}^\prime} M(v)\,,
\end{equation}
where
\begin{align}
{\bf v}\equiv  \frac{1}{2}  ( -{\bf q}+ {\bf q}^{\prime} - { \bf k} +{ \bf k}^{\prime}) \ .
\end{align}

Performing the angular integrations in both Eqs.~\eqref{eq:Md2} and \eqref{eq:Me2} we obtain    
\begin{equation}
M(t )= \frac{ 4 \pi} {t}  \int dr \  r \ \sin(tr) \veff  (r) \ .
\end{equation}
Note that, since $M(t)$ is real, $M_{e}^*=M_e$ and $M_{d}^*=M_d$. Hence, substituting Eqs.~(\ref{eq:Md2}) and (\ref{eq:Me2}) in Eq.~\eqref{eq:msquared}, we can write
\begin{align}
\frac{1}{ \nu}  \sum_{\sigma, \sigma^{\prime} \tau, \tau^{\prime} } 
|M|^2&=\frac{1}{V^2} \delta_{\mathbf{q}+\mathbf{q}^\prime-\mathbf{k}-\mathbf{k}^\prime} [\nu M(u)^2-2M(u)M(v) +\nu M(v)^2]\nonumber\\
&\equiv\frac{1}{V^2} \delta_{\mathbf{q}+\mathbf{q}^\prime-\mathbf{k}-\mathbf{k}^\prime} \mathcal{M}(u,v)\, .
\label{eq:Muv}
\end{align}

\section{The polarisation term} 

Substituting Eq.~(\ref{eq:Muv}) in Eq.~(\ref{sigma_p}) and carrying out the sum over $\mathbf{k}^\prime$ with the Kronecker delta function, the polarisation term $\Sigma_{2p1h}$ term 
can be cast in the form 
\begin{align}
\nonumber
\Sigma_{2p1h}  \left(    \bf {   k }, E \right)  & =  \frac{m}{V^2} \sum_{ { \bf q}, { \bf q} ^\prime}
   \frac{\mathcal{M}( { \bf u} , { \bf v} ) }   { { \bf q}^2 + { {\bf q} ^ \prime } ^2 -  ( { \bf q } + {\bf q}^{\prime} -  {\bf  k} )^2  -2m E   -  i \eta} \\
    & \times  n_{>} (\mathbf{q}) n_{>}( \mathbf{q}^\prime) n_{<} ( \mathbf{q}+\mathbf{q}^\prime-\mathbf{k})  \ , 
\end{align} 
where, owing to momentum conservation,  
$\mathbf{u}=\mathbf{q}-\mathbf{k}$ and $\mathbf{v}=\mathbf{q}^\prime-\mathbf{k}$.

In  the continuum limit $ \sum_{\mathbf{k}} \rightarrow \frac{V}{\left( 2 \pi \right)^3} \int d\mathbf{k} $, and
\begin{align}
\Sigma_{2p1h} \left( \bf {k}, E \right) =  \frac{m }{\left( 2\pi\right)^6}   
\int   &  d\mathbf{q}   \ d\mathbf{q}^\prime    \  \frac{\mathcal{M}(\mathbf{u},\mathbf{v})} {-2mE + \mathbf{q}^2 + {\mathbf{q}^\prime}^2  -  ( { \bf q } + {\bf q}^{\prime} -  {\bf  k} ) ^2   -i  \eta     }        \\ \nonumber 
 &  \times   n_{>} (\mathbf{q}) n_{>}( \mathbf{q}^\prime) n_{<} ( \mathbf{q}+\mathbf{q}^\prime-\mathbf{k})  \ . 
   \end{align}
%
The above equation can then be transformed using the variables $ \{ {\bf u}  , {\bf v}  \} $, with the result
 \begin{align}
\label{eq:sigma2p1h}
\Sigma_{2p1h} \left( \mathbf{k}, E \right) =  
 \frac{m }{\left( 2 \pi \right)^6}   \int   & d\mathbf{u} d\mathbf{v}    \frac {  \mathcal{M}(\mathbf{u}, \mathbf{v})  } { -2mE + \left| \mathbf{k} + \mathbf{u} \right |^2 + \left| \mathbf{k} + \mathbf{v}   \right |^2   -  \left |  \mathbf{k} + \mathbf{u} + \mathbf{v} \right |^2   - i \eta}  \nonumber    \\
     &\times   n_{>}   \left(   \mathbf{k} + \mathbf{u}      \right) n_{>} \left(   \mathbf{k} + \mathbf{v}    \right)  n_{<}  \left(   \mathbf{k} + \mathbf{u} + \mathbf{v}   \right)   \ . 
\end{align}
The denominator can be further rewritten using the relations
\begin{align}
&\left| \mathbf{k} + \mathbf{u} \right |^2 = \mathbf{k}^2 +\mathbf{u}^2 + 2 \mathbf{k}\cdot \mathbf{u}  \ ,  \nonumber\\
&\left| \mathbf{k} + \mathbf{v} \right |^2 = \mathbf{k}^2 +\mathbf{v}^2 + 2 \mathbf{k}\cdot \mathbf{v}    \ , \nonumber \\
&\left| \mathbf{k} + \mathbf{u} + \mathbf{v} \right |^2 = \mathbf{k}^2 +\mathbf{u}^2 + \mathbf{v}^2  + 2 \mathbf{k}\cdot\mathbf{u}+ 2 \mathbf{k}\cdot \mathbf{v} + 2\mathbf{u}\cdot \mathbf{v}\, .
\end{align}
Exploiting Eq.~(\ref{eq:imre}), the  imaginary  part of  Eq.~\eqref{eq:sigma2p1h} reads
\begin{align}
\label{eq:imse}
\IM  \Sigma_{2p1h} \left( \mathbf{k},  E \right)   =    \  \pi \  \frac{m }  {\left( 2 \pi \right)^6}    \int  & d\mathbf{u}  d\mathbf{v}  \,  \mathcal{M}(\mathbf{u}, \mathbf{v})   \delta \left ( 2mE - k^2 + 2 \mathbf{u} \cdot \mathbf{v} \right)  \times \nonumber  \\
&   n_{>}   (  \mathbf{k} + \mathbf{u} ) n_{>} (  \mathbf{k} + \mathbf{v} )  n_{<}  (  \mathbf{k} + \mathbf{u} + \mathbf{v}  )   \ , 
\end{align}
and the real part, obtained from a dispersion relations, turns out to be
\begin{align}
\nonumber
\RE  \Sigma_{2p1h} \left( \mathbf{k},  E \right)   & =  
\frac{1}{\pi}   \mathcal{P}\int_{-\infty} ^{\infty} \frac{\IM \Sigma_{2p1h} \left( \mathbf{k}, E ^{\prime}\right)} {E -E^{\prime}  } dE^{\prime} \\ 
& =   \frac{1}{\pi}   \mathcal{P}\int_{ \epsilon_F } ^{\infty} \frac{   \IM   \Sigma_{2p1h} \left( \mathbf{k}, E^{\prime}\right)} {E- E^{\prime}  } dE^{\prime} \ . 
\label{eq:Reself_p}
\end{align}

\section{The correlation term} 

Following the same steps described in the  previous section, the correlation term of the self energy $\Sigma_{2h1p}$ can be written in the form
\begin{equation}
\label{eq:sigma2h1p}
\begin{split}
\Sigma_{2h1p} \left( \mathbf{k}, E\right) = \frac{m}{\left( 2 \pi \right)^6}  \int    & d\mathbf{u} d\mathbf{v}  \frac { \mathcal{M}(\mathbf{u}, \mathbf{v})  } { 2mE  - \left| \mathbf{k} + \mathbf{u} \right |^2 - \left| \mathbf{k} + \mathbf{v}   \right |^2   + \left |  \mathbf{k} + \mathbf{u} + \mathbf{v} \right |^2   - i \eta}  \\
     &  \times   n_{<} ( \mathbf{k} + \mathbf{u} ) n_{<} (    \mathbf{k} + \mathbf{v} )  n_{>}  ( \mathbf{k} + \mathbf{u} + \mathbf{v} )    \ ,
\end{split}
\end{equation}
the imaginary part of which is given by
\begin{equation}
\label{eq:imseh}
\begin{split}
\IM   \Sigma_{2h1p} \left( \mathbf{k}, E \right)  & = 
 \  \pi  \ \frac{m}{\left( 2 \pi \right)^6} 
 \int  d\mathbf{u} d\mathbf{v} 	\,    \mathcal{M}(\mathbf{u},\mathbf{v})    \, \delta \left ( 2mE - k^2 + 2 \mathbf{u} \cdot \mathbf{v} \right)   \\
&  \times n_{<}  (  \mathbf{k} + \mathbf{u} ) n_{<} (  \mathbf{k} + \mathbf{v} )  n_{>}  (  \mathbf{k} + \mathbf{u} + \mathbf{v} )  \;.
\end{split}
\end{equation}
Using again a the dispersion relation, the real part can be evaluated from
\begin{align}
\nonumber
\RE  \Sigma_{2h1p} \left( \mathbf{k}, E \right)  & = \frac{1}{\pi}  
 \mathcal{P}\int_{-\infty} ^{\infty} \frac{ \IM   \Sigma_{2h1p} \left( \mathbf{k}, E^{\prime}\right)} {E- E^{\prime} } dE^{\prime}  \\ 
 & = \frac{1}{\pi}   \mathcal{P}\int^{ \epsilon_F} _{-\infty} \frac{   \IM   \Sigma_{2h1p} \left( \mathbf{k}, E^{\prime}\right)} {E- E^{\prime} } dE^{\prime}    \ .
\label{eq:Reself_h}
\end{align}

\section{Numerical evaluation of the  real part}

In order to evaluate the real part  of the self-energy using dispersion relations, one needs to integrate the off-shell imaginary parts of the polarisation and correlation terms  over a wide range of energies. 
Although 
 the imaginary part is a  decreasing function  of $|E|$, for $ E \rightarrow \infty $ the term $ \IM \Sigma_{2p1h}$ approaches zero very slowly,  and the computational effort 
 required for the evaluation of the real part  turns out to be huge, particularly for particle states  ($k > k_F$).
 This difficulty is not alleviated by the use of the subtracted dispersion relations. 
 
The above problem has been circumvented carrying out a direct calculation of  the real parts from the equations
 \begin{align}
 \RE  \Sigma_{2p1h} \left( \mathbf{k}, E \right) = 
 \lim_{\eta \rightarrow 0} 
\frac{m }{\left( 2\pi\right)^6}   
\int &
   d\mathbf{q}   \ d\mathbf{q}^\prime    \     
 {\mathcal{M}(\mathbf{u},\mathbf{v})}    \frac { \mathbf{q}^2 + {\mathbf{q}^\prime}^2  - {\mathbf{k}^\prime}^2  - 2 m E  }
 { (\mathbf{q}^2 + {\mathbf{q}^\prime}^2  - {\mathbf{k}^\prime}^2 - 2 m E )^2   +  \eta^2     }        \nonumber  \\
& \times   n_{>} (\mathbf{q}) n_{>}( \mathbf{q}^\prime) n_{<} ( \mathbf{q}+\mathbf{q}^\prime-\mathbf{k})  \ ,
  \end{align}
 \begin{align}
 \RE  \Sigma_{2h1p} \left( \mathbf{k}, E \right) = 
 \lim_{\eta \rightarrow 0} 
\frac{m }{\left( 2\pi\right)^6}   
\int &
   d\mathbf{q}   \ d\mathbf{q}^\prime    \     
 {\mathcal{M}(\mathbf{u},\mathbf{v})}    \frac { 2 m E  + {\mathbf{k}^\prime}^2   -\mathbf{q}^2 - {\mathbf{q}^\prime}^2     }
 { ( 2 m E  + {\mathbf{k}^\prime}^2   -\mathbf{q}^2 - {\mathbf{q}^\prime} )^2   +  \eta^2     }        \nonumber  \\
& \times   n_{<} (\mathbf{q}) n_{<}( \mathbf{q}^\prime) n_{<} ( \mathbf{q}+\mathbf{q}^\prime-\mathbf{k})  \ ,
  \end{align}
  where $ {\bf k} ^{\prime}  = {\bf q} + {\bf q}^{\prime} - {\bf k}$,  $ \mathbf{u}=\mathbf{q}-\mathbf{k} $ and $ \mathbf{v}=\mathbf{q}^\prime-\mathbf{k} $.

\chapter{The scattering amplitude in a Fermi fluid}
\label{app_ampl}

\section{The generalised Bethe-Salpeter equation}

In order to calculate  the scattering probability, the authors of Ref.\cite{JLTP1} have derived an expression for 
the scattering amplitude of two quasiparticles in a Fermi fluid. 
They followed closely Landau's original derivation of Ref.~\cite{Landau1959}, but were able to go beyond Landau's approximation of small momentum transfer in the collision. 
By using the diagrammatic techniques, 
they  obtained an integral equation for the scattering amplitude similar to the one yielding the $T$-matrix, discussed in Section~\ref{sec:lowdensity}.

\begin{figure}
  \subfloat[Partial sum of the irreducible vertex $\Gamma^{(1)}$]{%
	\begin{minipage}[c][1\width]{%
	   0.45\textwidth}
	   \centering%
	   \includegraphics[width=1.0\textwidth]{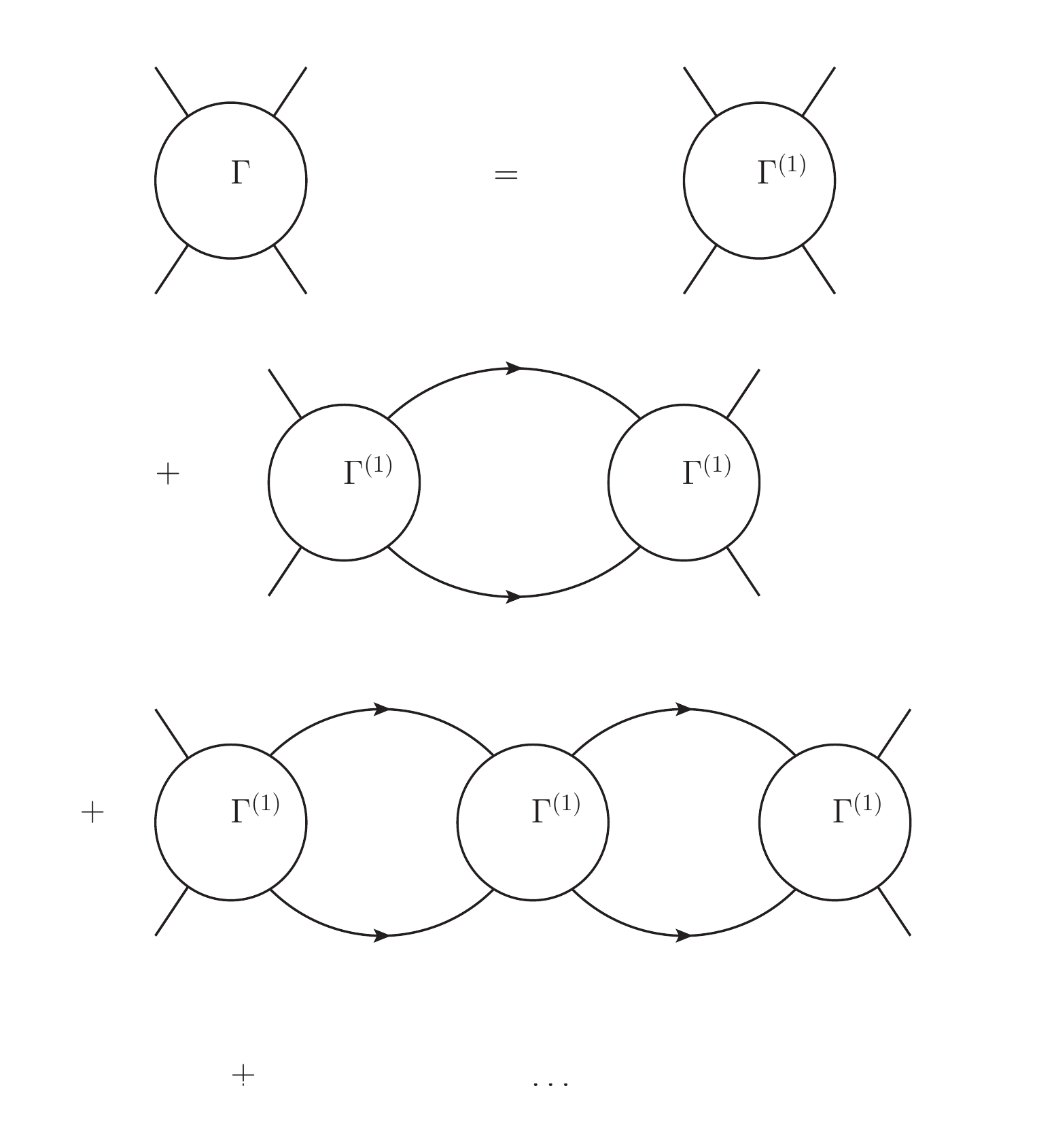}
	\end{minipage}}
	\qquad
  \subfloat[Complete sum of all the diagrams, defining an integral equation ]{%
	\begin{minipage}[c][1\width]{%
	   0.45\textwidth}
	   \centering%
	   \includegraphics[width=1.0\textwidth]{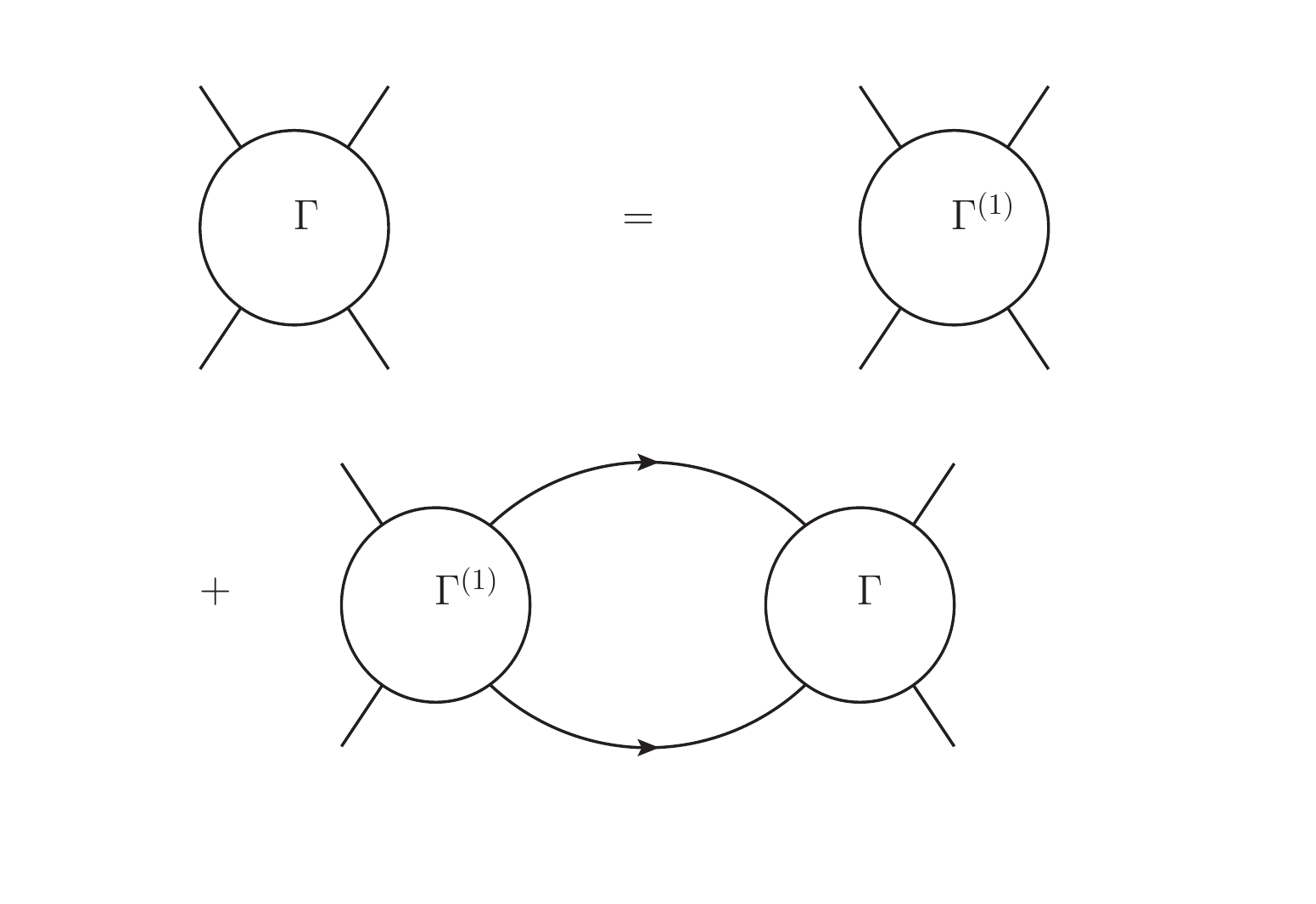}
	\end{minipage}}
\caption[Generalised Bethe Salpeter equation]{Diagrammatic representation of the generalised Bethe Salpeter equation}
\label{fi:scatampl}
\end{figure}

The diagrammatic representation of the so called generalised Bethe-Salpeter equation
is shown in Fig.~\ref{fi:scatampl}, where  the vertex $\Gamma$,  depicting the complete sum of  all  linked and topologically distinct 
diagrams contributing to the scattering amplitude, is obtained in terms of the irreducible vertex $\Gamma^{(1)}$, representing the sum of all diagrams that cannot be broken into two parts by cutting two internal lines.

The physical scattering amplitude  is directly related to the vertex $\Gamma$ through a volume factor and  the renormalisation constant of quasiparticle states.
The final integral equation for the scattering amplitude involves a generalization of Landau's $\mathpzc{f}$-function, suitable to describe processes taking place at arbitrary momentum transfer. 
Solving this integral equation, written in terms of the scattering amplitude  and the generalized function $\mathpzc{f}$,  involves severe difficulties. However, 
some explicit  solutions can be found for systems at  zero temperature.

\section{The  low-denstity hard-sphere system}

In Ref.~\cite{JLTP1}
the generalised  Bethe-Salpeter equation  has been solved for a low-density Fermi Gas of hard spheres at $T= 0$.
The solution includes corrections\textemdash linear in the parameter $c = k_F a $ \textemdash to the  free  two-particle scattering amplitude, which is in turn  related to the  $S$-wave scattering length $a$.
The resulting scattering amplitude  has been obtained for any values of the momenta, for particles in triplet (spin-symmetric, $\mathcal{A}_T$)    and singlet (spin-antisymmetric, $\mathcal{A}_S$) states. 

The transport coefficients have been derived in Ref.~\cite{JLTP2} using the same formalism. 
Here, we only outline the evaluation of the amplitude associated with scattering between particles on the Fermi surface, because this is the kinematical setup relevant for the  determination of transport properties.   

As pointed out in Chapter~\ref{chap3}, when all four momenta are on the Fermi surface, the variables that play a role in the description of the system are the two angles $\theta$ and $\phi$ defined in  Section~\ref{sec:scatprob}.
The  scattering probability is then given by
\begin{align}
\label{eq:JLTP_proba}
W (\theta , \phi ) = \frac{\pi}{ 16 \hbar }  \left \{   3 \left [ \mathcal{A}_T  (\theta, \phi) \right]^2 +  \left [ \mathcal{A}_S (\theta, \phi) \right]^2  \right \} \ ,
\end{align}
with the scattering amplitudes for the triplet and the singlet states given by 
\begin{align}
\label{eq:triplet}
\mathcal{A}_{T}  (\theta , \phi ) = 	\frac{4 \hbar^2 a  \ c }{m} \left [ U(\theta, \phi) - U (\theta, \phi + \pi) \right ] \ ,
\end{align}
\begin{align}
\label{eq:singlet}
\mathcal{A}_{S}  (\theta , \phi ) =  &  \	\frac{16 \pi \hbar^2 a }{m}  \  \left\{  1 +  \frac{c}{\pi} \left[  3 - \sin \frac{\theta}{2} \log \left |  \frac{1+ \sin \frac{\theta}{2 } }{1 -\sin \frac{\theta}{2 } } \right |   \right. \right. \\
\nonumber
&  \ \ \ \ \ \ \ \ \ \ \ \ \ \ \ \ \ +    \left.   \left. \frac{1}{4} U (\theta, \phi)  + \frac{1}{4} U (\theta, \phi + \pi) \right ]   \right \}  \  . 
\end{align}
The  function  $ U  (\theta , \phi ) $ appearing in the above equations reads
$$
U  (\theta , \phi ) =  \left [    \frac {  1 - \sin^2 \frac { \theta} {2}   \cos^2 \frac { \phi} {2}   }    {   \sin \frac { \theta} {2}   \cos \frac { \phi} {2}    }   \right ]  \log \left |  \frac{1+ \sin \frac{\theta}{2 } }{1 -\sin \frac{\theta}{2 } } \right |       \ . 
$$

Note that the singlet-state $\mathcal{A}_{S}$ features a contribution linear in the hard-core radius $a$, corrected by terms of order $c= k_F a$, while the corresponding triplet-state quantity, $\mathcal{A}_{S}$, 
does not exhibit  a contribution of zero order in $c$. This is consistent with the results of the partial-wave analysis, showing that the amplitude associated with the $\ell$-th partial wave  is proportional to $(ka)^{\ell+1}$, 
k being the momentum transfer. As a consequence, the $a$-dependence in  triplet states must be  at least quadratic.

The final expression for the scattering probability  for the hard-sphere system,  including corrections of order $c$, 
reads
\begin{align}
\label{eq:dgsfgW_oc2}
W (\theta , \phi ) =  
\frac{16 \pi^3 \hbar^3 a^2 }{m^2}  
 \left \{ 1 \frac{}{}   \right. &  + \left.    \frac{2 c}{	\pi}   \left[  3 - \sin \frac{\theta}{2} \log \left |  \frac{1+ \sin \frac{\theta}{2 } }{1 -\sin \frac{\theta}{2 } } \right |    \right. \right.  \nonumber \\
& +  \left . \left .   \frac{1}{4} U (\theta, \phi)  + \frac{1}{4} U (\theta, \phi + \pi) \right ]     + O (c)^2   \right \}   \ .
\end{align}

The  angular averages of the scattering probability, required for the calculation of the transport coefficients,
can be evaluate in closed form. The resulting expressions are  
\begin{align}
\label{eq:JLTP_wav}
\langle W \rangle = 
 \frac{32 \pi^3 \hbar^3 a^2}{m^2} \left [ 1 + \frac{c}{\pi}   \left ( 3 + \frac{\pi ^2} {4} - 4 G \right )  + O(c)^2 \right ] \ , 
 \end{align}
\begin{align}
\label{eq:JLTP_wav2}
\left \langle W \left [ (1 - \cos \theta) ^2 \sin\phi^2 \phi \right]  \right \rangle =
 \frac{512 \pi^3 \hbar^3 a^2}{15 m^2} \left [ 1 + \frac{c}{\pi}   \left ( \frac{111}{32} - \frac{75} {16} G  \right )  + O(c)^2 \right ] \ ,
 \end{align}
where $G$ is the Catalan's constant $G = 0.915996$ \cite{JLTP2}.

\backmatter

\cleardoublepage
\addcontentsline{toc}{chapter}{\listfigurename} 
\listoffigures

\cleardoublepage
\addcontentsline{toc}{chapter}{\bibname}
\bibliographystyle{babunsrt-fl}
\bibliography{bibliothesis}

\end{document}